\documentclass[nohyper,11pt,letterpaper]{JHEP3}
\usepackage[dvips]{epsfig}
\usepackage{amsmath}

\newcommand{\be}{\begin{equation}}
\newcommand{\ee}{\end{equation}}
\newcommand{\bea}{\begin{eqnarray}}
\newcommand{\eea}{\end{eqnarray}}
\newcommand{\ba}{\begin{eqnarray}}
\newcommand{\ea}{\end{eqnarray}}
\newcommand{\nn}{\nonumber \\}
\newcommand{\eqn}[1]{(\ref{#1})}
\newcommand{\beq}{\begin{equation}}
\newcommand{\eeq}{\end{equation}}
\newcommand{\beqa}{\begin{eqnarray}}
\newcommand{\eeqa}{\end{eqnarray}}
\newcommand{\beqar}{\begin{eqnarray*}}
\newcommand{\eeqar}{\end{eqnarray*}}

\newcommand{\reef}[1]{(\ref{#1})}
\newcommand{\ssc}{\scriptscriptstyle}
\newcommand{\eg}{{\it e.g.,}\ }
\newcommand{\ie}{{\it i.e.,}\ }

\newcommand{\tk}{\tilde{\bf{k}}}
\newcommand{\tom}{\tilde{\omega}}
\newcommand{\rhov}{\rho_v}
\newcommand{\dr}{\delta R}
\newcommand{\dphi}{\delta \phi}
\newcommand{\rv}{R_v}
\newcommand{\drv}{\dot{R}_{v}}
\def\nc {N_\mt{c}}
\def\nf {N_\mt{f}}
\def\ua {U(1)_\mt{A}}
\def\t6 {T_\mt{D6}}

\def\gym {g_\mt{YM}}

\newcommand{\te}{t_\mt{E}}
\newcommand{\td}{T_\mt{deconf}}
\newcommand{\tf}{T_\mt{fun}}
\newcommand{\leff}{g_\mt{eff}}

\newcommand{\mq}{M_\mt{q}}      % Quark mass
\newcommand{\mqs}{M_\mt{q}^*}      % Critical quark mass
\newcommand{\qc}{\langle \bar{\psi} \psi \rangle} % Quark condensate
\newcommand{\qcs}{\qc^*} %Critical quark condensate
\newcommand{\tq}{T_\mt{Dq}}      % Dq tension.
\newcommand{\R}{L} % AdS radius
\newcommand{\Y}{y} % Near-horizon sphere radius
\newcommand{\Z}{z} % Near-horizon distance to the horizon
\newcommand{\kk}{\mu} % Rescaling factor
\newcommand{\N}{{\cal N}} % Normalization factor
\newcommand{\mbar}{\bar{M}}
\newcommand{\rhomin}{{\rho_\mt{min}}}
\newcommand{\rhomax}{{\rho_\mt{max}}}
\newcommand\ict{I_{\mt{bound}}}
\newcommand{\gs}{g_\mt{s}}
\newcommand{\ls}{\ell_\mt{s}}
\newcommand{\ibk}{I_\mt{Dq}}
\newcommand{\ids}{I_\mt{D7}}

\newcommand{\ibulk}{I_\mt{bulk}}

\newcommand{\mt}[1]{\textrm{\tiny #1}}

\newcommand{\ieu}{I_\mt{E}}
\newcommand{\mc}{M_\mt{c}} %Constituent quark massnewcommand
\newcommand{\tildef}{\tilde{f}}

\newcommand{\fct}{F_{\ssc{\rm bound}}}

\newcommand{\trho}{\tilde{\rho}}  %dimensionful rho ie \tilde{rho}
\newcommand{\trhomax}{\tilde{\rho}_\mt{max}} %dimensionful rho max
\newcommand{\trhomin}{\tilde{\rho}_\mt{min}}  %dimensionful rho min
\newcommand{\tm}{\tilde{m}}  %dimensionful m ie m u_0
\newcommand{\tc}{\tilde{c}}  %dimensionful c ie c u_0^3
\newcommand{\om}{{u_0}}  %position of the horizon
\newcommand{\pmin}{\rho_\mt{min}}
\newcommand{\pmax}{\rho_\mt{max}}

\def\sac{\, , \,\,\,\,\,}

\newcommand{\al}{\alpha}
\newcommand{\lam}{\lambda}
\newcommand{\ga}{\gamma}

\newcommand{\labell}[1]{\label{#1}} %\qquad_{#1}}

\newcommand{\prt}{\partial}

\newcommand{\cv}{{c_\mt{V}}}
\newcommand{\vep}{\varepsilon}
\newcommand{\cc}{\langle \bar{\psi} \psi \rangle}

\newcommand{\tiL}{\tilde{L}}

\title{Thermodynamics of the brane}

\author{David Mateos,$^a$ Robert C. Myers$^{b,c,d}$ and Rowan M. Thomson$^{c,d}$\\
$^a$ Department of Physics, University of California, Santa Barbara, CA 93106-9530, USA\\
$^b$ Kavli Institute for Theoretical Physics, University of California, Santa Barbara, CA\\
\ \ 93106-4030, USA\\
$^c$ Perimeter Institute for Theoretical Physics,
Waterloo, Ontario N2L 2Y5, Canada \\
$^d$ Department of Physics and Astronomy, University of Waterloo,
Waterloo, Ontario\\
\ \ N2L 3G1, Canada\\

\\E-mail: \email{dmateos@physics.ucsb.edu,rmyers@perimeterinstitute.ca,
rthomson@perimeterinstitute.ca }}

\abstract{The holographic dual of a finite-temperature gauge theory
with a small number of flavours typically contains D-brane probes in
a black hole background. We have recently shown that these systems
undergo a first order phase transition characterised by a `melting'
of the mesons. Here we extend our analysis of the thermodynamics of
these systems by computing their free energy, entropy and energy
densities, as well as the speed of sound. We also compute the meson
spectrum for brane embeddings outside the horizon and find that
tachyonic modes appear where this phase is expected to be unstable
from thermodynamic considerations.}


\preprint{}

\begin{document}{\vskip 1cm}

%%%%%%%%%%%%%%%%%%%%%%%%%%%%%%%%%%%
%%%%%%%%%%%%%%%%%%%%%%%%%%%%%%%%%%%
%%%%%%%%%%%% INTRODUCTION %%%%%%%%%
%%%%%%%%%%%%%%%%%%%%%%%%%%%%%%%%%%%

\section{Introduction}
In a broad class of large-$\nc$, strongly coupled gauge theories
with a holographic dual, a small number of flavours of fundamental
matter, $\nf \ll \nc$, may be described by $\nf$ probe Dq-branes in
the gravitational background of $\nc$ black Dp-branes
\cite{flavour}. At a sufficiently high temperature $T$, the
background geometry contains a black hole \cite{witten}. It was
recently shown that these systems generally undergo a universal
first order phase transition characterised by a change in the
behaviour of the fundamental matter \cite{prl}.\footnote{Specific
examples of this transition were originally seen in
\cite{johanna,us} and aspects of these transitions in the D3/D7
system were independently studied in \cite{recent,recent2}.
Recently, similar holographic transitions have also appeared in a
slightly different framework \cite{recent8}.}

\FIGURE{
\includegraphics[width=0.99 \textwidth]{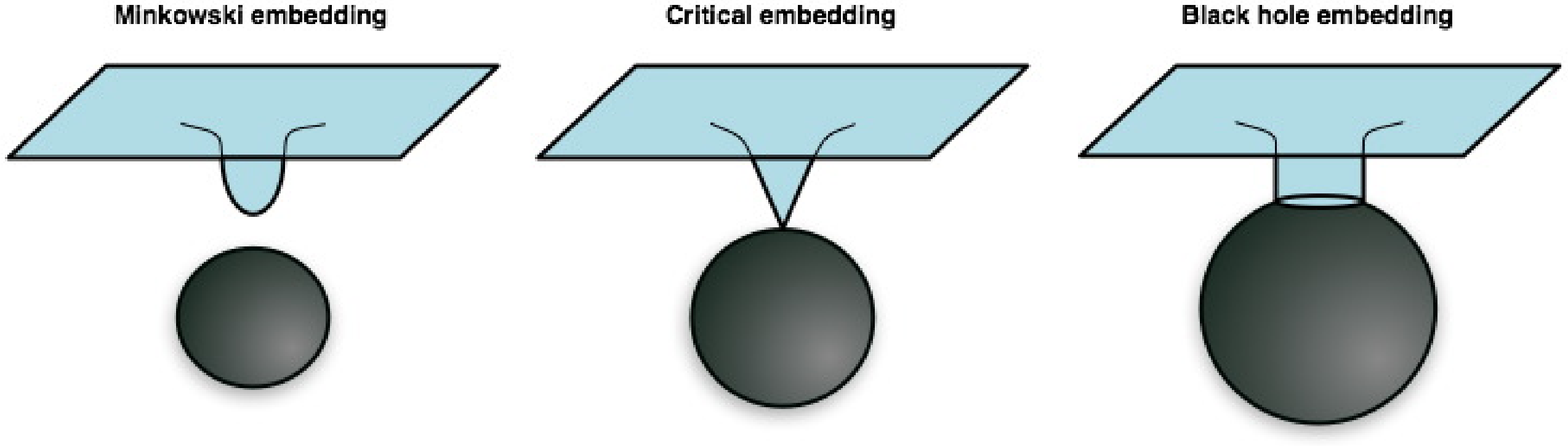}
\caption{Various Dq-brane configurations in a black Dp-brane
background with increasing temperature from left to right. For
low temperatures, the probe branes close off smoothly above
the horizon. For high temperatures, the branes fall through the
event horizon. In between, a critical solution exists in which the
branes just `touch' the horizon at a point.}
\label{embeddings}}
From the viewpoint of the holographic description, the basic physics
behind this transition is easily understood. Increasing the
temperature increases both the radial position and the energy
density of the event horizon in the Dp-brane throat. For a
sufficiently small temperature or a sufficiently large separation
for the Dq-branes, the probe branes are gravitationally attracted
towards the horizon but their tension is sufficient to balance this
attractive force. The probe branes then lie entirely outside of the
black hole in what we call a `Minkowski' embedding (see
fig.~\ref{embeddings}). However, above a critical temperature $\tf$,
the gravitational force overcomes the tension and the branes are
pulled into the horzion. We refer to such configurations where the
branes fall through the horizon as `black hole' embeddings. In
between these two phases, there exists a critical solution which
just `touches' the horizon. In \cite{prl}, we showed that in the
vicinity of this critical solution the embeddings show a
self-similar behaviour. As a result, multiple solutions of the
embedding equations exist for given temperature in a regime close to
$\tf$. Using thermodynamic considerations to select the true ground
state then reveals a first order phase transition at $\tf$, where
the probe branes jump discontinuously from a Minkowski to a black
hole embedding.

In the dual field theory,\footnote{Recall for these supersymmetric
field theories, the fundamental matter includes both fermions and
scalars, which we will refer to collectively as `quarks'.} this
phase transition is exemplified by discontinuities in, \eg the quark
condensate $\cc$ or the contribution of the fundamental matter to
the energy density. However, the most striking feature of this phase
transition is found in the spectrum of the mesons, \ie the
quark-antiquark bound states. The latter correspond to excitations
supported on the probe branes -- see, \eg
\cite{us-meson,holomeson,holomeson2}. In the low-temperature or
Minkowski phase, the mesons are stable (to leading order within the
approximations of large $\nc$ and strong coupling) and the spectrum
is discrete with a finite mass gap. In the high-temperature or black
hole phase, stable mesons cease to exist. Rather one finds a
continuous and gapless spectrum of excitations \cite{melt,spectre}.
Hence the first order phase transition is characterised by the
dissociation or `melting' of the mesons.

This physics is particularly interesting in theories that exhibit a
confinement/deconfinement phase transition. The dual description of
the confining, low-temperature phase  involves a horizon-free
background. At a temperature $\td$ the theory undergoes a phase
transition at which the gluons and the adjoint matter become
deconfined, at which point the dual background develops a black hole
horizon \cite{witten}. However, if the mass of the fundamental
matter is large enough, the branes remain outside the horizon and
therefore mesonic bound states survive for temperatures $\td < T <
\tf$. At $T=\tf$ the branes finally fall into the horizon, \ie the
mesons  melt. This physics is in qualitative agreement with that
observed in QCD for heavy-quark mesonic bound states. For example,
lattice calculations suggest that charmonium states such as the
$J/\psi$ meson melt at temperatures between $1.65 \td$
%RCM wording on Tc below changed
\cite{ccbar,summary} and $2.1 \td$ \cite{newcc}, while lattice
results for the QCD deconfinement temperature are in the range: $\td
\simeq 151$ to $192$ MeV \cite{deconfT}. Although the holographic
description may provide some useful geometric intuition for this
phenomenon, there are also some caveats that we will discuss in due
course.

An overview of the paper is as follows:  In section \ref{black}, we
review the throat geometries for black Dp-branes which are dual to
$(p+1)$-dimensional super-Yang-Mills (SYM) at finite temperature
\cite{itz}. Section \ref{self} reviews and expands on the
self-similar behaviour of the embeddings near the critical solution
for general Dp/Dq systems, as originally presented in \cite{prl}. In
the subsequent detailed discussion of the thermodynamics, we focus
our attention on the D3/D7 \cite{johanna} and D4/D6 \cite{us} cases
for concreteness. In section \ref{D3D7phase}, we compute the free
energy, entropy and energy densities, as well as the speed of sound
for the D3/D7 system. We also study the meson spectrum on the
Minkowski embeddings in this section. This spectrum is related to
the dynamical stability, or lack thereof, of this phase, as we find
that tachyonic modes appear where thermodynamic considerations
indicate that these embeddings are unstable. Section \ref{D4D6phase}
repeats the salient calculations for the D4/D6 system. Then section
\ref{discuss} concludes with a discussion of results. Finally there
are several appendices containing various technical details.
Appendix \ref{approxSol} provides an analytic description of the
D7-brane embeddings at very high and very low temperatures. Then
appendix \ref{entropy7} presents some of the details of the
calculation of the entropy density contributed by the D7-branes.
Appendix \ref{kinks} discusses the appearance of the `swallow tail'
form in the plots of the free energy, \eg fig.~\ref{freeEnD3D7}.
Appendix \ref{constitution} provides a calculation of the
constituent quark mass in the low temperature phase of the
fundamental matter. Finally, appendix \ref{holo4} discusses the
holographic renormalization of the D4-brane background.

\section{Black Dp-branes}\label{black}

In this section we briefly review the relevant aspects of the throat
geometries and thermodynamics of black Dp-branes. This will be of
use in subsequent sections, in particular, in sections
\ref{D3D7phase} and \ref{D4D6phase}, where we specialise to black D3-
and D4-brane backgrounds, respectively.

\subsection{Supergravity Background}

The supergravity solution corresponding to the decoupling limit of
$\nc$ coincident black Dp-branes is, in the string frame (see, \eg
\cite{johnson} and references therein),
\beqa
 ds^2 &=& H^{-\frac{1}{2}} \left( -f
dt^2 +  dx_{\it p}^2 \right) + H^{\frac{1}{2}} \left( \frac{du^2}{f}
+ u^2 d\Omega_{\it 8-p}^2 \right) \,, \nonumber\\
e^\Phi &=&  H^{\frac{3-p}{4}} \sac\qquad  C_{01\ldots p} = H^{-1}
\,, \labell{metric} \eeqa
where $H(u)= (\R/u)^{7-p}$ and $f(u) =1-(u_0/u)^{7-p}$. The horizon
lies at $u=u_0$.  The length scale $\R$ is defined in terms of the
string coupling constant $\gs$ and the string length $\ell_s$:
\beq \R^{7-p}=  \gs \nc\, (4\pi
\ell_s^2)^{\frac{7-p}{2}}\,\Gamma\!\left({\scriptstyle
\frac{7-p}{2}}\right)/4\pi\ \, . \labell{L} \eeq
For the special case $p=3$, $\R$ is the radius of curvature for the
AdS$_5\times S^5$ geometry appearing in eq.~\reef{metric}.

According to the general gauge/gravity duality of \cite{itz}, type
II string theory in these backgrounds is dual to the super-Yang-Mills
$SU(\nc)$ gauge theory on the $(p+1)$-dimensional worldvolume of the
Dp-branes. For general $p$ ($\ne3$), the gauge theory is distinguished
from the conformal case $(p=3)$ by the fact that the Yang-Mills
coupling $\gym$ is dimensionful. The holographic dictionary provides
\beq  \gym^2 = 2\pi \gs (2\pi\ls)^{p-3}\,. \labell{gym}\eeq
Hence there is a power-law running of the dimensionless
effective coupling with the energy scale $U$:
\beq \leff^2 = {\gym^2\nc}\,U^{p-3} \,,\labell{couple} \eeq
where $U=u/\al'$ by virtue of  the usual energy/radius
correspondence. The absence of conformal invariance for the general
case is manifested in the dual geometry by the radial variation of
both the string coupling and the spacetime curvature. The
supergravity solution \reef{metric} is a trustworthy background
provided that both the curvatures and string coupling are small.
Hence in these general dualities, the supergravity description is
limited to an intermediate regime of energies in the field theory or
of radial distances in the background. This restriction is
succinctly expressed in terms of the effective coupling
\reef{couple} as \cite{itz}:
\beq 1\ll \leff \ll \nc^{\frac{4}{7-p}}\,.\labell{range}\eeq
Hence the field theory is always strongly coupled where the dual
supergravity description is valid.

With the event horizon at $u=u_0$, Hawking radiation appears in the
background with a temperature fixed by the surface gravity
$T=\kappa/2\pi$. This temperature is identified with that of the
dual $(p+1)$-dimensional gauge theory. In the geometry \eqn{metric},
the temperature can also be determined by demanding regularity of
the Euclidean section obtained through the Wick rotation $t
\rightarrow i\te$. Then $\te$ must be periodically identified with a
period $\beta$ where
\beq \frac{1}{\beta}= T = \frac{7-p}{4\pi  \R} \left( \frac{u_0}{\R}
\right)^{\frac{5-p}{2}} \,. \labell{beta} \ee

In some cases, one periodically identifies some of the Poincar\'e
directions $x_{\it p}$ in order to render the theory effectively
lower-dimensional at low energies; a prototypical example is that of
a D4-brane with one compact space direction -- see, \eg
\cite{witten, us}. Under these circumstances a different background
with no black hole may describe the low-temperature physics, and a
phase transition at $T=\td$ may occur \cite{witten}. In the gauge
theory this is typically a confinement/deconfinement phase
transition for the gluonic (or adjoint) degrees of freedom.
Throughout this paper we assume that $T>\td$, in which case the
appropriate gravitational background has an event horizon, as in
eq.~\eqn{metric}.

\subsection{Thermodynamics}\label{house}

Now as alluded to above, with the Wick rotation $t \rightarrow
i\te$, the Euclidean path integral yields a thermal partition
function. Further the Euclidean black hole is interpreted as a
saddle-point in this path integral and so the gravity action
evaluated for this classical solution is interpreted as the leading
contribution to the free energy, \ie $I_\mt{E}=\beta F$ -- see, \eg
\cite{hawk}. Hence to study the gauge theory thermodynamics
holographically, one needs to evaluate the supergravity action
$\ieu$ for the Euclidean version of the above backgrounds
\reef{metric}. This suffers from IR (large radius) divergences, but
these may be regulated by adding appropriate boundary terms to the
action. These boundary terms were originally found for
asymptotically AdS backgrounds, such as the black D3-brane, in
\cite{ct,ct1}. As we discuss in appendix \ref{holo4}, similar
surface terms should exist in the general gauge/gravity dualities to
complete the holographic description. Here we simply comment that
for the black D4-brane, which is the relevant background in section
\ref{D4D6phase}, we are guided in the construction of these
counterterms by considering the M5-brane counterpart in M-theory. In
any event, after including the appropriate boundary terms, the
Euclidean action is finite.\footnote{For the above backgrounds
\reef{metric} describing the gauge theory on flat $p$-dimensional
space, the action still contains an IR divergence, namely a factor
of the spatial volume $\widetilde{V}_x = \int d^p x$. In the
following, we divide all extensive thermodynamic quantities by
$\widetilde{V}_x$ so that we are really looking at densities, \eg
eq.~\reef{Fbulk} really gives the free energy per unit $p$-volume.
When we refer to contributions from the brane probes, the relevant
volume factor is instead that of the defect on which the fundamental
matter lives, $V_x = \int d^dx$.\label{foot1}} Then with $F= T
I_\mt{E}$ and standard thermodynamic relations, various thermal
quantities can be determined. For example, the entropy $S$ and the
energy $E$ are computed as:
\be S= - \frac{\prt F}{\prt T}  \sac\qquad E = F + T S \,.
\labell{thermiden}\ee

For the black D3-brane background, the length scale \eqn{L} is given by
$L^4=4\pi g_s \nc \ell_s^4$, and the free energy is
\beq F = -\frac{\pi^6 L^8}{16 G}\,T^4=-\frac{\pi^2}{8}\nc^2\,T^4 \,,
\labell{free3}\eeq
where $G$ is the ten-dimensional Newton's constant. In terms of the
string length and coupling, the latter is given by:
\be 16 \pi G = (2\pi)^7 \ls^8\, g_s^2 \,. \labell{NewtG}\ee
For the black D4-brane geometry we have $L^3=\pi g_s \nc \ell_s^3$ and
\beq F = -\frac{2^{10} \pi^7 L^9}{3^7 G}\,T^6
=-\frac{2^5\pi^2}{3^7}\,\lam\, \nc^2\,T^6\,, \labell{free4}\eeq
where as usual $\lam=\gym^2\nc$ denotes the 't Hooft coupling. (The
reader is referred to appendix \ref{holo4} for further discussion of
this case.) In general, the free energy for a general black Dp-brane
geometry can be written as \cite{itz,polpee}
\beq F \sim \nc^2 T^{p+1} \leff(T)^{\frac{2(p-3)}{5-p}} \, ,
\labell{Fbulk} \eeq
where
\be \leff ^2(T) = \lam T^{p-3} = \gym^2 \nc T^{p-3} \labell{geff}
\ee
is the effective coupling \reef{couple} evaluated at the temperature
scale $U=T$. In eq.~\reef{Fbulk}, $\nc^2$ reflects the number of
degrees of freedom in the $SU(\nc)$ gauge theory while $T^{p+1}$ is
the expected temperature dependence for a $(p+1)$-dimensional theory.
However, the dependence on $\leff$ is a prediction of the holographic
framework for the strongly coupled gauge theory. Note that for the
conformal case ($p=3$), but only for this case, this factor is simply unity
and so the thermodynamic results can compared to those
calculated at weak coupling \cite{3/4}.

Another quantity that is often studied in the context gauge/gravity
duality is the speed of sound, \eg
\cite{ss1,ss2M,ss2D4,ss3deviate,ss4}. While this quantity can be
inferred from the pole structure of certain correlators
\cite{ss1,ss2M}, it can also be derived from the
thermal quantities discussed above, with
\beq v_s^2=\frac{\prt P}{\prt E}=\frac{\prt P}{\prt T}\,\left(
\frac{\prt E}{\prt T}\right)^{-1}=\frac{S}{\cv}\,.\labell{deaf}\eeq
Here we have used the fact that for a system without a chemical
potential, the pressure and free energy density are identical up to
a sign, \ie $P=-F$. Hence $\prt P/\prt T=-\prt F/\prt T=S$. Also we
use $\cv$ to denote the heat capacity (density), \ie $\cv\equiv\prt
E/\prt T$. From eqs.~\reef{Fbulk} and \reef{geff}, one finds the
simple result that for the strongly coupled gauge theory in $(p+1)$
dimensions
\beq v_s^2=\frac{5-p}{9-p}=\left\lbrace
 \begin{matrix}
  \ 1/3&\ \ {\rm for}\ p=3\,,\cr
  \ 1/5&\ \ {\rm for}\ p=4\,.\cr
 \end{matrix}
\right. \labell{speeed}\eeq
We see above that the conformal result $v_s^2=1/p$ is only achieved
for $p=3$ \cite{ss1,ss2M}, as expected. We note, however, that
the $p=1$ and 4 backgrounds are related through a simple chain of
dualities to the AdS$_4$ and AdS$_7$ throats of M2- and M5-branes,
respectively. Hence for these specific cases with $v_s^2=1/2$ and
1/5, the speed of sound reflects the conformal nature of the
holographic theories dual to these M-theory backgrounds \cite{ss2M}.

%%%%%%%%%%%%%%%%%%%%%%%%%%%%%%%%%%%%%%%%%%%%%%%%%%%%%%%%%%%%%%%%%%%%%%%%%
%%%%%%%%%%%%%%%%%%%%%%%%%%%%%%%%%%%%%%%%%%%%%%%%%%%%%%%%%%%%%%%%%%%%%%%%%

\section{Criticality, scaling, and phase transitions in Dp/Dq
systems}\label{self}

We now turn to the systems of interest in this paper: Configurations
of probe Dq-branes in the backgrounds of black Dp-branes.  The
addition of the probes in the gravitational description is dual to
the addition of matter in the fundamental representation in the
gauge theory \cite{flavour}. This section is mainly a review of
\cite{prl} that includes some details that were omitted in that
reference. We describe the embedding of the Dq-brane, study the
critical behaviour and analyse the nature of the phase transition
for general $p$ and $q$. The latter involves extending the Euclidean
techniques of the previous section to the worldvolume action of the
Dq-brane, to study the thermal properties of the fundamental matter.
This discussion naturally leads to sections \ref{D3D7phase} and
\ref{D4D6phase}, where we provide a detailed analysis of the D3/D7
and D4/D6 brane systems.

\subsection{Dp/Dq brane intersections}

Consider a configuration of $\nc$ coincident black Dp-branes
intersecting $\nf$ coincident Dq-branes along $d$ spacelike
directions. In the limit $\nf \ll \nc$ the Dq-branes may be treated
as a probe in the Dp-brane geometry  \reef{metric}, wrapping an
$S^n$ inside the $S^{8-p}$. We will assume that the Dq-brane also
extends along the radial direction, so that $q=d+n+1$.  The
corresponding gauge theory now contains fundamental matter
propagating along a $(d+1)$-dimensional defect. To ensure stability,
we will consider Dp/Dq intersections which are supersymmetric at
zero temperature. Generally this means that we are interested in
$q=p+4,$ $p+2$ or $p$, as studied in \cite{holomeson,holomeson2}. In
this case, the Ramond-Ramond field sourced by the Dp-branes does not
couple to the Dq-brane. For the two cases of special interest here,
the D3/D7 and the D4/D6 systems, one has $n=3$ and $n=2$
respectively. If the appropriate direction along the D4-brane is
compactified, then both cases can effectively be thought of as
describing the dynamics of a four-dimensional gauge theory with
fundamental matter.

\subsection{Critical behaviour}\label{critikue}

To uncover the critical behaviour of the Dp/Dq brane system, we
study the behaviour of the probe brane near the horizon, following
\cite{frolovnew} closely -- see also \cite{frolov}. First it is
useful to adapt the $S^{8-p}$ metric in \reef{metric} to the probe
brane embedding, and so we write
\be d\Omega_{\it 8-p}^2 = d\theta^2 + \sin^2 \theta \, d\Omega_{\it
n}^2 + \cos^2 \theta \, d\Omega_{\it 7-p-n}^2  \,.
 \labell{sphere8}\ee
As described above, the Dq-brane wraps the internal $S^n$ with
radius $\sin\theta$ in this line element. Now we zoom in on the near
horizon geometry with the coordinates
\beq u = u_0 +  \pi T \Z^2 \sac \theta = \frac{\Y}{\R}  \left(
\frac{\R}{u_0} \right)^\frac{p-3}{4} \sac \tilde{x} = \left(
\frac{u_0}{\R} \right)^\frac{7-p}{4} x
 \,, \labell{coord} \eeq
with $T$ the temperature defined in \reef{beta}. With these
coordinates, the event horizon is at $\Z=0$. Further $\Y=0$ denotes
the axis running orthogonally to the Dq-brane from the Dp-branes.
Expanding the metric \reef{metric} to lowest order in $\Z$ and $\Y$
gives Rindler space together with some spectator directions:
\be ds^2 = - (2\pi T)^2 \Z^2 dt^2 + d\Z^2 + d\Y^2  + \Y^2
d\Omega_{\it n}^2 + d\tilde{x}_{\it d}^2 + \cdots \,.
\labell{nearGeom}\ee
The Dq-brane lies at constant values of the omitted coordinates, so
these play no role in the following. The Dq-brane embedding is
specified by a curve $(\Z(\sigma), \Y(\sigma))$ in the
$(\Z,\Y)$-plane. Since the dilaton approaches a constant near the
horizon, up to an overall constant the Dq-brane (Euclidean) action
is simply the volume of the brane, namely
\beq
 \ibulk \propto \int d\sigma
\sqrt{\dot{\Z}^2 + \dot{\Y}^2} \, \Z \Y^n \,,
\eeq
where the dot denotes differentiation with respect to $\sigma$ and
the reason for the subscript `bulk' will become clear shortly. This
is precisely the action considered in ref.~\cite{frolovnew}. In the
gauge $\Z=\sigma$ the equation of motion takes the form
\beq
\Z \Y
\ddot{\Y} + (\Y \dot{\Y} - n \Z) (1 + \dot{\Y}^2) = 0 \,, \labell{eom}
\eeq
while the gauge choice $\Y = \sigma$ yields
\beq
\Y \Z
\ddot{\Z} + (n\Z \dot{\Z} -  \Y) (1 + \dot{\Z}^2) = 0 \,. \labell{eom2}
\eeq

The two types of embeddings described in the introduction for the
full background extend to this near-horizon geometry
\reef{nearGeom}. Hence the solutions again fall into two classes:
`black hole' and `Minkowski' embeddings -- see fig.
\ref{embeddings}. Black hole embeddings are those for which the
brane falls into the horizon, and may be characterised by $\Y_0$,
the size of the $S^n$ there, which is also the size of the induced
horizon on the Dq-brane worldvolume. The appropriate boundary
condition is $\dot{\Y} = 0 , \Y = \Y_0$ at $\Z=0$. Minkowski
embeddings are those for which the brane closes off smoothly above
the horizon. These are characterised by the distance of closest
approach to the horizon, $\Z_0$, and satisfy the boundary condition
$\dot{\Z} =0, \Z = \Z_0$ at $\Y=0$. There is a simple limiting
solution for the equations of motion \reef{eom}: $\Y= \sqrt{n}
\,\Z$. This critical solution just touches the horizon at the point
$\Y=\Z=0$, and so it lies between the above two classes. Note that
this point is a singularity in the induced metric of the Dq-brane.

The equation of motion \eqn{eom} enjoys a scaling symmetry: If
$\Y=f(\Z)$ is a solution, then so is $\Y=f(\kk \Z)/ \kk$ for any
real positive $\kk$. This transformation rescales $\Z_0\rightarrow
\Z_0/ \kk$ for Minkowski embeddings, or $\Y_0 \rightarrow \Y_0/\kk$
for black hole embeddings, which implies that all solutions of a
given type can be generated from any other one by this scaling
transformation.

Consider now a solution very close to the critical one, $\Y(\Z) =
\sqrt{n} \, \Z + \xi (\Z)$. Linearising the equation of motion
\eqn{eom}, one finds that for large $\Z$ the solutions are of the
form $\xi (\Z) = \Z^{\nu_\pm}$, with
\be \nu_\pm = - \frac{n}{2} \pm \frac{\sqrt{n^2 - 4(n+1)}}{2} \,.
\labell{exponents} \ee
If $n\leq 4$, these exponents have non-vanishing imaginary parts,
which leads to oscillatory behaviour. It appears that one can also
get real exponents with $n \geq 5$. However, we will show below that
no such systems are realized in superstring theory. Hence we will
only work with $n\leq 4$ in the following. In this case it is
convenient to write the general solution as
\be \Y =  \sqrt{n} \,
\Z + \frac{T^{-1}}{(T\Z)^{\frac{n}{2}}} \Big[ a \sin (\alpha \log T
\Z) + b \cos (\alpha \log T \Z) \Big] \,,
\ee
where $\alpha=\sqrt{4(n+1)-n^2}/2$ and $a,b$ are dimensionless
constants determined by $\Z_0$ or $\Y_0$. It is easy to show that
under the rescaling discussed above, these constants transform as
\be
\begin{pmatrix}
a \cr b
\end{pmatrix}
\rightarrow \frac{1}{\kk^{\frac{n}{2}+1}}
\begin{pmatrix}
\cos (\alpha \log \kk) & \sin (\alpha \log \kk) \cr
-\sin (\alpha \log \kk) & \cos (\alpha \log \kk)
\end{pmatrix}
\begin{pmatrix}
  a\cr
  b
\end{pmatrix} \,.
\labell{transf}
\ee
This result implies that the solutions exhibit discrete
self-similarity and yields critical exponents that characterise the
near-critical behaviour. We refer the reader to
\cite{frolovnew,frolov} for details but emphasise that this
behaviour depends only on the dimension of the sphere.  Hence it is
universal for all Dp/Dq systems (with $n\le4$).

Each near-horizon solution gives rise to a global solution when
extended over the full spacetime \reef{metric}. Each of these
embeddings is characterised two constants, which can be read off
from its asymptotic behaviour and which can be interpreted as the
quark mass $\mq$ and (roughly) the quark condensate $\qc$ in the
dual field theory -- see below. Both of these quantities are fixed
by $\Z_0$ or $\Y_0$. As we will see, the values corresponding to the
critical solution, $\mqs$ and $\qcs$, give a rough estimate of the
point at which a phase transition occurs.

\subsubsection{Real scaling exponents?}

From eq.~\reef{exponents}, we see that the exponents will be real if
the dimension of the internal sphere wrapped by the Dq-brane is
sufficiently large, \ie if $n \geq 5$. This would be interesting
because, whereas the oscillatory behaviour for $n\le4$ leads to a
first order phase transition, as we show below, real exponents would
seem to lead to a second order phase transition. However, we will
now argue that (under the same assumption to guarantee stability as
above) no such analysis can be applied for the Dp/Dq systems that
actually arise in superstring theory.

Choosing a value of $n$, the dimension of the internal sphere,
places restrictions on the allowed values of both $p$ and $q$. The
internal $S^n$ is a subspace of the spherical part of the geometry
\reef{metric} and hence we must have $p< 8-n$. We have taken a
strict inequality here, \ie we do not consider $p=8-n$, because the
size of the $n$-sphere must vary to have nontrivial embeddings and
so it can not fill the entire internal (8$-p$)-sphere. Given that
$p\ge0$,\footnote{No black brane geometry exists for a Euclidean
D(--1)-brane.} we need only consider $n=5,6,7$.

Next, we note that by T-dualising along the $p$ directions common to
both sets of branes, the brane configuration is reduced to a
D0/D$q'$ intersection, where $q'=n+1+(p-d)$. Given the previous
restriction on $n$, we must have $q' \geq 6$. Now, if we require as
above that the intersection be supersymmetric at zero temperature
(for stability), then we must have $q'=8$. Hence the only brane
configurations of interest are T-dual to the D0/D8 system. However,
these configurations are those in which string creation arises
through the Hanany-Witten effect \cite{amiwit}. In particular, as
discussed in \cite{amiwit2}, the background Ramond-Ramond field of the
Dp-branes will induce a nontrivial worldvolume gauge field on the
Dq-brane. While this does not rule out the possibility of
interesting embeddings and a possible (second order) phase
transition, it certainly indicates that the present analysis (with
no worldvolume gauge fields) does not apply to these systems.
For this reason, in the remainder of this paper we will concentrate
on Dp/Dq systems with $n\leq 4$.

\subsection{Phase Transitions}\label{transit}

In order to study the global solutions corresponding to the near
horizon solutions of the previous subsection it is convenient to
introduce an isotropic, dimensionless radial coordinate $\rho$
through
\be
\left( u_0 \rho
\right)^{\frac{7-p}{2}} = u^{\frac{7-p}{2}} + \sqrt{ u^{7-p} -
u_0^{7-p}} \,. \labell{rho}
\ee
Note that the horizon is at $\rho=1$. Following the discussion in
the previous subsection,\footnote{Above, we pointed out that our
present analysis does not apply to Dp/Dq systems T-dual to
D0/D8-branes. Systems T-dual to D0/D0 systems would be trivial for
the present purposes as $n=0$. Hence those T-dual to the D0/D4 or
D3/D7 system are the only other possibility with a supersymmetric
limit.} we assume that the Dp/Dq system under consideration is
T-dual to the D3/D7 one, in which case $(p-d) +( n+1) =4$. Then the
Euclidean Dq-brane action density of $\nf$ coincident Dq-branes in
the black Dp-brane background is
\beq
\frac{\ibulk}{\N} =
\int_{\rhomin}^\infty d\rho \left( \frac{u}{u_0 \rho} \right)^{d-3}
\left( 1 - \frac{1}{\rho^{2(7-p)}} \right) \rho^n
 {(1-\chi^2)}^{\frac{n-1}{2}} \sqrt{ 1-\chi^2 + \rho^2
\dot{\chi}^2  } \,,
\labell{action}
\eeq
where $\chi=\cos \theta$, \mbox{$\dot{\chi} =d\chi/d\rho$} and we
have introduced the normalisation constant
\be \N = \frac{\nf \tq  u_0^{n+1} \Omega_n}{4T}\,. \labell{N} \ee
Here, $\tq=1/{(2\pi \ell_s)}^q g_s \ell_s$ is the Dq-brane tension
and $\Omega_n$ is the volume of a unit $n$-sphere. Up to a numerical
constant of $O(1)$, the normalisation factor is found to be
\be
\N \sim \nf \nc T^d \leff(T)^{\frac{2(d-1)}{5-p}}  \,, \labell{N2}
\ee
where $\leff(T)$ is the effective coupling \eqn{geff} and we have
used the standard gauge/gravity relations \eqn{L} and \reef{gym}.

The equation of motion that follows from \eqn{action} leads to the
large-$\rho$ behaviour\footnote{Here we assume $n>1$. Otherwise the
term multiplied by $c$ is $\log \rho/\rho$.}
\be \chi = \frac{m}{\rho} + \frac{c}{\rho^{n}} + \cdots \,.
\labell{asymp} \ee
Holography relates the dimensionless constants $m, c$ to the quark
mass and condensate by\footnote{Note that the factor of $\nf$ in the
second equation was missing in refs.~\cite{prl,viscosity}.}
\beqa \mq &=& \frac{u_0 m}{2^{\frac{9-p}{7-p}}\pi \ell_s^2}\sim
\leff(T)^{\frac{2}{5-p}}\,T\,m \,,\labell{mc}\\
\langle {\cal O}_m \rangle& =& - \frac{2 \pi \ell_s^2 (n-1) \Omega_n
\nf \tq u_0^n c} {4^{\frac{n}{7-p}}} \sim
-\nf\,\nc\,\leff^{\frac{2(d-2)}{5-p}}\,T^d\,c\,. \labell{donc} \eeqa
Here $\mq$ is the mass of the fields in the fundamental
hypermultiplets, both the fermions $\psi$ and the scalars $q$. The
operator ${\cal O}_m$ is a supersymmetric version of the quark
bilinear, and it takes the schematic form
\be {\cal O}_m = \bar{\psi} \psi + q^\dagger\Phi q+ \mq q^\dagger q \,,
\labell{oops} \ee
where $\Phi$ is one of the adjoint scalars. We will loosely refer to
its expectation value as the `quark condensate'. A detailed
discussion of this operator, including a precise definition, can be
found in appendix A of ref.~\cite{finite}.

Eq.~\eqn{mc} implies the relation $m^{(5-p)/2} = \mbar/T$ between
the dimensionless quantity $m$, the temperature $T$ and the mass
scale
\be \bar{M} = \frac{7-p}{2^{\frac{9-p}{7-p}}\pi \R} \left(
\frac{2\pi \ell_s^2 \mq}{L} \right)^{\frac{5-p}{2}} \sim
\frac{\mq}{\leff(\mq)} \,. \labell{deep} \ee
Up to numerical factors, this scale is the mass gap in the discrete
meson spectrum at temperatures well below the phase transition
\cite{us-meson,holomeson,holomeson2,us}. We shall see below that it
is also the scale of the temperature of the phase transition for the
fundamental degrees of freedom, $\tf \sim \mbar$, since the latter
takes place at $m \sim 1$.

The key observation \cite{frolov} is that the values $(m,c)$ of a
near-critical solution are linearly related to the integration
constants fixing the corresponding embedding in the near-horizon
region. Combining this with the transformation rule \eqn{transf} for
the near-horizon constants $(a,b)$ and eliminating $\kk$, we deduce
that $(m-m^*) / \Z_0^{\frac{n}{2}+1}$ and $(c-c^*) /
\Z_0^{\frac{n}{2}+1}$ are periodic functions of $(\alpha /2 \pi)
\log \Z_0$ with unit period for Minkowski embeddings, and similarly
with $\Z_0$ replaced by $\Y_0$ for black hole embeddings. This is
confirmed by our numerical results, which will be discussed in the
next sections and are illustrated for the D3/D7 brane system in
figure \ref{mfunc}.

The oscillatory behaviour of $m$ and $c$ as functions of $\Z_0$ or
$\Y_0$ implies that for a fixed value of $m$ near the critical
value, several consistent Dq-brane embeddings are possibile with
different values of $c$. Alternatively, one finds the quark
condensate is not a single-valued function of the quark mass.
Physically, the preferred solution will be the one that minimises
the free energy density of the Dq-brane, $F=T \ibk$. As with the
bulk action, the Dq-action \eqn{action} contains large-radius
divergences, as can be seen by substituting the asymptotic behaviour
\eqn{asymp} in eq. \eqn{action}. It therefore needs to be
regularised and renormalised. We can achieve the former by replacing
the upper limit of integration by a finite ultraviolet cut-off
$\rhomax$. Then in analogy to the holographic renormalisation of the
supergravity action \cite{ct,ct1},  boundary `counter-terms' $\ict$
are added to the brane action $\ibulk$, such that the renormalised
brane energy $\ibk=\ibulk + \ict$ is then finite as the cut-off is
removed, $\rhomax \rightarrow \infty$ \cite{karch1}. The latter
method applies directly to asymptotic AdS geometries, but it can be
easily extended to the D4/D6 system, as discussed below. We expect
that a similar procedure can be developed for any Dp/Dq system for
which there is a consistent gauge/gravity duality. (In any event,
the brane action can also be regulated by subtracting the free
energy of a fiducial embedding.) The details for the D3/D7 and D4/D6
cases are discussed in the following sections and the results are
presented in figures \ref{freeEnD3D7} and \ref{freeEnD4D6},
respectively. In both cases, we see that as the temperature is
increased, a first order phase transition occurs by discontinuously
jumping from a Minkowski embedding (point A) to a black hole
embedding (point B). We emphasise again that this first order
transition is a direct consequence of the multi-valued nature of the
physical quantities brought on by the critical behaviour described
in the previous section. It may be possible to access this
self-similar region by super-cooling the system (although most of
the other solutions in this region are dynamically unstable -- see
below).

It is interesting to ask if the strong coupling results obtained
here could in principle be compared with a weak coupling
calculation. It follows from our analysis that the free energy
density takes the form $F=\N T f(m^2)$, where the function $f$ can
only depend on even powers of $m$ because of the reflection symmetry
$\chi \rightarrow -\chi$. The limit $m \rightarrow 0$ may be
equivalently regarded as a zero quark mass limit or as a
high-temperature limit. In this limit the brane lies near the
equatorial embedding $\chi=0$, which slices the horizon in two equal
parts. In general $f(0)$ is a non-zero numerical constant; in the
D3/D7 case, for example, a straightforward calculation yields
$f(0)=-1/2$. Given eq.~\reef{N2}, we have that at strong coupling
the free energy density scales as
\be F \sim \nf\, \nc\, T^{d+1}\, \leff(T)^{\frac{2(d-1)}{5-p}}  \,.
\labell{F} \ee
The temperature dependence is that expected on dimensional grounds
for a $d$-dimensional defect, and the $\nf \nc$ dependence follows
from large-$N$ counting rules. However, the dependence on the
effective 't Hooft coupling indicates that this contribution comes
as a strong coupling effect, without direct comparison to any weak
coupling result. The same is true for other thermodynamic quantities
such as, for example, the entropy density $S=-\partial F/\partial
T$. We remind the reader that the background geometry makes the
leading contribution to the free energy density \eqn{Fbulk}, which
corresponds to that coming from the gluons and adjoint matter.
Recall that only for $p=3$ is the effective coupling factor absent
in eq.~\reef{Fbulk}. Only in this case the string coupling result
differs from that at weak coupling by a mere numerical factor of 3/4
\cite{3/4}. For the fundamental matter, a similar circumstance
arises for $d=1$, as would be realized with the D1/D5, D2/D4 or
D3/D3 systems. In these special cases, the strong and weak coupling
calculations for the fundamental matter could in principle be
compared. Hence the D3/D3 system is singled out since such a
comparison can be made for both the adjoint and fundamental sectors.

\section{The D3/D7 system}
\label{D3D7phase}

Here we will specialise the above discussion to the D3/D7 system.
This intersection is summarised by the array
\begin{equation}
\begin{array}{ccccccccccc}
   & 0 & 1 & 2 & 3 & 4& 5 & 6 & 7 & 8 & 9\\
\mbox{D3:} & \times & \times & \times & \times & & &  &  & & \\
\mbox{D7:} & \times & \times & \times & \times & \times  & \times & \times & \times &  &   \\
\end{array}\labell{D3D7}
\end{equation}
Of course, this is an interesting system because both the gluons and
the fundamental fields in the gauge theory propagate in $3+1$
dimensions.

\subsection{D7-brane  embeddings}\label{bedtime}

In the D3/D7 brane system with the radial coordinate defined in
\reef{rho},
\beq
(u_0 \rho)^2 = u^2 + \sqrt{u^4-u_0^4}\, ,
\labell{change}
\eeq
the background metric \reef{metric} becomes
\beq
ds^2 = \frac{1}{2} \left(\frac{u_0 \rho}{L}\right)^2
\left[-{f^2\over \tilde f}dt^2 + \tilde{f} dx^2_3 \right]
+ \frac{L^2}{\rho^2}\left[ d\rho^2 +\rho^2 d\Omega_5^2  \right] \,,
\labell{D3geom}
\eeq
where
\be
f(\rho)= 1- \frac{1}{\rho^4} \sac \tilde{f}(\rho)=1+\frac{1}{\rho^4} \,.
\ee
The coordinates $\{ t, x^i \}$ parametrise the intersection, while
$\{\rho, \Omega_5\}$ are spherical coordinates on the
456789-directions transverse to the D3-branes. As in
eq.~\reef{sphere8}, it is useful to adapt the metric on the
five-sphere to the D7-brane embedding. Since the D7-brane spans the
4567-directions, we introduce spherical coordinates $\{r,
\Omega_3\}$ in this space and $\{R, \phi\}$ in the 89-directions.
Denoting by $\theta$ the angle between these two spaces we then
have:
\beq
\rho^2=r^2+R^2 \sac r=\rho\sin\theta \sac
R=\rho\cos\theta \,,
\labell{coord2}
\eeq
and
\beqa
d\rho^2+\rho^2d\Omega^2_5&=& d\rho^2+\rho^2\left(d\theta^2+\sin^2\theta
\, d\Omega^2_3+ \cos^2\theta \, d\phi^2\right) \, \labell{met2} \\
&=& dr^2 + r^2d \Omega^2_3 + dR^2+R^2d\phi^2 \,.
\labell{met1}
\eeqa

Describing the profile in terms of $\chi(\rho) = \cos \theta (\rho)$
simplifies the analysis -- note that $\chi=R/\rho$. With this
coordinate choice, the induced metric on the D7-brane becomes
\beq ds^2 = \frac{1}{2} \left(\frac{u_0 \rho}{L}\right)^2
\left[-{f^2\over \tilde f}dt^2 + \tilde{f} dx_3^2 \right] +
\left(\frac{L^2}{\rho^2}+\frac{L^2 \dot{\chi}^2}{1-\chi^2} \right)
d\rho^2 +L^2(1-\chi^2)\, d\Omega_3^2 \,, \labell{induce} \eeq
where, as above, $\dot{\chi} = d \chi /d\rho$. Since we are studying
static embeddings of the probe brane, the equation of motion for
$\chi(\rho)$ can be derived equally well from the Lorentzian or
Euclidean action. Here we proceed directly to the latter because it
is relevant for the thermodynamic calculations in the following. The
Euclidean D7-brane action density is
\beq \frac{\ibulk}{\N} = \int d\rho  \left(
1-\frac{1}{\rho^8}\right) \rho^3 (1-\chi^2) \sqrt{1-\chi^2+\rho^2
\dot{\chi}^2} \, ,\labell{act2s} \eeq
where
\be {\cal N} = \frac{2\pi^2\nf T_\mt{D7} u_0^4}{4T}
=\frac{\lambda\nf\nc}{32}\,T^3\labell{ND3D7} \ee
is the normalisation constant defined in \reef{N}. Recall from
footnote \ref{foot1} that $\ibulk$ denotes a density because we have
divided out the volume $V_x$. The equation of motion for
$\chi(\rho)$ is then
\beq
\partial_\rho \left[\left( 1-\frac{1}{\rho^8}\right)
\frac{\rho^5 (1-\chi^2) \dot{\chi}}{\sqrt{1-\chi^2+\rho^2
\dot{\chi}^2}} \right]+\rho^3 \left(
1-\frac{1}{\rho^8}\right)\frac{3 \chi (1-\chi^2)+2 \rho^2 \chi
\dot{\chi}^2}{\sqrt{1-\chi^2+\rho^2 \dot{\chi}^2}}=0 \,,
\labell{psieom} \eeq
which implies that the field $\chi$ asymptotically approaches zero as
\beq \chi = \frac{m}{\rho} + \frac{c}{\rho^3}+\cdots \,.
\labell{asympD7} \eeq
The dimensionless constants $m$ and $c$ are related to the quark
mass and condensate through eqs.~\reef{mc} and \reef{donc} with
$p=3$ and $n=3$:
\beqa \mq &=& \frac{u_0 m}{2^{3/2}\pi \ell_s^2} =
\frac{1}{2}\sqrt{\lambda}\,T\,m\,,\labell{mqD3D7}\\
\langle {\cal O}_m \rangle &=& - 2^{3/2}\pi^3 \ell_s^2 \nf T_\mt{D7}
u_0^3\, c = -\frac{1}{8}\sqrt{\lambda}\,\nf\,\nc\,T^3\,c\,.
\labell{cqD3D7} \eeqa
In this case $m=\bar{M}/T$ and eq.~\eqn{deep} takes the form
\be \bar{M} = \frac{\sqrt{2} (2\pi \ell_s^2 \mq)}{\pi L^2} = \frac{2
\mq}{\sqrt{\lambda}}  = \frac{M_\mt{gap}}{2\pi} \,,
\labell{mbarD3D7} \ee
where $\lambda = \gym^2 \nc = 2\pi g_s \nc$ is the 't Hooft
coupling. In the last equality, we are relating $\bar{M}$ to the
meson mass gap in the D3/D7 theory at zero temperature \cite{us-meson}.

The equation of motion \eqn{psieom} can be recast in terms of the
$R$ and $r$ coordinates, related to the $\rho$ and $\theta$
coordinates via \reef{coord2}:
\beq
\partial_r\left[ r^3 \left(1-{1\over (r^2+R^2)^4} \right)
{\partial_r R \over \sqrt{1+(\partial_r R)^2}} \right]  = 8 {r^3 R
\over (r^2 +R^2)^5} \sqrt{1+(\partial_r R)^2} \,, \labell{eomR} \eeq
where the embedding of the D7-brane is now specified by $R=R(r)$.
Asymptotically,
\beq
R(r) = m  + \frac{c}{r^2} + \cdots \, . \labell{asympD7R}
\eeq

\FIGURE{
\includegraphics[width=0.5 \textwidth]{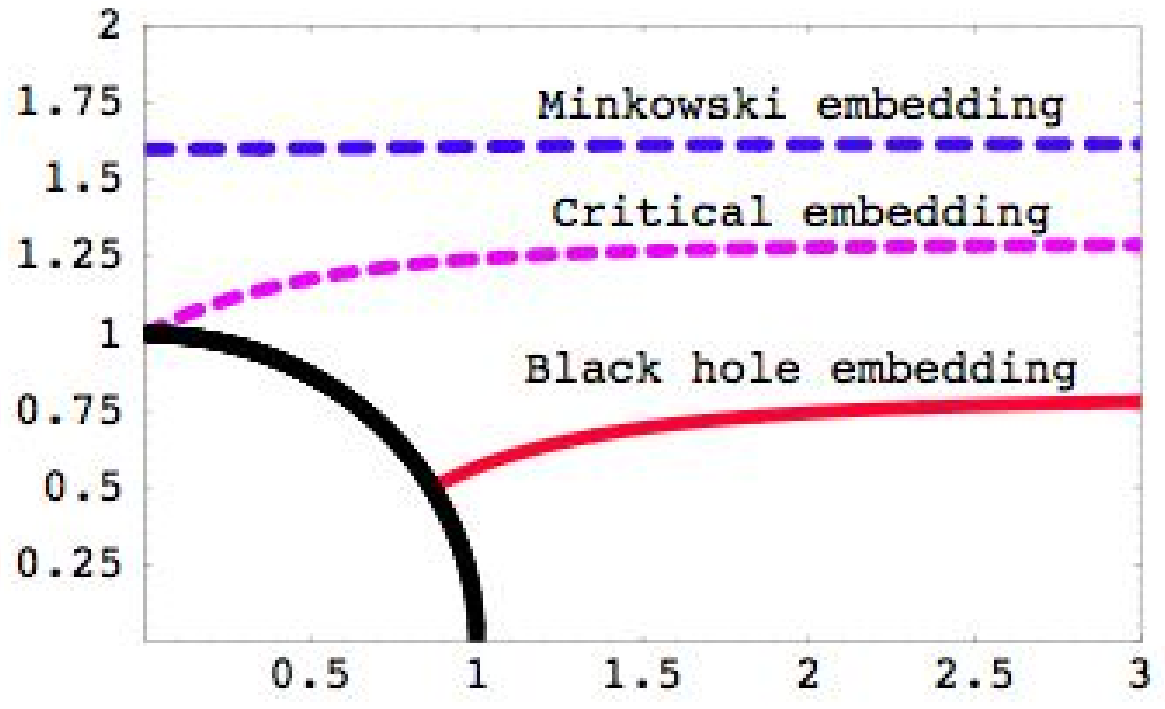}
\caption{Profiles  various D7-brane embeddings in a D3-brane
background in the $(R,r)$-plane. The circle represents the horizon
at $\rho=1$. \label{embeddings3} }}
In the limits of large and small $m$ we were able to find
approximate analytic solutions for the embeddings -- see discussion
below and Appendix \ref{approxSol}.  However, for arbitrary $m$ we
were unable to find an analytic solution of eq.~\reef{psieom} or
\reef{eomR} and so we resorted to solving these equations
numerically. It was simplest to solve for Minkowski embeddings using
the $(R,r)$ coordinates with equation of motion \reef{eomR} while
the $(\chi, \rho)$ coordinates were best suited to the black hole
embeddings. Our approach was to specify the boundary conditions at a
minimum radius and then numerically integrate outward. For the black
hole embeddings, the following boundary conditions were specified at
the horizon $\rhomin=1$: $\chi = \chi_0$ and $d\chi /d\rho =0$ for
$0\leq \chi_0 <1$. For Minkowski embeddings, the following boundary
conditions were specified at $r_{min} =0$ (\ie at the axis
$\chi=1$): $R=R_0$ and $\partial_r R =0$ for $R_0
> 1$. In order to compute the constants $m,c$ corresponding to each
choice of boundary conditions at the horizon, we fitted the
solutions to the asymptotic form \reef{asympD7} for $\chi (\rho)$ or
\reef{asympD7R} for $R(r)$. A few characteristic profiles are shown
in fig.~\ref{embeddings3}.

Recall that, as elucidated in section \ref{critikue}, the black hole
and Minkowski embeddings are separated by a critical solution which
just touches the horizon. This critical embedding is characterised
by certain critical values of the integration constants, $m^\ast$
and $c^\ast$. For Minkowski embeddings near the critical solution,
fig.~\ref{mfunc} shows plots of $(m-m^\ast)/(R_0-1)^{5/2}$ and
$(c-c^\ast)/(R_0-1)^{5/2}$ versus $\sqrt{7}\log (R_0-1) /4 \pi$. In
this regime, we may relate the boundary value to that in the near
horizon analysis with $R_0-1\simeq z_0$. Here our numerical results
confirm that, near the critical solution, $(m-m^\ast)/z_0^{5/2}$ and
$(c-c^\ast)/z_0^{5/2}$ are both periodic functions of $\sqrt{7}\log
(z_0) /4 \pi$ with unit period, as discussed above in section
\ref{critikue}. This oscillatory behaviour of $m$ and $c$ as
functions of $z_0$ (or $y_0$) implies that the quark condensate is
not a single-valued function of the quark mass and this is clearly
visible in our plots of $c$ versus $T/\mbar=1/m$, displayed in
figure \ref{cond}. By increasing the resolution in these plots, we
are able to follow the two families of embeddings spiralling in on
the critical solution, the behaviour predicted by the near-horizon
analysis. Thermodynamic considerations will resolve the observed
multi-valuedness by determining the physical solution as that which
minimizes the free energy density of the D7-branes. As discussed in
section \ref{transit}, since the physical parameters are
multi-valued, we can anticipate that there will be a first order
phase transition when the physical embedding moves from the
Minkowski branch to the black hole branch. We will proceed to
computing the free energy density in the next subsection. The
position of the resulting phase transition is indicated in the
second plot of fig.~\ref{cond}.
\FIGURE{
\includegraphics[width=\textwidth]{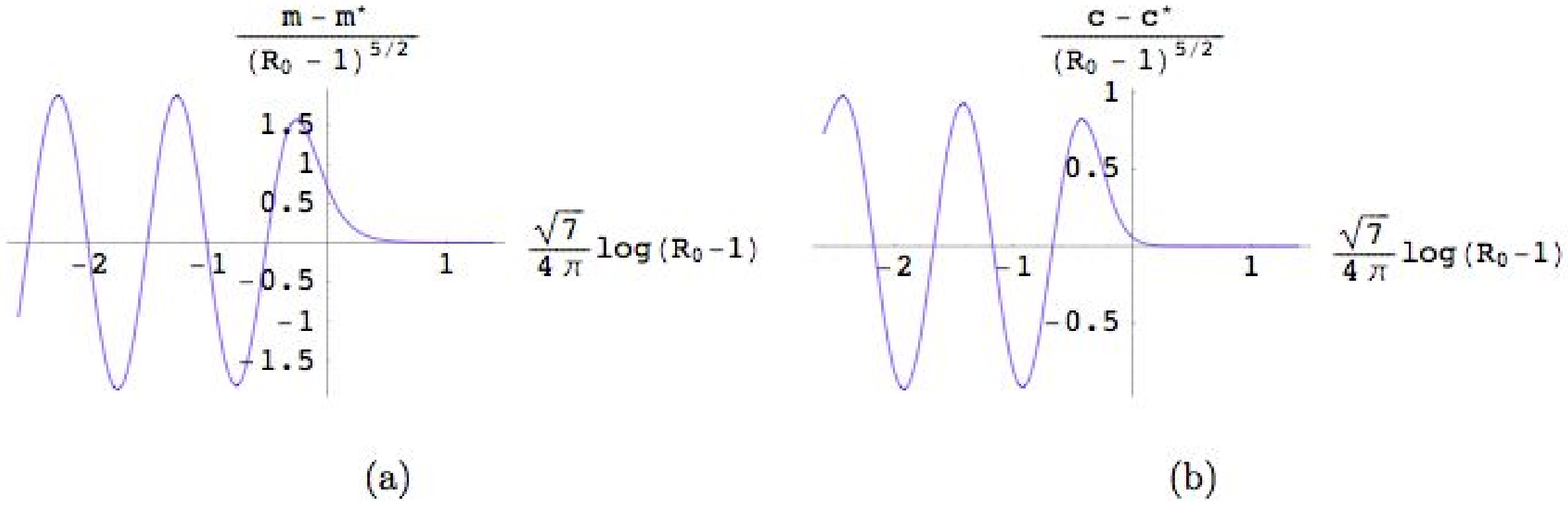} 
\caption{Quark mass (a) and condensate (b) as a function of the
distance to the horizon $R_0-1$ for D7-brane Minkowski embeddings in
a D3-brane background. Note that near the horizon $R_0-1 \sim z_0$.
} \label{mfunc}}
\FIGURE{
\includegraphics[width=\textwidth]{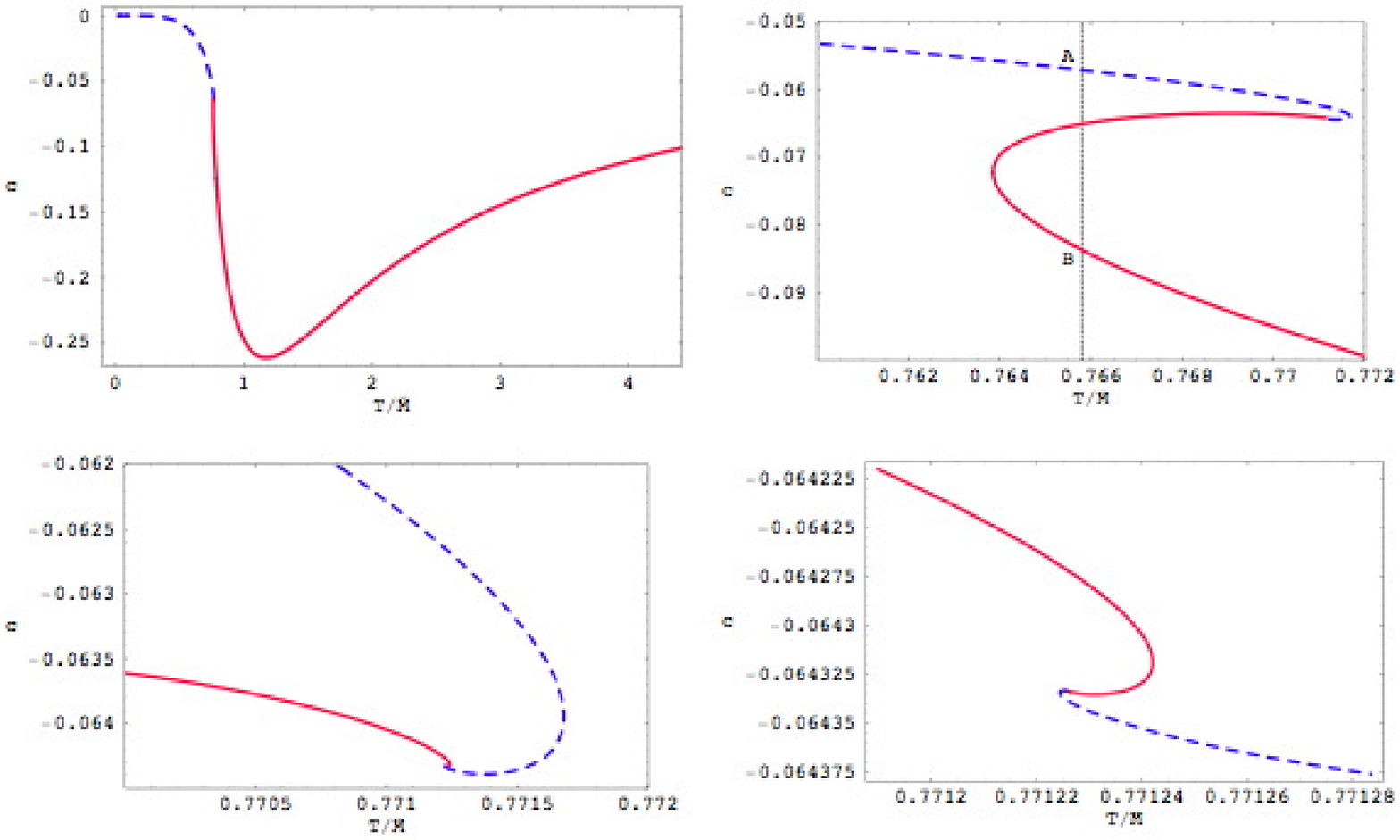} 
\caption{Quark condensate $c$ for a D7 in a D3 background versus
$T/\mbar$. The blue dashed (red continuous) curves correspond to the
Minkowski (black hole) embeddings.  The dotted vertical line
indicates the precise temperature of the phase transition. }
\label{cond}}

\subsection{D7-brane thermodynamics}\label{therm77}

Having discussed the embeddings of the D7-brane in the black
D3-brane geometry, we proceed to compute the free energy, entropy
and energy densities associated with the D7-brane, or equivalently,
the fundamental fields.  We start with the Euclidean D7-brane action
\eqn{act2s}. Using the asymptotic behaviour \reef{asympD7}, we see
that the action contains a UV divergence, since
\beq
\frac{\ibulk}{\N} \simeq \int_{\rhomin}^{\rhomax}
d\rho \, \rho^3 \simeq \frac{1}{4} \rhomax^4 \labell{diverge}
\eeq
diverges as the regulator is removed, \ie $\rhomax \to \infty$.

This kind of problem is well-known in the context of the AdS/CFT
correspondence and was first resolved for the gravity action by
introducing boundary counter-terms, which depend only on the
intrinsic geometry of the boundary metric \cite{ct,ct1}. These ideas
can be generalized to other fields in an AdS background, such as a
scalar \cite{scale} -- for a review, see \cite{revue}. The latter
formed the basis for the renormalization of probe brane actions in
\cite{karch1}, where the brane position or profile is treated as a
scalar field in an asymptotically AdS geometry. That is, one
implicitly performs a Kaluza-Klein reduction of the D7 action to
five dimensions so that it appears to be a complicated nonlinear
action for a scalar field $\chi$ propagating in a five-dimensional
(asymptotically) AdS geometry. One then introduces boundary
counter-terms which are local functionals (polynomials) of the
scalar field (and boundary geometry) on an asymptotic regulator
surface. These terms are designed to remove the bulk action
divergences that arise as the regulator surface is taken off to
infinity, as in eq.~\reef{diverge}. The D3/D7 system is explicitly
considered in ref.~\cite{karch1}, which also introduces a finite
counterterm that ensures that the brane action vanishes for the
supersymmetric embedding of a D7-brane in an extremal D3-background,
\ie eq.~\reef{metric} with $u_0=0$ and $p=3$. In the calculation of
\cite{karch1} the D7-brane embedding is specified as $\theta(\rho)$,
but this is easily converted to a counter-term action for
$\chi(\rho)$ using the obvious coordinate/field redefinition:
$\frac{\pi}{2}-\theta=\arcsin\chi\simeq\chi+1/6\,\chi^3+\cdots$. The
final result is
\beq
{\ict \over \N} = -{\R^4 T \over u_0^4} \int d t_\mt{E} d^3 x
\sqrt{\det \gamma}\left(1-2\chi^2+\chi^4\right) \,, \labell{bact}
\eeq
where this boundary action is evaluated on the asymptotic regulator
surface $\rho=\rhomax$ introduced above. The boundary metric
$\gamma$ at $\rho=\rhomax$ in the (effective) five-dimensional
geometry is given by
\beq
ds^2 (\gamma) ={1\over2}\left({u_0 \rhomax\over L}\right)^2\left(
{f(\rhomax)^2\over\tildef(\rhomax)}\,d t_\mt{E}^2+\tildef(\rhomax)dx_3^{\,2}\right)
\labell{boundmet}
\eeq
and so $\sqrt{\ga}={u_0^4
\rhomax^4}f(\rhomax)\tildef(\rhomax)/4L^4$. Evaluating the
counter-term action \reef{bact} with an asymptotic profile as in
eq.~\reef{asymp}, one finds
\beq
{\ict \over \N} =-{1\over 4} \left[(\rhomax^2-m^2)^2 -4 mc\right]. \labell{bact2}
\eeq
Here we have divided out the volume factor $V_x$ -- see footnote
\ref{foot1}. Comparing eqs.~\reef{diverge} and \reef{bact2}, one
sees that the leading divergence proportional to $\rho^4_\mt{max}$
cancels in the sum of $I_\mt{D7}=\ibulk+\ict$. As a further check,
one can consider the supersymmetric limit  $u_0 \rightarrow 0$, in
which one must work with a rescaled coordinate $\trho = u_0 \rho$,
since the change of variables \eqn{change} is not well defined at
$u_0=0$. In this limit $\chi= u_0 m/\trho=\sqrt{2} \, 2\pi \ell_s^2
\mq/\trho$ is an exact solution, and one can easily verify that for
this configuration $\ids=\ibulk+\ict=0$.

In order to produce a finite integral which is more easily evaluated
numerically, it is useful to incorporate the divergent terms in the
boundary action \reef{bact2} into the integral in eq.~\reef{act2s}
using
\beqa
\rhomax^4&=&\int_{\rhomin}^{\rhomax}d\rho\,4\rho^3+\rhomin^4\ ,\nonumber\\
\rhomax^2&=&\int_{\rhomin}^{\rhomax}d\rho\,2\rho+\rhomin^2\ .\labell{junk9}
\eeqa
Then the total action may be written as
\beq
{\ids \over \N} = G(m) -{1\over 4}\left[(\rhomin^2-m^2)^2-4mc\right]\ ,
\labell{tact}
\eeq
where $G(m)$ is defined as
\beq
G(m) = \int_{\rhomin}^\infty
d\rho\,\left[\rho^3
\left(1-{1\over\rho^8}\right)\left(1-\chi^2\right)
\left(1-\chi^2+\rho^2\,\dot{\chi}^2\right)^{1/2}-\rho^3+m^2\rho\right].
\labell{integral}
\eeq
Note that the upper bound for the range of integration has been set
to infinity, since the integral above is finite.

From these expressions, the free energy density is given by $F= T
I_{D7}$. Now using our numerical results, the free energy density is
shown as a function of the temperature in the first two plots in
fig.~\ref{freeEnD3D7}. The second of these shows the classic
`swallow tail' form, typically associated with a first order phase
transition. To our best numerical accuracy, the phase transition
takes place at $\tf/\bar{M} = 0.7658$ (or $m=1.306$), where the free
energy curves for the Minkowski and black hole phases cross. The
fact that the transition is first order is illustrated by
fig.~\ref{cond}, which shows that the quark condensate makes a finite
jump at this temperature between the points labelled A and B.
Similar discontinuities also appear in other physical quantities,
like the entropy and energy density, as we now calculate.

Given the free energy density, a standard identity \reef{thermiden}
yields the entropy density as
\beq
S = -\frac{\partial F}{\partial T} = -\pi L^2
\frac{\partial F}{\partial u_0} \,, \labell{entropy}
\eeq
where we have used the expression $u_0 = \pi \R^2 T$ from
eq.~\reef{beta}. Evaluating this expression requires a
straightforward but somewhat lengthy calculation, which we have
relegated to appendix \ref{entropy7}. The final result is
\beq
{S \over \N} = -4 G(m)+(\rhomin^2-m^2)^2 -6mc \,. \labell{sss6}
\eeq
Comparing eqs.~\reef{tact} and \reef{sss6}, we see that the entropy
and free energy densities are simply related as
\beq S=- \frac{4 F}{T} \left( 1 +
\frac{2\,\N\,mc}{4F/T} \right)
\,.\labell{curious} \eeq
The first term above can be recognized as the behaviour expected for
a conformal system, \ie a system for which $F\propto T^4$. Hence the
second term can be interpreted as summarising the deviation from
conformal behaviour. We note that, as illustrated in
fig.~\ref{cond}, $c$ vanishes in both the limits $T\rightarrow0$ and
$T\rightarrow\infty$ and so the deviation from conformality is
reduced there. More precisely, using the results from appendix
\ref{approxSol} we see that $c \sim m$ at high temperature and $c
\sim 1/m^5$ at low temperature. Together with \eqn{ND3D7} this
implies that the deviation from conformality scales as $\bar{M}^2
/T^2$ at high temperature. Conformality is also restored at low
temperatures but only because both $S$ and $F/T$ approach zero more
quickly than $T^3$. That is, $S\sim T^7 / \bar{M}^4$ as
$T\rightarrow0$.

Finally, the thermodynamic identity $E=F+TS=T(I_\mt{D7}+S)$ gives
the contribution of the D7-brane to the energy density:
\beq
{E \over \N T} = -3 G(m)+ {3 \over 4}
\left[(\rhomin^2-m^2)^2 -{20 \over 3}mc\right].
\labell{energy}
\eeq
We evaluated both the expressions \reef{sss6} and \reef{energy}
numerically and plotted $S$ and $E$ in fig.~\ref{freeEnD3D7}. In
both cases, the phase transition is characterised by a finite jump
in these quantities, as illustrated by the second plot in each case.
However, these plots also show that there is a large rise in, say,
the entropy density in the vicinity of $\tf$ and that the jump
associated with the phase transition only accounts for roughly
$3\%$ of this total increase.

We close with a few observations about these results. First, recall
from \reef{ND3D7} that $\N \sim \lambda \nc \nf T^3$ so that the
leading contribution of the D7-branes to all the various
thermodynamic quantities will be order $\lambda \nc \nf$, in
comparison to $\nc^2$ for the usual bulk gravitational
contributions. As noted in \cite{prl,viscosity}, the factor of
$\lambda$ represents a strong coupling enhancement over the
contribution over a simple free-field estimate for the $\nc \nf$
fundamental degrees of freedom. We return to this point below in
section \ref{discuss}.

Next, note that in order for the entropy $S=- \prt F/ \prt T$ to be
positive, the free energy $F$, or equivalently the action $\ids$,
must always be a decreasing function of the temperature. This means
that the apparent `kinks' in the plot of these quantities versus the
temperature are true mathematical kinks and not just very rapid turn
overs. An analytic proof of this fact is given in appendix
\ref{kinks}.

Finally, from the plots of the energy density one can immediately
read off the qualitative behaviour of the specific heat $c_V = \prt
E/ \prt T$. In particular, note that this slope must become negative
as the curves spiral around near the critical solution. Hence the
corresponding embeddings are thermodynamically unstable. Examining
the fluctuation spectrum of the branes, we will show that a
corresponding dynamical instability sets in at precisely the same
points. One may have thought that these phases near the critical
point could be accessed by `super-cooling' the system but this
instability severely limits the embeddings which can be reached with
such a process.
\FIGURE{
\includegraphics[width=\textwidth]{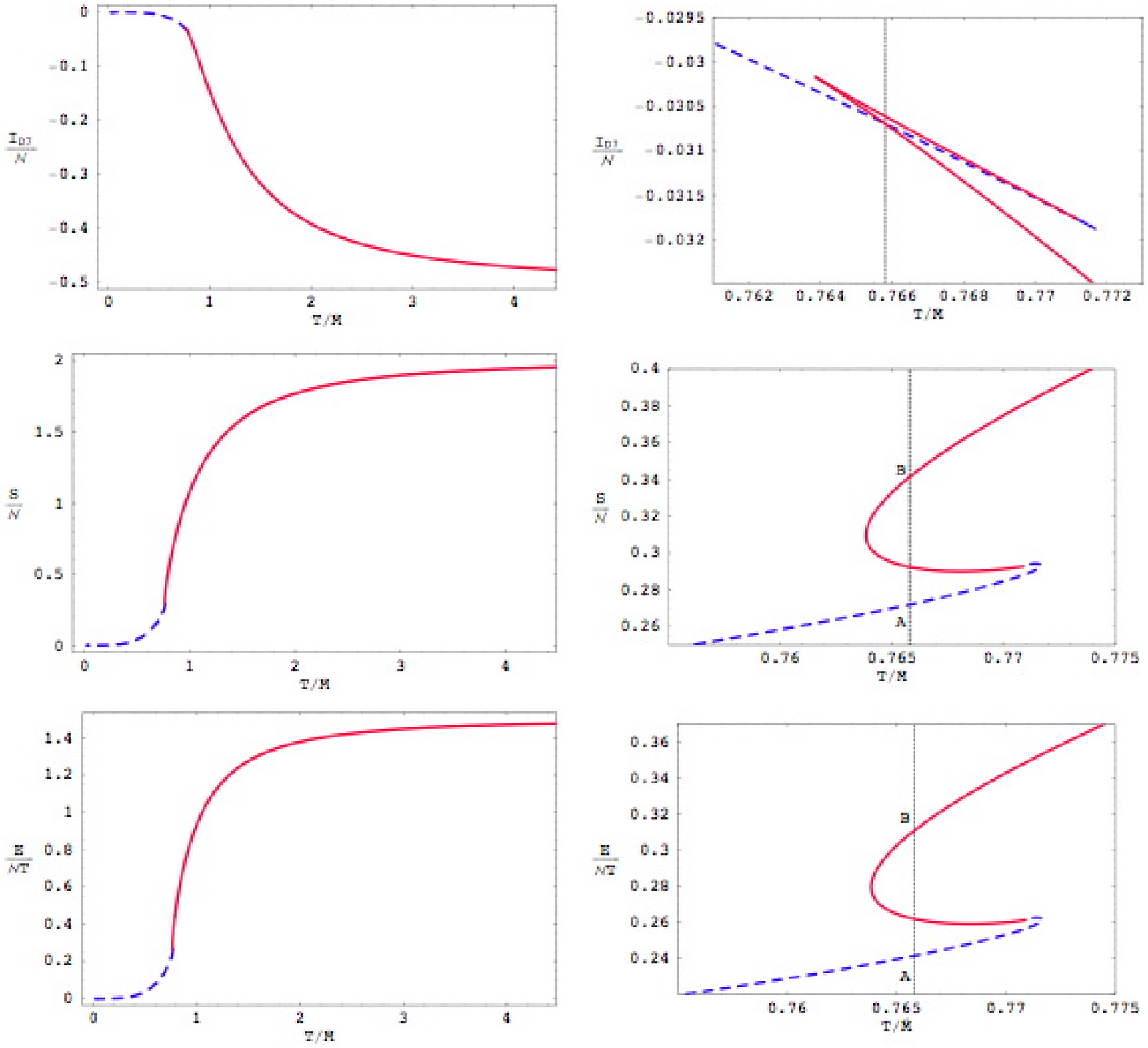} 
\caption{Free-energy, entropy and energy densities for a D7-brane in
a D3-brane background; note that $\N \propto T^3$. The blue dashed
(red continuous) curves correspond to the Minkowski (black hole)
embeddings.  The dotted vertical line indicates the precise
temperature of the phase transition. } \label{freeEnD3D7}}

\subsubsection{Thermodynamic expressions for large $T/\mbar$}

With precisely $m=0$, $\chi(\rho)=0$ is an exact solution. We denote
this solution as the equatorial embedding, since the D7-brane
remains at the maximal $S^3$ for all values of $\rho$. This
embedding describes the infinite-temperature limit for massive
quarks (or massless quarks for any temperature), \ie $T/\mbar
\rightarrow \infty$. For $T/\mbar \gg 1$ or $m\ll1$, approximate
analytic solutions for the D7-brane profile can be found by
perturbing around the equatorial embedding, as discussed in appendix
\ref{approxSol}. The final result is given in eq.~\reef{psitilde}.
In the notation of the appendix, the integral \reef{integral} can be
expressed as
\beqar
G(m) &=& \int _1 ^\infty {dx \over 2} \left[x \left(1-{1\over x^4} \right)
\left( 1-{3 \over 2}m^2 \tilde{\chi}^2+ 2 x^2 \left( \partial_x m \tilde{\chi}
\right)^2 \right) -x +m^2 \right] \nonumber \\
&=& -{1 \over 4} +m^2 G_2
\eeqar
where we have introduced
\beq
G_2 \equiv \int_1 ^\infty {dx \over 2} \left[x \left(1-{1\over x^4}
\right)\left( -{3 \over 2} \tilde{\chi}^2+ 2 x^2 \left( \partial_x
\tilde{\chi} \right)^2 \right) +1 \right] \simeq 0.413893 \, .
\eeq
We were only able to evaluate this integral numerically.

We are now in a position to evaluate the various thermal quantities
given by eqs.~\reef{tact}, \reef{sss6} and \reef{energy} in this
limit. We find
\beqar {I_\mt{D7}\over \N} &\simeq & -{1\over 2} + \left(G_2
+\tilde{c}+{1\over 2} \right) \left({\mbar \over T} \right)^2 -{1
\over 4} \left({\mbar \over T} \right)^4+\cdots
\,,\\
%\simeq  -{1\over 2} + 0.456946 \left({\mbar \over T} \right)^2 -{1 \over 4}
%\left({\mbar \over T} \right)^4 \\
%
{S\over \N} &\simeq & 2 + \left(-4 G_2 -6\tilde{c}-2\right) \left({\mbar \over T}
\right)^2 + \left({\mbar \over T} \right)^4+\cdots \,, \\
%\simeq 2 -0.91389 \left({\mbar \over T} \right)^2 + \left({\mbar \over T} \right)^4 \\
%
{E\over \N T} &\simeq & {3 \over 2} + \left(-3 G_2
-5\tilde{c}-{3\over 2}\right) \left({\mbar \over T} \right)^2 +{3
\over 4}  \left({\mbar \over T} \right)^4+\cdots \, ,
%\simeq {3 \over 2} -0.456944 \left({\mbar \over T} \right)^2 + {3 \over 4}
%\left({\mbar \over T} \right)^4
\labell{scale}
\eeqar
using  $\rhomin =1$ for black hole embeddings and $\tilde{c} \simeq
-0.456947$ from eq.~\reef{tildec}. In this high temperature limit,
the quark mass is negligible and so the first term in these
expressions could be characterised as conformal behaviour. The
remaining contributions are small corrections indicating a deviation
from this simple behaviour generated by the finite quark mass. This
is essentially the form expected in the high $T$ limit in finite
temperature field theory -- for example, see \cite{lands} and the
references therein.

\subsubsection{Thermodynamic expressions for small $T/\mbar$}

Turning to the opposite, low-temperature limit, \ie $T/\mbar \ll 1$,
the D7-branes lie on flat embeddings far from the event horizon, \ie
$\chi\simeq R_0/\rho$ to leading order. One can calculate
perturbative improvements to this simple embedding -- see appendix
\ref{approxSol} -- but it suffices to determine the leading
thermodynamic behaviour. We find that
\beq
G(m) = \int_0^\infty dr \left[r^3\left(1-\frac{1}{\rho^8}\right)
\sqrt{1+\left(\partial_r R \right)^2} +(r+R\, \partial_r R)(m^2-\rho^2)
\right] \simeq \frac{1}{12} \frac{1}{m^4} \,. \\
\eeq
Then using $R_0 \simeq m$ and $c\simeq -1/6m^5$, the thermal
densities become
\beq {I_\mt{D7}\over \N} \simeq  -{1\over 12} \left({T\over \mbar }
\right)^4 \,,\qquad
{S\over \N} \simeq \frac{2}{3} \left({T \over \mbar} \right)^4
\,,\qquad
{E\over \N T} \simeq   \frac{7}{12} \left({T \over \mbar}
\right)^4\,. \eeq
Hence these contributions are going rapidly to zero. Note that they
still contain the same normalization constant \reef{ND3D7} and so
these densities are still proportional to $\lambda\nf\nc$. At low
temperature, one might have expected that the thermodynamics of the
fundamental matter is dominated by the low lying-mesons, \ie the
lowest energy excitations in the fundamental sector, and so that the
leading contributions are proportional to $\nf^2$, reflecting
the number of mesonic degrees of freedom. Such contributions to the
thermal densities will arise in the gravity path integral in
evaluating the fluctuation determinant on the D7-brane around the
classical saddle-point. As indicated by the $\nc$ and $\lambda$
factors, the leading low-temperature contributions above come from
the interaction of the (deconfined) adjoint fields and the
fundamental matter.

\subsubsection{Speed of sound}

As mentioned in section \ref{house}, the speed of sound is another
interesting probe of the deconfined phase of the strongly coupled
gauge theories. In this section, we calculate the effect of
fundamental matter on the speed of sound. From eq.~\eqn{deaf}, we
must evaluate the D7-branes contribution to the total entropy
density and the specific heat. The first is already given by
eq.~\reef{sss6} and we denote this contribution as $S_7$ in the
following. From eq.~\reef{energy}, the energy density can be written
as $E=-3F-2{\cal N}Tmc$. Then recalling ${\cal N}\propto T^3$ from
eq.~\reef{ND3D7}, the D7-brane contribution to the specific heat can
be written as
\beq \cv_7= \frac{\prt E}{\prt T} = 3 S_7-\al\, \frac{\prt}{\prt T}
\, ( T^4\, mc) \,, \labell{cv7}\eeq
where we have introduced the dimensionless constant
$\al\equiv\lambda\nf\nc/16$. From the black D3 background, the free
energy of the adjoint fields is given in eq.~\reef{free3}. It
follows then that the adjoint contributions to the entropy and
specific heat are:
\beq S_3=-{\pi^2\over2}\nc^2\,T^3\,,\qquad \cv_3=3\,S_3\,.\eeq
Combining all of these results, we can now calculate the speed of
sound
\beqa
v_s^2&=&{S\over \cv}={S_3+S_7\over\cv_3+\cv_7}\nonumber\\
&=&{S_3+S_7\over 3S_3+3S_7-\al\,\prt_T\!\left(T^4\,
mc\right)}\nonumber\\
&\simeq&{1\over3}\left[1+{\al\over 3\,S_3}\, \prt_T\!\left(\,T^4\,
mc\,\right)\right] \,. \labell{speedE} \eeqa
Note all of our brane calculations are to first-order in an
expansion in $\nf/\nc$ and hence we have applied the Taylor
expansion in the last line above, reflecting this perturbative
framework \eg $\cv_7/\cv_3\ll 1$. Now using various expressions
above, as well as $m=2M_q/\sqrt{\lam}T$ and
$\vep\equiv{\lam\over2\pi}{\nf\over\nc}$, we may write the final
result as
\beq \delta v_s^2\equiv
v_s^2-{1\over3}\simeq{\vep\over 12\pi}\left(mc+{1\over 3} mT{\prt
c\over\prt T}\right)\,.\labell{final}\eeq
This expression indicates that the D7-brane produces a small
deviation away from the conformal result, $v_s^2={1\over3}$.

The result of numerically evaluating $\delta v_s^2$ as a function of
the temperature is given in fig.~\ref{speedy}. We see that $\delta
v_s^2$ is negative. That is, the fundamental matter reduces the
speed of sound. Following the discussion below eq.~\reef{curious}
one finds that $\delta v_s^2 \sim T^4 / \mbar^4$ at low temperature
and $\delta v_s^2 \sim \mbar^2/T^2$ at high temperature. Thus we see
again that the deviation from conformal behaviour vanishes for large
and small $T$. We also note that $\delta v_s^2$ is largest near the
phase transition, where it makes a discrete jump. Since we are
working in a perturbative framework, eq.~\reef{final} is only valid
when this deviation is a small perturbation. By assumption $\vep
\propto\nf/\nc\ll 1$ and so this is guaranteed provided the last
factor in \eqn{final} is not large. This is indeed satisfied for the
thermodynamically favoured embeddings, as illustrated in
fig.~\ref{speedy}. Similar deviations have been investigated in
\cite{ss3deviate} for other gauge/gravity dualities.

In fig.~\ref{speedy}, we have also continued $\delta v_s^2$ on the
disfavoured embeddings beyond the phase transition and we see that
it diverges (towards $-\infty$) at precisely the points where, \eg
the energy density curve turns around -- see fig.~\ref{freeEnD3D7}.
That is, $\cv_7$ diverges at these points, so that the perturbative
derivation of eq.~\reef{final} breaks down. Hence our perturbative
framework does not allow us to investigate interesting effects, as
seen in \cite{ss4}.
\FIGURE{
\includegraphics[width=\textwidth]{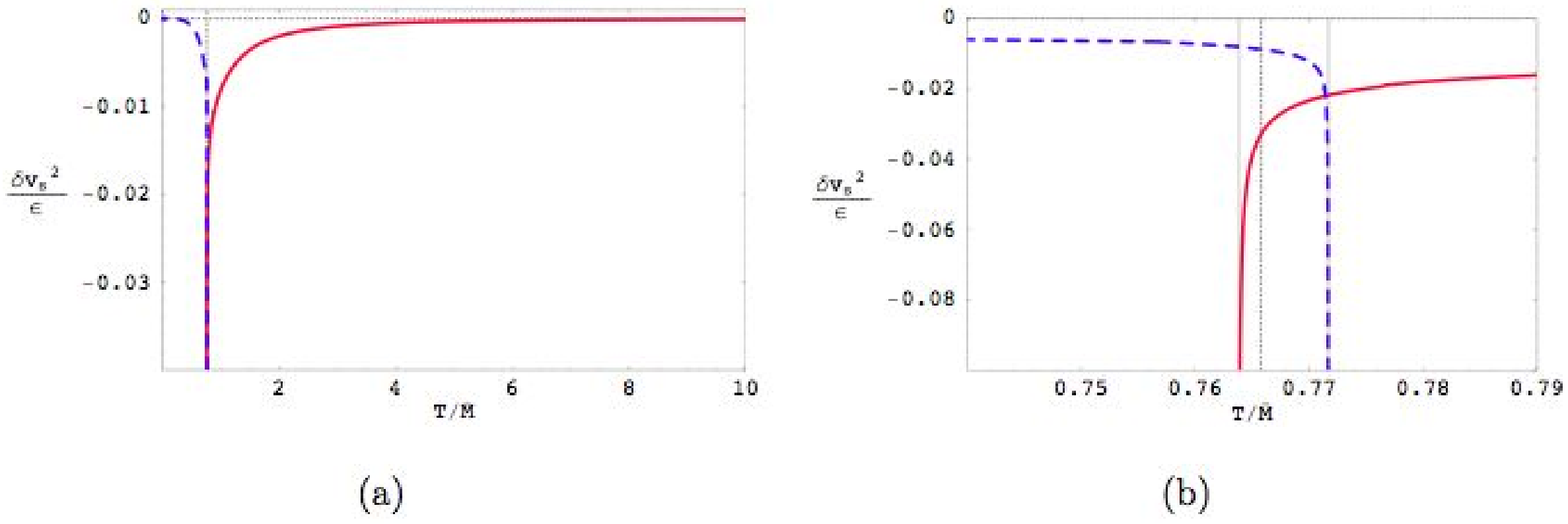} 
\caption{The deviation of the speed of sound from the conformal
value. (a) In the limits $T \to 0$ and $T \to \infty$, $\delta v_s^2
\to 0$.  (b) The temperature of the phase transition is marked by
the dashed vertical line. Note there is a finite discontinuity in
the speed of sound at the phase transition.  If we follow the black
hole branch (red line) or the Minkowski branch (dotted blue line)
past the phase transition, we find that $\delta v_s^2$ diverges.}
\label{speedy}}

We see from eq.~\eqn{final} that, for massive quarks, the deviation
from the conformal result is of proportional to $\nf/\nc$, as
expected from large-$\nc$ counting rules. However,
if $\mq=0$ then the result above vanishes, and so $\delta v_s^2
= O (\nf^2/\nc^2)$ at least. Presumably, this additional
suppression is due to the fact that for massive quarks conformal
invariance is broken explicitly at the classical level, whereas if
$\mq=0$ it is broken only at the quantum mechanical level by the
non-vanishing beta function of the theory in the presence of
fundamental matter. This is proportional to $\nf/\nc$, leading to an
additional suppression. In the gravitational description this is
most easily understood at zero temperature. In this case the
D3-brane background is exactly $AdS_5 \times S^5$, and the
isometries of the first factor correspond to the conformal group in
four dimensions. Adding D7-brane probes with non-zero quark mass
breaks the conformal isometries, and hence this effect is visible at
order $\nf/\nc$. Instead, if $\mq=0$ then the branes' worldvolume is
$AdS_5 \times S^3$, which preserves all the $AdS$ isometries. Hence
in this case one must go beyond the probe approximation to see the
breaking of conformal invariance, \ie beyond $O(\nf/\nc)$.

\subsection{Meson spectrum}

As discussed earlier, introducing the D7-brane probes into the black
D3-brane geometry corresponds to adding dynamical quarks into the
gauge theory.  The resulting theory has a rich spectrum of mesons,
\ie quark-antiquark bound states.  Since the mesons are dual to open
strings with both ends on the D7-brane, the mesonic spectrum can be
found by computing the spectrum of D7-brane fluctuations.  For
temperatures below the phase transition, $T< \tf$, corresponding to
Minkowski embeddings of the D7-branes, we expect the spectrum to
exhibit a mass gap and be discrete, just as found at $T=0$
\cite{us-meson,holomeson,holomeson2}; this is confirmed by our
calculations below -- similar calculations have also appeared
recently in \cite{melt}. For temperatures above the phase
transition, corresponding to black hole embeddings, the spectrum
will be continuous and gapless. Excitations of the fundamental
fields in this phase are however characterised by a discrete
spectrum of quasinormal modes, in analogy with \cite{quasi}.
Investigations of the black hole phase appear elsewhere
\cite{melt,spectre}.

\subsubsection{Mesons on Minkowski embeddings}

In this section we compute the spectrum of low-lying mesons
corresponding to fluctuations of the D7-brane in the black D3-brane
geometry \reef{D3geom}. The full meson spectrum would include
scalar, vector and spinor modes. For simplicity, we will only focus
on scalar modes corresponding to geometric fluctuations of the brane
about the embeddings determined in section \ref{bedtime}. We will
work with the $(R,r)$ coordinates introduced in eq.~\reef{coord2},
in which case the background embedding is given by $R=R_v(r)$,
$\phi=0$, where the subscript $v$ now indicates that this is the
`vacuum' solution. Explicitly, we consider small fluctuations $\dr,
\dphi$ about the background embedding:
\beq
R=\rv (r)+\dr \, , \quad \phi = 0 + \dphi.
\eeq
The pullback of the bulk metric \reef{D3geom} to this embedding is
\beqar
ds^2 &=& \frac{1}{2} \left(\frac{u_0 \rho}{L}\right)^2
\left[-{f^2\over \tilde f}dt^2 + \tilde{f} dx_3^2 \right]
+ \frac{L^2}{\rho^2}\left[ (1+\drv^2 )dr^2 +r^2 d \Omega_3^2
+2 ( \partial_a \dr) \dot{\rv} dx^a d r \right] \\
&&+ \frac{L^2}{\rho^2}\left[ (\partial_a \dr)(\partial_b \dr)dx^a
dx^b+ (\rv +\dr)^2 (\partial_a \dphi)(\partial_b \dphi)dx^a dx^b
\right] \,, \eeqar
where the indices $a,b$ run over all D7 worldvolume directions.
Using the DBI action, the D7-brane Lagrangian density to quadratic
order in the fluctuations is
\beqa \mathcal{L} &=& \mathcal{L}_0 -T_\mt{D7}  \frac{u_0^4}{4} r^3 \sqrt{h}
\sqrt{1+\drv^2}\left\{\frac{1}{2}
\frac{L^2}{\rhov^2}\left(1-\frac{1}{\rhov^8} \right) \sum_a g^{aa}
\left(\frac{(\partial_a \dr)^2}{1+\drv^2} +\rv^2 (\partial_a
\dphi)^2 \right)
\right.\nonumber \\
&& + \left. \frac{4 \rv \drv \partial_r (\dr)^2}{\rhov^{10}
(1+\drv^2)}+\frac{4(\dr)^2}{\rhov^{10}} -
\frac{40\rv^2(\dr)^2}{\rhov^{12}} \right\} \,, \labell{lagFluc} \eeqa
where  $\mathcal{L}_0$ is the Lagrangian density for the vacuum
embedding:
\beq
\mathcal{L}_0 = - T_\mt{D7} \frac{u_0^4}{4} r^3 \sqrt{h}  \sqrt{1+\drv^2}
\left(1-\frac{1}{\rhov^8} \right).
\eeq
Here $\rhov ^2 = r^2 +\rv^2 $ and $h$ is the determinant of the
metric on the $S^3$ of unit radius.  The metric $g_{ab}$ in the
first line of \reef{lagFluc} is the induced metric on the D7-brane with the
fluctuations set to zero:
\beq
ds^2(g) = \frac{1}{2} \left(\frac{u_0 \rho_v}{L}\right)^2
\left[-{f^2\over \tilde f}dt^2 + \tilde{f} dx_3^2 \right]
+ \frac{L^2}{\rho_v^2}\left[ (1+\drv^2 )dr^2 +r^2 d \Omega_3^2 \right] \,.
\eeq
Note that integration by parts and the equation of motion for $\rv$
allowed terms linear in $\dr$ to be eliminated from the Lagrangian
density. The linearised equation of motion is
\beqar
\partial_a \left[\frac{L^2 r^3 f \tilde{f} \sqrt{h}}{\rhov^2 \sqrt{1+\drv^2}}
 g^{aa}
\partial_a(\dr) \right] = 8 \sqrt{h} \left[ \frac{r^3}{\rhov^{10}} \sqrt{1+\drv^2}
\left(1- \frac{10 \rv^2 }{\rhov^{2}}\right) -
\partial_r \left(\frac{r^3 \rv \drv}{\rhov^{10}\sqrt{1+\drv^2}}
\right)\right] \dr \eeqar
for $\dr$ and
\beq
\partial_a \left[{r^3 f \tilde{f}  \sqrt{h}}\rv^2 {\sqrt{1+\drv^2}}
\frac{L^2}{\rhov^2}
g^{aa} \partial_a(\dphi) \right] = 0
\eeq
for $\dphi$. Summation over the repeated index $a$ is implied.

We proceed by separation of variables, taking
\beq
\dphi = \mathcal{P}(r) \, \mathcal{Y}^{\ell_3} (S^3)\, e^{-i \omega t} e^{i{\bf k}
\cdot {\bf x}} \, , \quad \dr = \mathcal{R}(r)\, \mathcal{Y}^{\ell_3} (S^3) \,
e^{-i \omega t} e^{i{\bf k} \cdot {\bf x}} \,,
\eeq
where $\mathcal{Y}^{\ell_3} (S^3)$ are spherical harmonics on the
$S^3$, satisfying $\nabla^2_{S^3} \mathcal{Y}^{\ell_3} =
-{\ell_3} ({\ell_3}+2)\,\mathcal{Y}^{\ell_3}$.  The equation
of motion for the angular fluctuations becomes
\beq
\partial_r \left[ \frac{r^3f \tilde{f}\rv^2}{\sqrt{1+\drv^2}}\partial_r \mathcal{P}
\right] + r^3 f \rv^2 \sqrt{1+\drv^2} \left[\frac{2}{\rhov^4} \left(
\frac{\tilde{f}^2}{f^2} \tom^2 -\tk^2 \right)-\frac{\ell_3 (\ell_3+2)}{r^2}
\tilde{f} \right] \mathcal{P} = 0\, ,
\labell{phiEom}
\eeq
while for the radial fluctuations we have:
\beqar
&&\partial_r \left[ \frac{r^3f \tilde{f}}{(1+\drv^2)^{3/2}}\partial_r
\mathcal{R}  \right] + \frac{r^3 f}{\sqrt{1+\drv^2}} \left[
\frac{2}{\rhov^4} \left( \frac{\tilde{f}^2}{f^2} \tom^2 - \tk^2 \right)
 - \frac{\ell_3 ( \ell_3+2)}{r^2}\tilde{f}   \right]\mathcal{R} \\
&& \qquad \qquad = 8\left[ \frac{r^3}{\rhov^{10}}  \sqrt{1+\drv^2}\left(
1-\frac{10\rv^2}{\rhov^{2}} \right)
- \partial_r \left( \frac{r^3 \rv \drv}{\rhov^{10} \sqrt{1+\drv^2}}
\right) \right]  \mathcal{R} . \labell{radEom}
\eeqar
In these equations, $\tom$ and $\tk$ are dimensionless and are
related to their dimensionful counterparts via
\beq
\omega^2 = \tom^2 {u_0^2  \over  L^4 }  = \tom^2  \pi^2 T^2 =
\tom^2 {\pi^2  \mbar^2\over m^2}  \,, \labell{tom}
\eeq
and analogously for ${\bf k}$.

We solve these equations using the shooting method.  For each choice
of the three-momentum $\tk$, the angular momentum $\ell_3$, and the embedding $\rv(r)$
(corresponding to one value of quark mass and chiral condensate) we
solve these equations numerically, requiring that with $r_{min}\to 0$,
$\mathcal{P}(r_{min}) =r_{min}^{\ell_3}$ and $\partial_r \mathcal{P} (r_{min}) =\ell_3 r_{min}^{\ell_3-1}$  for
the $\delta \phi$ fluctuations and $\mathcal{R} (r_{min}) =r_{min}^{\ell_3}$ and
$\partial_r \mathcal{R} (r_{min}) =\ell_3 r_{min}^{\ell_3-1}$  for $\delta R$.  Then, as
$\mathcal{P} (r) \sim Ar^{\ell_3} +Br^{-\ell_3-2}$ and $\mathcal{R} (r) \sim Cr^{\ell_3}
+Dr^{-\ell_3-2}$ for some constants $A,B,C,D$ as $r \to \infty$, we tune
$\tom^2$ to find solutions which behave as $r^{-\ell_3-2}$  asymptotically.
\FIGURE[b!]{
\includegraphics[width= 0.9 \textwidth]{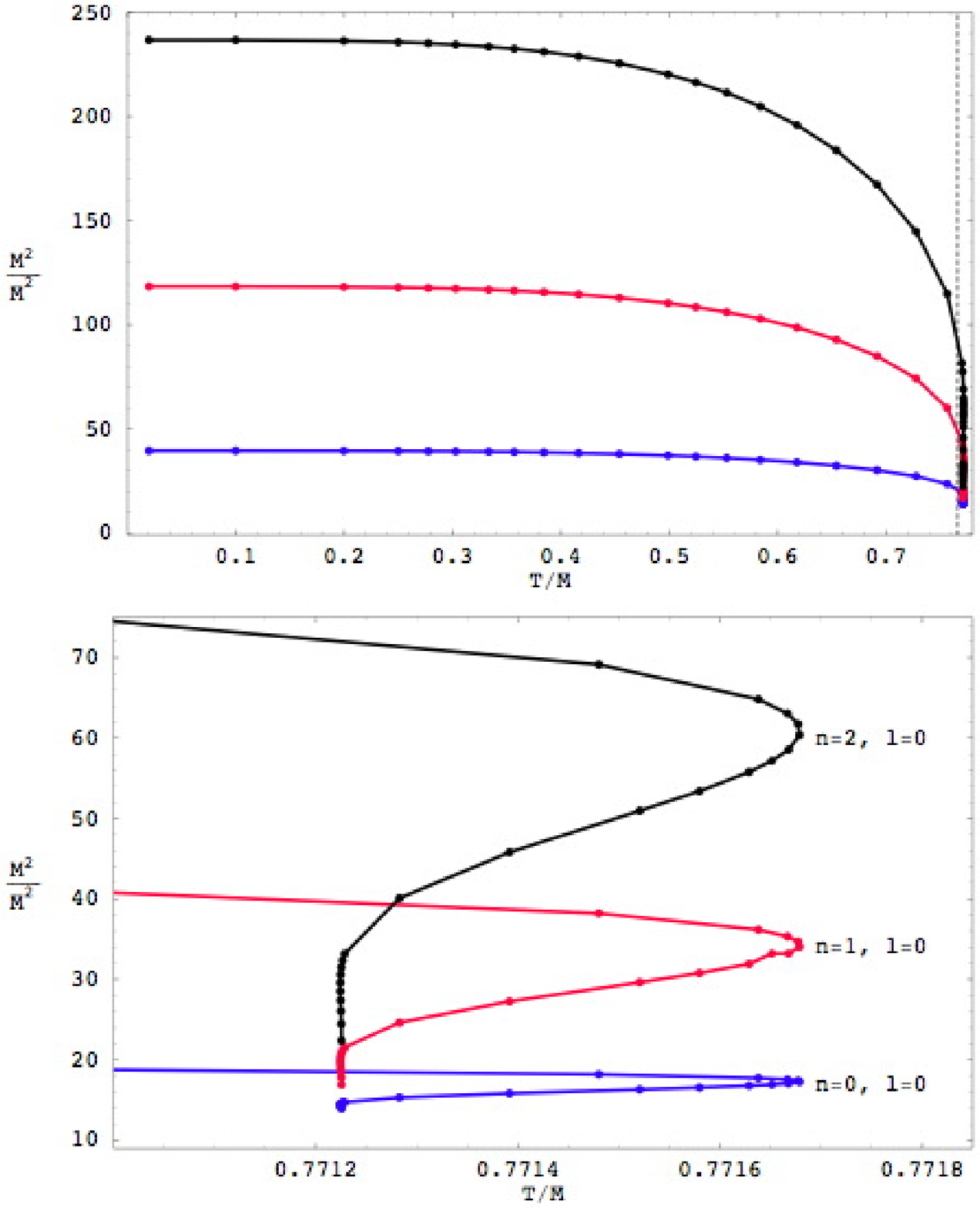} 
\caption{Mass spectrum $M^2=\omega^2|_{k=0}$ for the $\dphi$
fluctuations for Minkowski embeddings in the D3/D7 system.}
\label{mesonMassesPhiW} }
\FIGURE[h!]{
\includegraphics[width= 0.9 \textwidth]{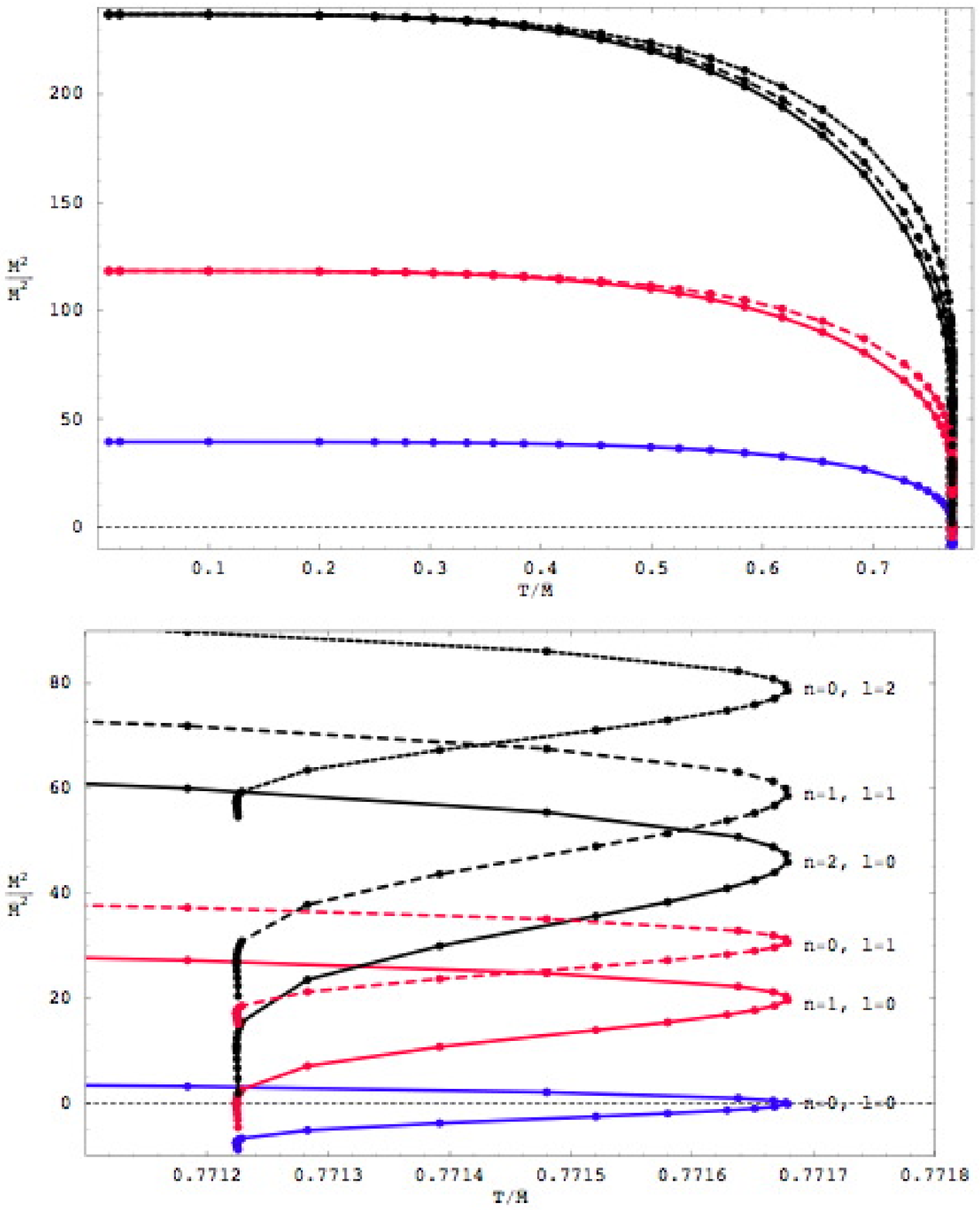} 
\caption{Mass spectrum $M^2=\omega^2|_{k=0}$ for the $\dr$
fluctuations for Minkowski embeddings in the D3/D7 system.  Note
that some of the modes are tachyonic.} \label{mesonMassesRW} }

At finite temperature, the system is no longer Lorentz invariant and
so one must consider the precise definition for the meson masses. We
define the `rest mass' of the mesons as the energy $\omega$ with
vanishing three-momentum ${\bf k}$ in the rest-frame of the
plasma.\footnote{Note that this definition differs from
\cite{johanna,recent} which choose $M^2=-{\bf k}^2$ with $\omega=0$.
The latter might better be interpreted as the low-lying masses of a
confining theory in 2+1 dimensions, in analogy to, \eg
\cite{witten,glueball}.} Thus, solving the equations of motion
\reef{phiEom} and \reef{radEom} with $\tk =0$ yields the
dimensionless constants $\tom^2$, which then give the rest masses
through \eqn{tom}.

Plots of the mass spectrum for these modes are given in
figs.~\ref{mesonMassesPhiW} and \ref{mesonMassesRW}. Note that in
the zero-temperature limit, the $\delta R$ and $\dphi$ spectra
coincide with those previously calculated for the supersymmetric D3
background \cite{us-meson,holomeson,holomeson2}. In particular,
using \eqn{mbarD3D7}, the lightest meson in both spectra has a mass
squared matching $M_\mt{gap}^2= 4\pi^2 \mbar^2\simeq 39.5\,\mbar^2$.
The degeneracy between the two different modes arises because
supersymmetry is restored at $T=0$ and both types of fluctuations
are part of the same supermultiplet
\cite{us-meson,holomeson,holomeson2}. At finite $T$, this degeneracy
between $\delta R$ and $\dphi$ modes is broken. For example, at the
phase transition, the mass of the lightest meson is roughly 25\% and
50\% of its zero-temperature value in the $\delta R$ and $\dphi$
spectra, respectively. The supersymmetric spectrum also showed an
unexpected degeneracy in that it only depended on the combination
$n+\ell_3$, where $n$ and $\ell_3$ are the radial and angular
quantum numbers characterising the individual excitations
\cite{us-meson}. Fig.~\ref{mesonMassesRW} illustrates that this
degeneracy is broken at finite temperature, where the masses are
shown for all the modes with $n+\ell_3=1$ and 2. However, this
breaking is not large except near the phase transition.

Both figures show that in general the meson masses decrease as the
temperature increases. As noted above, the thermal shift of the
meson rest mass may be of the order of 25 to 50 percent at the phase
transition. This reduction must reflect in part the decrease in the
constituent quark mass, discussed in appendix \ref{constitution}.
However, the lowering of the meson masses is actually small relative
to that seen for the constituent quark mass. As seen in figure
\ref{constQmass}, at the phase transition, the latter has fallen to
only 2\% of its $T=0$ value. However, the thermal shift of the
mesons becomes even more dramatic near the critical solution. In
particular, embeddings with $R_0 \in (1,1.07)$ possess tachyonic
$\dr$ fluctuations. Note that $R_0=1$ corresponds to the critical
solutions and the phase transition occurs at $R_0\simeq 1.15$, \ie
this is the minimum value of $R_0$ for which the thermodynamically
preferred embedding is of Minkowski type. As discussed above, the
embeddings are not unique in the vicinity of the critical solution
and so physical quantities spiral in on their critical values. As
observed at the end of section \ref{therm77}, the spiralling of the
energy density leads to a negative specific heat and indicates an
instability. It is satisfying to note in the second plot of
fig.~\ref{mesonMassesRW} that the lowest-lying $\dr$-mode becomes
tachyonic at precisely the point where the first turn-around in the
spiral occurs (with $T$). Hence a dynamical instability is appearing
in the Minkowski embeddings, in precise agreement with the
thermodynamic considerations. In fact the second lowest-lying
$\dr$-mode becomes tachyonic at the second turn-around and it seems
to suggest that at the $i$'th turn of the spiral, the $\dr$-mode with
$n=i-1,\ell_3=0$ becomes tachyonic. We have found no other evidence
of instabilities in other modes.
%R
In particular, we have made a detailed examination of the spin-one
mesons corresponding to excitations of the worldvolume gauge field.
In this case, the observed behaviour is very similar to that of the
pseudoscalar $\delta\phi$ modes. It is not surprising that a
dynamical instability manifests itself in these $\dr$-modes, since
in the region near the critical solution, the nonuniqueness that
brings about the phase transition arises precisely because the
branes have slightly different radial profiles $R(r)$.

While a dynamical instability set in for the Minkowski embeddings,
in agreement with the thermodynamic analysis, it is interesting that
this point is away from the phase transition. In particular, the
Minkowski embeddings with $R_0 \in [1.07,1.15]$, namely those
between the point at which the phase transition takes place and the
first turn-around, do not exhibit any tachyonic modes. Thus these
embeddings are presumably meta-stable and might be reached through
super-cooling.

\FIGURE[h]{
\includegraphics[width=\textwidth]{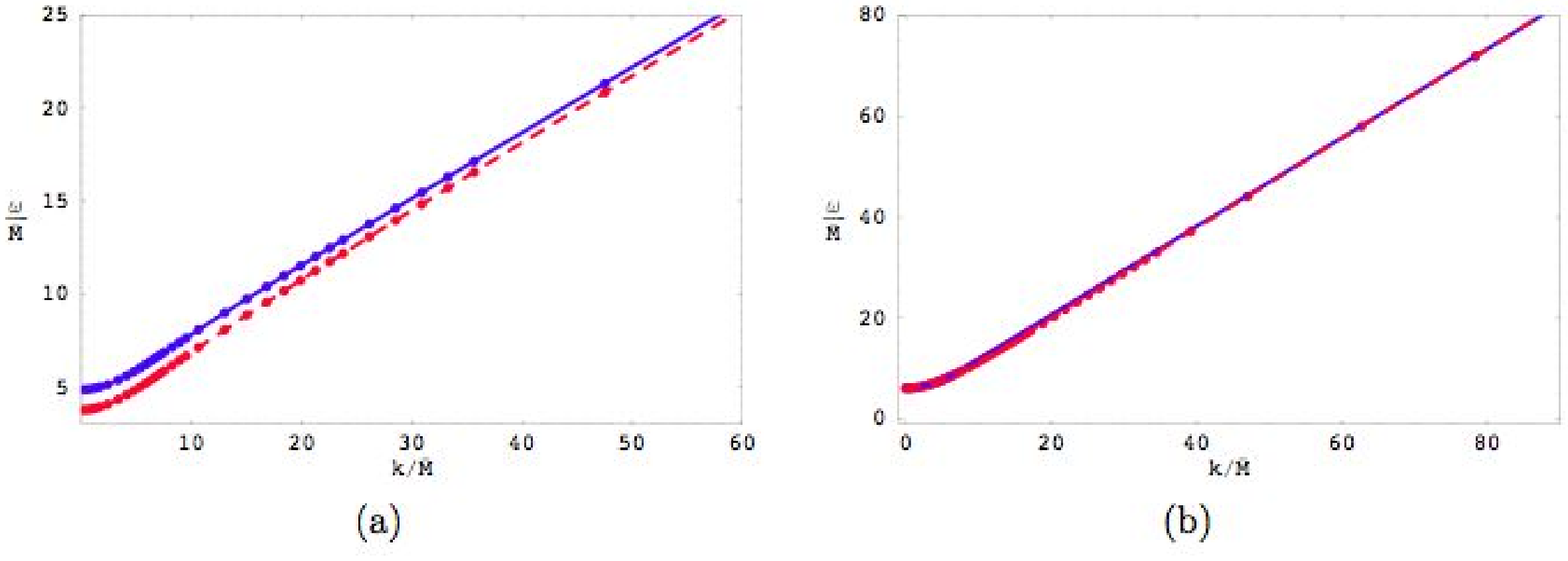} 
\caption{Dispersion relation $\omega(k)$ for  Minkowski D7-brane
embeddings with (a) $R_0=1.20$ ($m=1.32$) and (b) $R_0=2.00 $
($m=2.00$) in a D3-brane background. The solid blue line corresponds
to $\dphi$ fluctuations, whereas the red dashed line corresponds to
$\dr$ fluctuations.} \label{dispersion} }
We have also made some preliminary investigations of these low-lying
mesons moving through the thermal plasma and numerical results are
shown in figure \ref{dispersion}. For non-relativistic motion (small
three-momenta), we expect that the dispersion relation takes the
form
\beq \omega (k)  \simeq M_0 +\frac{{\bf k}^2}{2M_{kin}}
\label{disp1} \,,
\eeq
where $M_0=M_0(T)$ is the rest mass calculated above and
$M_{kin}=M_\mt{kin}(T)$ is the effective kinetic mass for a moving
meson. Although $M_\mt{kin}(T)$ is not the same as $M_0(T)$, for low
temperatures the difference between the two quantities is expected
to be small. For example, fitting the small-$\tk$ results for $\tom$
for the lowest $\delta R$-mode at $T/\mbar=0.5$ (or $R_0=2$) yields
\beq \frac{\omega}{\mbar} = 6.084+0.076 \frac{{\bf k}^2}{\mbar^2
}+\cdots\,. \label{fit1} \eeq
Hence in this case, we find $M_0/\mbar \simeq 6.084$ and
$M_\mt{kin}/\mbar\simeq6.579$. Recall that at $T=0$, we would have
$M_0=M_\mt{kin}=M_\mt{gap}=2\pi\mbar\simeq6.283\mbar$ and so both
masses have shifted by less than 5\%. Note that while the rest mass
has decreased, the kinetic mass has increased. The latter is perhaps
counter-intuitive as it indicates it is actually easier to set the
meson in motion through the plasma than in vacuum. From a gravity
perspective, it is perhaps less surprising as the Minkowski branes
are bending towards the black hole horizon and so these fluctuations
experience a greater redshift than in the pure AdS$_5\times S^5$
background.

\FIGURE[h]{
\includegraphics[width= 0.5 \textwidth]{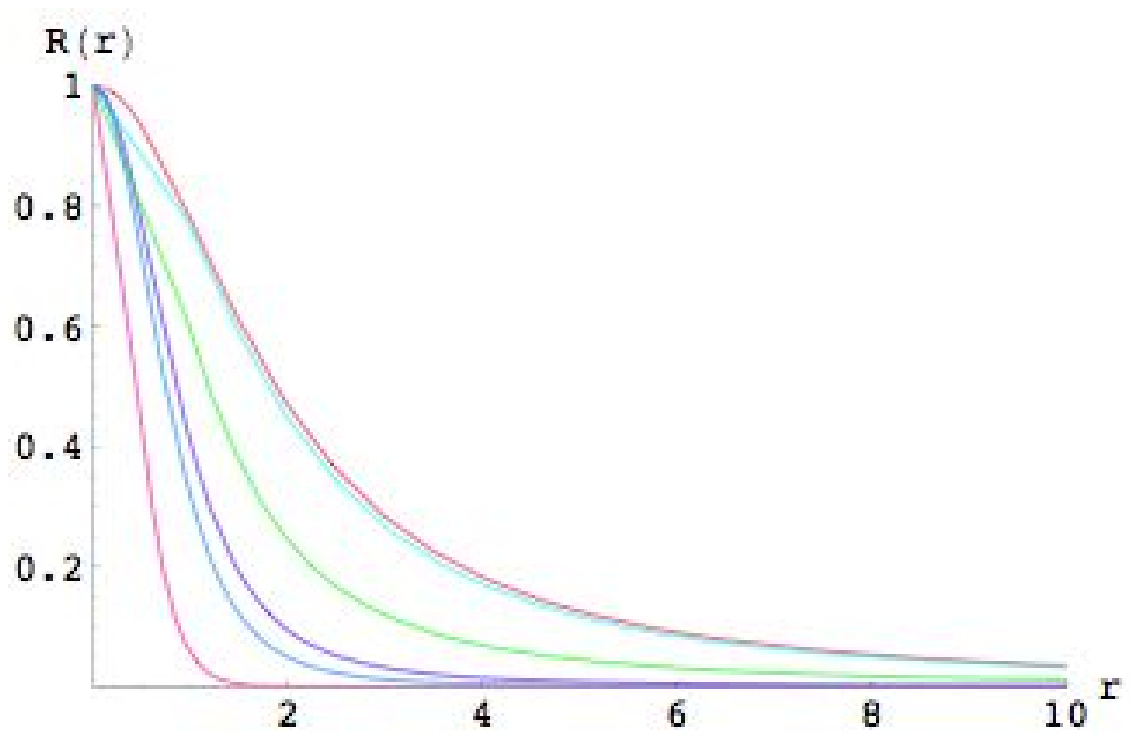} 
\caption{The radial profile for $\delta R$ with $R_0=2$ with various
spatial momentum. From top to bottom, the profiles correspond to:
$k=$ 0, 4.96, 18.81, 31.4, 39.2, 62.7, 94.1.} \label{profileq} }
Examining the regime of large three-momenta, we find that $\omega$
grows linearly with $k$. Naively, one might expect that the constant
of proportionality should be one, \ie the speed of light. However,
one finds that
\beq \frac{\omega}{\mbar} =
v_m\,\frac{k}{\mbar}+\frac{M_1}{\mbar}+O\left(\frac{\mbar}{k}\right)\,.
\labell{fast}\eeq
with $v_m<1$, as illustrated in fig.~\ref{dispersion}. There our
numerical results show that for $R_0=1.2$ ($m=1.32$), $v_m \simeq
0.353$ and $M_1/\mbar \simeq 4.14$ for $\delta R$ and $v_m \simeq
0.350$ and $M_1/\mbar \simeq 4.71$ for $\delta \phi$, while for
$R_0=2$ ($m=2$) $v_m \simeq 0.884$ and $M_1/\mbar \simeq 2.61$ for
either type of fluctuation.  Note that in fig.~\ref{dispersion}b
the dispersion relations $\omega(k)$ for $\dr$ and $\dphi$ are
nearly coincident for all $k$ because supersymmetry is being
restored at low temperatures. Our results show that the strongly
coupled plasma has a significant effect on reducing the maximum
velocity of the mesons. This effect is easily understood from the
perspective of the dual gravity description. The mesonic states have
a radial profile which is peaked near $R_0$, the minimum radius of
the Minkowski embedding, as illustrated in fig.~\ref{profileq}, and
so we can roughly think of them as excitations propagating along the
bottom of the D7-brane. At large $k$, the speed of these signals
will be set by the local speed of light
\beq c =\left. \sqrt{-\frac{g_{tt}}{g_{zz}}}\right|_{r=R_0} =
\frac{f(R_0)}{\tilde{f}(R_0)}\,.\labell{seer}\eeq
The latter gives $c \simeq 0.349$ for $R_0=1.2$  and $c \simeq
0.882$ for $R_0=2$, both of which closely match our results for
$v_m$ given above. It is interesting that at finite temperature as
$k$ increases, the radial profiles of the mesonic states seem to
become more peaked towards $R_0$, as illustrated in
fig.~\ref{profileq}. Recall that at $T=0$, these profiles are
invariant under boosts in the gauge theory directions. Finally we
note that we did not discover any simple relation between $M_1$ in
eq.~\reef{fast} and $M_0$ and $M_{kin}$ in eq.~\reef{disp1}.

Note that with the approximations made here, our analysis reveals no
dragging forces on these low-lying mesons from the thermal bath. We
expect that these would only appear through string-loop effects,
which in particular would include the Hawking radiation of the
background black hole. This would parallel the similar findings for
the drag force experienced by large-$J$ mesons composed of heavy
quarks \cite{messydrag} and by heavy quarks themselves
\cite{herzog,drag}. These large-$J$ mesons also exhibited a maximum
velocity similar to the effect discussed above \cite{messydrag}.

\section{The D4/D6 system}\label{D4D6phase}

We now turn to the D4/D6 system, described by the array
\begin{equation}
\begin{array}{ccccccccccc}
   & 0 & 1 & 2 & 3 & 4& 5 & 6 & 7 & 8 & 9\\
D4 & \times & \times & \times & \times & \times & &  &  & & \\
D6 & \times & \times & \times & \times &   &\times &\times & \times &  &   \\
\end{array}\labell{D4D6}
\end{equation}
In the decoupling limit, the resulting gauge theory is
five-dimensional super-Yang-Mills coupled to fundamental
hypermultiplets confined to a four-dimensional defect. In order to
obtain a four-dimensional gauge theory at low energies, one may
compactify $x^4$, the D4-brane direction orthogonal to the defect,
on a circle. If periodic boundary conditions for the adjoint
fermions are imposed, then supersymmetry is preserved and the
four-dimensional theory thus obtained is non-confining. In this case
the appropriate dual gravitational background at any temperature is
\eqn{metric} with $x^4$ periodically identified. Instead, if
antiperiodic boundary conditions for the adjoint fermions are
imposed, then supersymmetry is broken and the four-dimensional
theory exhibits confinement \cite{witten} and spontaneous chiral
symmetry breaking \cite{us}. The holographic description at
zero-temperature consists then of D6-brane probes in a horizon-free
background, whose precise form is not needed here. At a temperature
$\td$ set by the radius of compactification, the theory undergoes a
first order phase transition at which the gluons and the adjoint
matter become deconfined. In the dual description the
low-temperature background is replaced by \eqn{metric}. If $\td <
\tf$, the D6-branes remain outside the horizon in a Minkowski
embedding, and quark-antiquark bound states survive \cite{us}. As
$T$ is further increased up to $\tf$ a first order phase transition
for the fundamental matter occurs.

Below we study the thermodynamic and dynamical properties of the
D6-branes in the black D4 background appearing above the
deconfinement phase transition. Along the way we will have to
introduce boundary terms to regulate the D6-brane brane action.

\subsection{D6-brane embeddings}

As in section \ref{transit}, we begin by transforming to the
coordinate system with radial coordinate $\rho$ defined in
\reef{rho}, which is better adapted to study the brane embeddings in
the background.  For $p=4$, the radial coordinate is then
\beq (u_0 \rho)^{3/2}=u^{3/2}+\sqrt{u^3-u_0^3} \,, \labell{radius}
\eeq
and the black D4-brane metric is
\beq ds^2 = \frac{1}{2} \left(\frac{u_0 \rho}{L}
\right)^{3/2}\left[ -\frac{f^2}{\tilde{f}} dt^2+ \tilde{f}
dx_4^2\right] + \left(\frac{L}{u_0 \rho} \right)^{3/2}
\frac{u_0^2 \tilde{f}^{1/3}}{2^{{1/3}}}
\left[d\rho^2 +\rho^2 d\Omega_4^2 \right] \,, \labell{D4geom} \eeq
where we now have $f(\rho) = 1-1/\rho^3$ and $\tilde{f}(\rho) =
1+1/\rho^3$. From eq.~\reef{beta}, the temperature is given by
\beq T={3\over4\pi}\left({\om\over L^3}\right)^{1/2}\,.
\labell{period}\eeq
We also have the holographic relations for the dual five-dimensional
gauge theory
\beq  L^3=\pi \gs \nc\ls^3\ ,\qquad \gym^2=4\pi^2\gs\ls \,,
\labell{relatif}\eeq
where we remind the reader that the Yang-Mills coupling $\gym$ is
now dimensionful.

The D4/D6 intersection is described by the array \eqn{D4D6}. In
analogy to the D3/D7 case,  we introduce spherical coordinates
$\{r, \Omega_2\}$ in the 567-directions, and polar coordinates $\{R,\phi \}$
on the $89$-plane.  Computing boundary terms is also facilitated by
introducing an angular coordinate between the $r$ and $R$ directions
so that we have, as before,
\be \rho^2 = r^2 +R^2 \sac r = \rho \sin \theta \sac R = \rho \cos
\theta \,, \ee
and
\beqa
d\rho^2 +\rho^2 d\Omega_4^2 &=&
d\rho^2 +\rho^2(d\theta^2+\sin ^2 \theta \, d\Omega_2^2
+ \cos ^2 \theta \, d\phi^2) \\
&=& dr^2 + r^2 d\Omega_2^2+ dR^2 +R^2 d\phi^2  \, .
\eeqa
Following our analysis for the D3/D7 system, we choose coordinates
on the brane such that asymptotically the metric naturally splits
into a product of the form D4-throat$\times S^2$. We describe the
embedding of the D6-brane in terms of $\chi(\rho)=\cos\theta(\rho)$
-- note then that $\chi=R/\rho$. Later, we will have to regulate the
Euclidean D6-brane by adding local counter-terms written in terms of
this `field.' The induced metric on the D6-brane is then
\beq ds^2 = \frac{1}{2} \left(\frac{u_0 \rho}{L}
\right)^{3/2}\left[ -\frac{f^2}{\tilde{f}} dt^2+\tilde{f}
dx_3^2\right] + \left(\frac{L}{u_0 \rho} \right)^{3/2}
\frac{u_0^2 \tilde{f}^{1/3}}{2^{{1/3}}} \left[\left(
1+\frac{\rho^2 \dot{\chi}^2}{1-\chi^2}\right) d\rho^2 +\rho^2
(1-\chi^2) d\Omega_2^2 \right] \,,
\eeq
where, as usual, $\dot{\chi}=d\chi /d\rho$.  The D6-brane action
takes the form
\beq \frac{\ibulk}{\N} = \int d\rho \, \rho^2 \,
\left(1-\frac{1}{\rho^6}\right) \sqrt{(1-\chi^2)(1-\chi^2+\rho^2
\dot{\chi}^2)} \,, \labell{D6action} \eeq
where $\N$ is given by  \reef{N} with $n=2$:
\be \N = \frac{\pi}{T}\nf T_\mt{D6} u_0^3=
\frac{2^2}{3^6}\,\nf\,\nc\,\leff(T)^4\,T^3\,, \ee
where $\leff(T)^2=\gym^2\nc T$. The resulting equation of motion is
\beq
\partial_\rho \left[ \rho^4 \left( 1-\frac{1}{\rho^6}\right)
\frac{\sqrt{1-\chi^2}\dot{\chi}}{\sqrt{1-\chi^2+\rho^2
\dot{\chi}^2}} \right] + \rho^2 \left(1-\frac{1}{\rho^6}\right) \chi
\left[\sqrt{\frac{1-\chi^2+\rho^2
\dot{\chi}^2}{1-\chi^2}}+\sqrt{\frac{1-\chi^2}{1-\chi^2+\rho^2
\dot{\chi}^2}}\right]=0 \labell{resulting} \,,
\eeq
and $\chi$ asymptotically approaches zero as
\beq
\chi = \frac{m}{\rho} + \frac{c}{\rho^2} + \cdots \,, \labell{asympD6}
\eeq
with $m$ and $c$ related to the quark mass and condensate via
eqs.~\reef{mc} and \reef{donc} with $p=4, n=2$:
\beqa \mq &=& \frac{u_0 m}{2^{5/3}\pi
\ell_s^2}={2^{1/3}\over3^2}\,\leff(T)^2\,T\,m \,, \labell{mc4}\\
\langle {\cal O}_m \rangle &=& -2^{5/3}\pi^2 \ell_s^2 \nf T_\mt{D6}
u_0^2 c = -\frac{2^{5/3}}{3^4}\,\nf\,\nc\,\leff(T)^2 T^3 c \,.\labell{donc4}
\eeqa
In this case, we may write $m=\mbar^2/T^2$ with
\be \mbar^2 = \frac{9}{2^{1/3} }
\left(\frac{\mq}{\leff(\mq)}\right)^2 \simeq
7.143\left(\frac{\mq}{\leff(\mq)}\right)^2\,. \ee
The scale $\mbar$ is again related to the mass gap in the meson
spectrum of the D4/D6 system at zero temperature. For either
background, the latter must be determined numerically. In the case
of the supersymmetric background, one finds
\cite{holomeson,holomeson2}:
\beq m_\mt{gap}^2 = 8 \pi^2\,(1.67)\,
\left(\frac{\mq}{\leff(\mq)}\right)^2 \simeq
131.9\left(\frac{\mq}{\leff(\mq)}\right)^2 \ \longrightarrow\
\frac{\mbar}{m_\mt{gap}}\simeq 0.233\,. \labell{mbarD4D6}\ee
One finds essentially the same result for the confining D4
background \cite{us}. The similarity of these results is probably a
reflection of the underlying supersymmetric structure of the
five-dimensional gauge theory. In the confining theory, the lowest-lying
meson is a pseudo-Goldstone boson, whose mass is determined by
the Gell-Mann--Oakes--Renner relation, and the latter linear form
extrapolates directly to the supersymmetric result at large $\mq$
\cite{us}.

The equation of motion \eqn{resulting} can of course be recast in
terms of the $R, r$ coordinates as
\beq
\partial_r \left[r^2 \left(1-\frac{1}{\rho^6} \right)
\frac{\partial_r R}{ \sqrt{1+\left( \partial_r R \right)^2}}
\right] = 6\, \frac{r^2}{\rho^8}\, R \sqrt{1+\left(
\partial_r R \right)^2} \,, \labell{D6eomRr}
\eeq
which is again suitable to study the Minkowski embeddings.

For arbitrary $m$ we solved for the D6-brane embeddings numerically.
Black hole embeddings are most simply described in the $\chi, \rho$
coordinates and we used boundary conditions at the horizon:
$\chi(\rho=1)=\chi_0$ and $\dot{\chi}|_{\rho=1}=0$ for various $0
\leq \chi_0 <1$.  For Minkowski embeddings, we used the $R,r $
coordinates and the boundary conditions at the axis were:
$R(r=0)=R_0$ and $\partial_r R|_{r=0}=0$ for $R_0
>1$.  We computed $m$ and $c$ by fitting
the numerical solutions to the asymptotic forms of $\chi$ and $R$
given above. In particular, we produced plots of $c$ versus
$T/\mbar$, as shown in figure \ref{d4d6cPlot}. Again by increasing
the resolution, we are able to follow the two families of embeddings
spiralling in on the critical solution. However, thermodynamic
considerations indicate that a phase transition occurs at the point
indicated in the second plot.
\FIGURE{
\includegraphics[width=\textwidth]{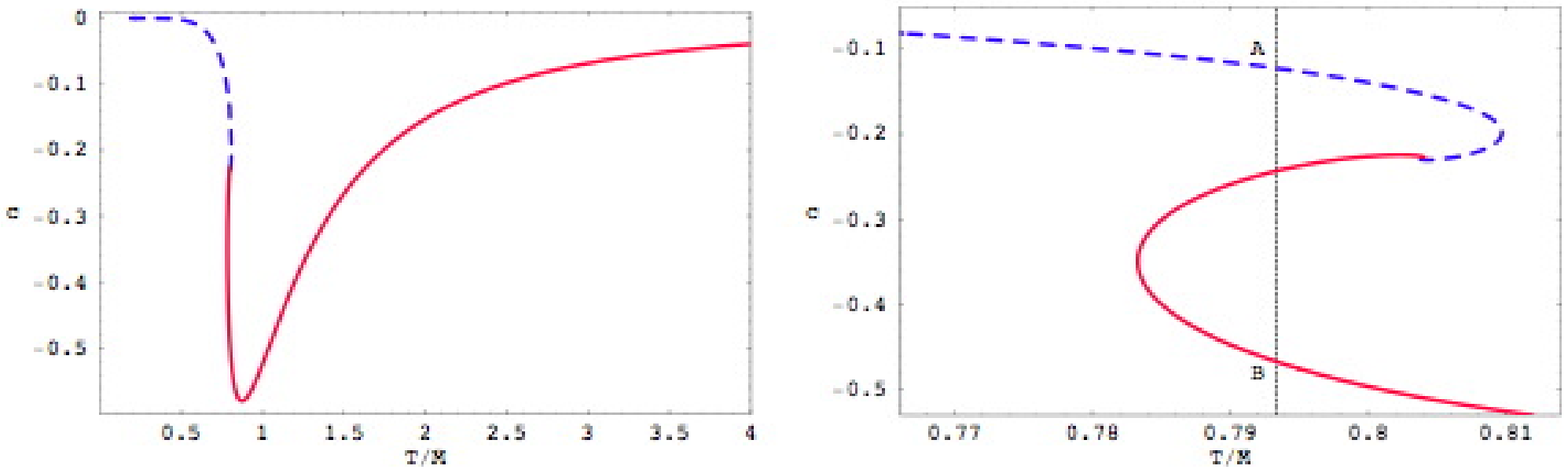} 
\caption{Quark condensate $c$ versus temperature $T/\mbar$ for a
D6-brane in a D4-brane background. The dotted vertical line
indicates the precise temperature of the phase transition. }
\label{d4d6cPlot}}

\subsection{D6-brane thermodynamics}

As with the D3/D7 system, we wish to compute the contribution of the
fundamental matter to the free energy, entropy and energy densities.
That is, we will calculate the contributions of the D6-brane to the
Euclidean path integral. This requires that we regularise and
renormalise the D6-brane action. We will do this by constructing the
appropriate counterterms.

Using the asymptotic behaviour \reef{asympD6} in  \eqn{D6action} we
find that the D6 action contains a UV divergence, since
\beq
\frac{\ibulk}{\N} \simeq \int^\rhomax d\rho \, \rho^2 \, \simeq \frac{\rhomax}{3}
\eeq
diverges for $\rhomax \to \infty$.  We expect the counter-terms that
must be supplemented to have the form $\int d t_\mt{E} d^3 x
\sqrt{\det \gamma}\left(a+b\chi^2+c\chi^4\right)$. In the present
case, we might expect to pick additional factors of $e^\chi$ and
$e^\Phi$. In any event, we would choose the constants to eliminate
the divergence. Further for a supersymmetric embedding, we should be
able to construct the counter-term action so that the total brane
action vanishes.

We take as our ansatz for the counter-terms:
\beq I_\mt{bound} = 4 \pi L^3 T_\mt{D6} \int d t_\mt{E} d^3 x \sqrt{
\det \gamma}\,e^{2\sigma+B\Phi}F(\chi) \Big\vert_{\rho=\pmax} \,,
\labell{bact1d4d6} \eeq
where $B$ and $F(\chi)$ are a dimensionless constant and functional
of $\chi$, both to be determined. We have also defined
$e^{2\sigma}\equiv g_{\theta\theta}$; this factor naturally appears
in the measure as it is proportional to the asymptotic volume of the
internal $S^2$. Now the boundary metric $\gamma_{ij}$ at
$\rho=\pmax$ in the effective five-dimensional (brane) geometry is
given by
\beq
ds^2 (\gamma) =\frac{1}{2}\left({u_0 \pmax\over L}\right)^{3/2}\left(
{f^2(\pmax)\over\tilde{f}(\pmax)}\,d t_\mt{E}^2+\tilde{f}(\pmax)dx_3^2\right)
\labell{boundmetd4d6}
\eeq
and so $\sqrt{\det \gamma}={1\over 4} \left({u_0\pmax\over
L}\right)^3f(\pmax)\tilde{f}(\pmax)$. In this coordinate system we have
\beq e^{2\Phi} = \frac{1}{2} \left(\frac{u_0 \rho}{L} \right)^{3/2}
\tilde{f} = e^{6 \sigma} \labell{fieldsd4d6} \,. \eeq
Now evaluating the counterterm ansatz \reef{bact1d4d6} with the
supersymmetric background ($u_0=0$) with the profile\footnote{See the
discussion below \eqn{bact2}.}
$\chi = m u_0/\trho$, one
finds that the leading divergences cancel if $B=-2/3$ and
$F(0)=-1/3$. One also finds that a complete cancellation occurs if
we choose
\beq
F(\chi)=-{1\over3}(1-\chi^2)^{3/2}\ .\labell{bfuncd4d6}
\eeq
Thus, the complete counter-term action can be chosen as either of
the following:
\beqa I_\mt{bound}&=&-{4\pi\over3} L^3T_\mt{D6} \int d t_\mt{E} d^3
x \sqrt{ \det \gamma} \,e^{2\sigma-2\Phi/3}
(1-\chi^2)^{3/2}\Big\vert_{\rho=\pmax} \,,
\labell{bact3d4d6}\\
{I'}_\mt{bound}&=&-{4\pi\over3} L^3T_\mt{D6} \int  d t_\mt{E} d^3 x
\sqrt{ \det \gamma} \,e^{2\sigma -2\Phi/3}\left( 1
-{3\over2}\chi^2\right)\Big\vert_{\rho=\pmax} \,.
\labell{bact4d4d6}\eeqa
In the second expression, we have kept only the terms which
contribute to the divergence in the small $\chi$ expansion -- the
next term of ${\cal O} (\chi^4)$ vanishes as
$\pmax\rightarrow\infty$. Computationally, this seems like the
easier action with which to work; note however that the first form has the
nice property that, even with finite $\rhomax$, it produces a
precise cancellation  for the supersymmetric configuration, \ie
$I_\mt{D6}=I_\mt{bulk}+I_\mt{bound}=0$.

Proceeding with ${I'}_\mt{bound}$ and using \reef{asympD6}, the
boundary term evaluates to
\beq {I'}_\mt{bound} = -\frac{4\pi T_\mt{D6}}{3 T}  \left( \rhomax^3
-\frac{3}{2} m^2 \rhomax-3 mc \right) \,, \labell{baced4d6}
\eeq
where we have divided out the spatial volume $V_x$ -- see footnote
\ref{foot1}. The total action may then be written as:
\beq \frac{I\mt{D6}}{\N} = G(m) -\frac{1}{3}
\left(\rhomin^3-\frac{3}{2} m^2 \rhomin -3 m c \right) \,, \eeq
where the integral is defined as
\beq G(m) = \int_\rhomin^\infty d\rho \left[\rho^2 \left(
1-\frac{1}{\rho^6}\right) \sqrt{(1-\chi^2)(1-\chi^2+\rho^2
\dot{\chi}^2)} -\rho^2 +\frac{m^2}{2} \right] \,.
\labell{integrald4d6} \eeq
Of course, the free energy follows from this as $F=T I_\mt{D6}$ and
then one can compute the entropy $S=-\partial F / \prt T$ and the energy
$E=F+TS$.  For the computation of the entropy, one must split the
free energy into  bulk and boundary terms and evaluate the action of
the derivative on each of the terms, just as was done for the D3/D7
case.  We do not present all the details of the calculation here but
simply give the final result:
\beq \frac{S}{\N}= -6 \, G(m) + 2 \left(\rhomin^3-\frac{3}{2} m^2
\rhomin -4 m c \right)\, , \eeq
where the integral $G$ was defined in \reef{integrald4d6}.  The
contribution of the D6-brane to the energy then follows as
\beq
\frac{E}{\N T}=-5 \, G(m)+\frac{5}{3} \left(\rhomin^3
-\frac{3}{2} m^2 \rhomin -\frac{21}{5} mc \right) \,.
\eeq
\FIGURE{
\includegraphics[width=\textwidth]{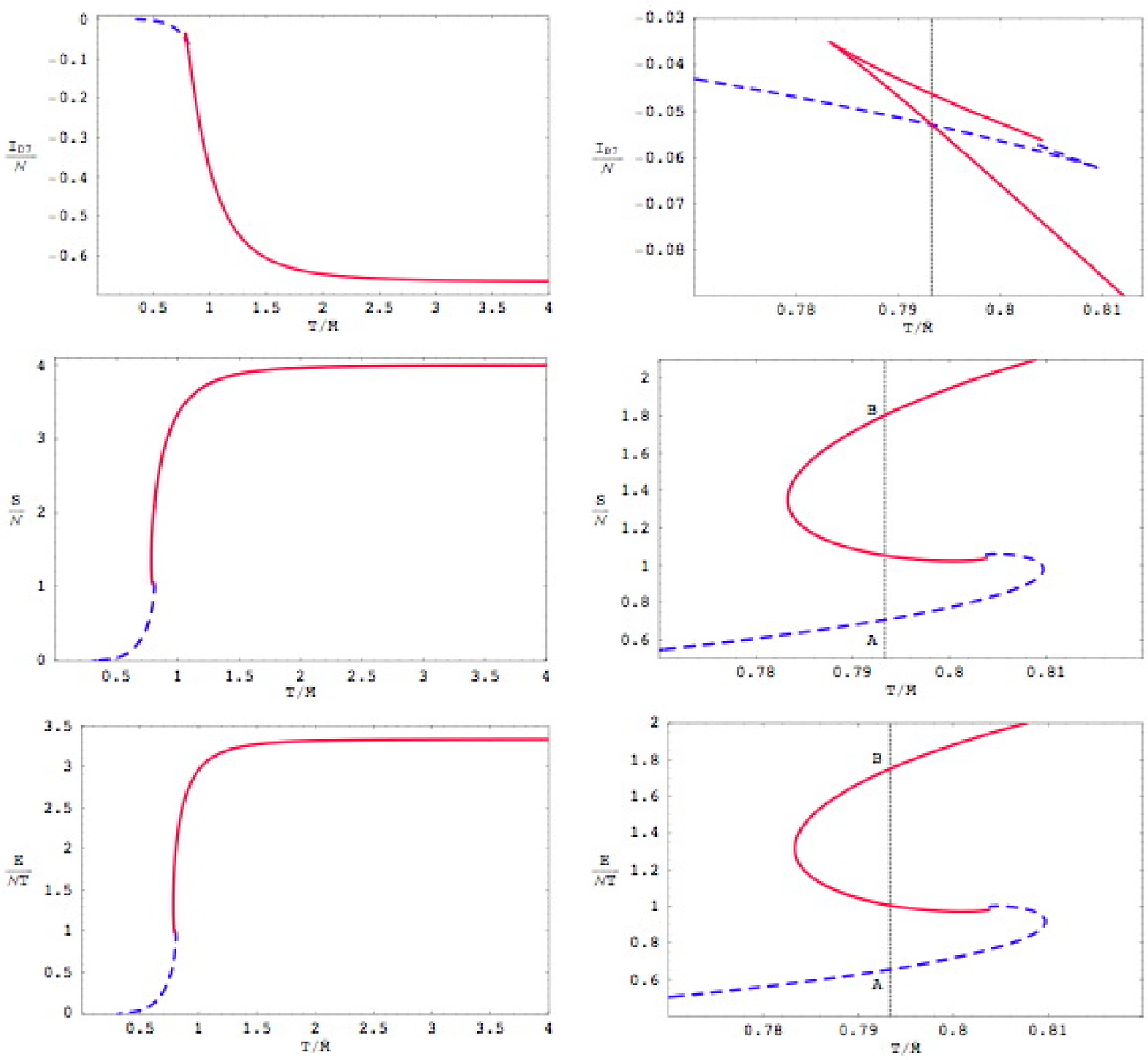} 
\caption{Free-energy, entropy and energy densities for a  D6-brane
in a D4-brane background. The blue dashed (red continuous) curves
correspond to the Minkowski (black hole) embeddings.  The dotted
vertical line indicates the precise temperature of the phase
transition.}
\label{freeEnD4D6}}

Using our numerical results, these thermodynamic quantities are
plotted in fig.~\ref{freeEnD4D6}. Again the free energy density
shows the classic `swallow tail' form and, to our best numerical
accuracy, a first order phase transition takes place at $\tf/\bar{M} =
0.7933$ (or $m=1.589$), where the free energy curves for the
Minkowski and black hole phases cross. The fact that the transition
is first order is illustrated by the entropy and energy densities, which
make a finite jump at this temperature between the points labelled A
and B.

\subsection{Meson spectrum for Minkowski embeddings}

The meson spectrum corresponding to fluctuations of the D6-brane in
the black D4-brane geometry is computed in the same way as for
D3/D7.  We focus here on Minkowski embeddings for which the spectrum
is discrete and stable. The excitations of the black hole embeddings
will be described by a spectrum of quasinormal modes, as discussed
elsewhere \cite{melt,spectre}.

We consider small fluctuations $\dr,\dphi$ about the fiducial
embedding, which we now denote by $\rv$, so that the D6-brane
embedding is specified by $R=\rv(r)+\dr(x^a)$ and $\phi =
0+\dphi(x^a)$, where $\rv(r)$ satisfies \reef{D6eomRr}. The pull-back
of the bulk metric \reef{D4geom} is then
\beq
P[G]_{ab}= g_{ab} + \left(\frac{L}{u_0 \rho} \right)^{3/2}
\frac{u_0^2 \tilde{f}^{1/3}}{2^{1/3}}\left\{
\drv \left[(\partial_a \dr) \delta_b^r +(\partial_b \dr)
\delta_a^r \right]+(\partial_a \dr) \partial_b \dr +R^2
(\partial_a \dphi) \partial_b \dphi \right\} \,,
\eeq
where the metric $g$ is given by
\beq
ds^2(g) = \frac{1}{2} \left(\frac{u_0 \rho }{L}\right)^{\frac{3}{2}}
\left[-\frac{f^2}{\tilde{f}}dt^2 + \tilde{f} dx_3^2 \right] +
\left(\frac{L}{u_0 \rho }\right)^{\frac{3}{2}}\frac{u_0^2
\tilde{f}^{\frac{1}{3}}}{2^{\frac{1}{3}}} \left[(1+ \drv^2)dr^2
+r^2 d\Omega_2^2 \right]
\labell{d6induced}
\eeq
and, as usual, $\rho^2 =r^2+R^2$.  The DBI action yields the
D6-brane Lagrangian density to quadratic order in the fluctuations
$\dr, \dphi$:
\beqa
\mathcal{L} &=& \mathcal{L}_0 -T_\mt{D6} \frac{u_0^3}{4} r^2
\sqrt{h}  \sqrt{1+\drv^2} \left\{ f \tilde{f} \left( \frac{L}{u_0\rhov}\right)^{3/2}
\frac{u_0^2 \tilde{f}^{1/3}}{2^{4/3}}
\sum_a g_v^{aa} \left[\frac{(\partial_a \dr)^2}{1+\drv^2}
+R^2 (\partial_a \dphi)^2 \right]  \right. \nn
&&+ \left.  \frac{3 \rv \drv \partial_r(\dr)^2}{\rhov^8
(1+\drv^2)}+\frac{3 (\dr)^2}{\rhov^8} -\frac{24 \rv^2
(\dr)^2}{\rhov^{10}}  \right\} \,, \labell{lagd4d6fluc}
\eeqa
where $h$ is the determinant of the metric on the $S^2$ of unit
radius, $\rhov^2= r^2+\rv^2$, and  $\mathcal{L}_0$ is the Lagrangian
density for the vacuum embedding:
\beq
\mathcal{L}_0 = -T_\mt{D6} \frac{u_0^3}{4} r^2 \sqrt{h}
\sqrt{1+\drv^2}\left(1-\frac{1}{\rhov^6} \right).
\eeq
Note that terms linear in $\dr$ were eliminated from the Lagrangian
density $\mathcal{L}$ by integration by parts and by using the
equation of motion \reef{D6eomRr} for $\rv$. Since we are retaining
terms only to quadratic order in the fluctuations, the metric $g_v$
in \reef{lagd4d6fluc} is \reef{d6induced} with $R=\rv$ and the
functions $f$ and $\tilde{f}$ in  \reef{lagd4d6fluc} and subsequent
expressions are evaluated at $\rhov$.

The linearised equations of motion for the fluctuations are then
\beq
\partial_a \left[f\tilde{f}
\left( \frac{L}{u_0\rhov}\right)^{3/2} \frac{u_0^2
\tilde{f}^{1/3}}{2^{1/3}}  r^2\sqrt{h}\rv^2
\sqrt{1+\drv^2}\, g_v^{ab}\, \partial_b \dphi \right]=0
\eeq
for $\dphi$, and
\beqar && \partial_a \left[f \tilde{f} \left(
\frac{L}{u_0\rhov}\right)^{3/2} \frac{u_0^2
\tilde{f}^{1/3}}{2^{1/3}}  \frac{r^2\sqrt{h}
}{\sqrt{1+\drv^2}}\, g_v^{ab}\, \partial_b \dr \right]\nn
&&\qquad  \qquad = 6 \frac{r^2}{\rhov^8} \sqrt{h} \sqrt{1+\drv^2} \left(
1-\frac{8 \rv^2}{\rhov^{2}} \right)\dr -6 \sqrt{h}
\partial_r \left( \frac{r^2}{\sqrt{1+\drv^2}} \frac{\rv
\drv}{\rhov^8} \right) \dr \eeqar
for $\dr$. Proceeding via separation of variables, we take
\beq \dphi = \mathcal{P}(r) \mathcal{Y}^{\ell_2}(S^2)\, e^{-i\omega t}e^{i {\bf k} \cdot {\bf
x}},\quad \dr = \mathcal{R}(r) \mathcal{Y}^{\ell_2}(S^2)\,e^{-i\omega t}e^{i {\bf k} \cdot {\bf
x}} \,\eeq
where $ \mathcal{Y}^{\ell_2}(S^2)$ are spherical harmonics on the
$S^2$ of unit radius satisfying
$\nabla^2_{S^2} \mathcal{Y}^{\ell_2} = -\ell_2 (\ell_2+1) \mathcal{Y}^{\ell_2}$.
We obtain the radial differential equation
\beq
\partial_r \left[\frac{r^2 f \tilde{f}\rv^2}{\sqrt{1+\drv^2}}
\partial_r \mathcal{P} \right] + f \rv^2 \sqrt{1+\drv^2}
\left[2^{2/3}\frac{r^2}{\rhov^3}\tilde{f}^{1/3} \left(
\frac{\tilde{f}^2}{f^2} \tom^2-\tk^2 \right) - \ell_2(\ell_2+1)
\tilde{f} \right]\mathcal{P}=0\labell{eomPhid4d6}
\eeq
for $\dphi$ and
\beqar && \partial_r \left[
\frac{r^2 f \tilde{f}}{(1+\drv^2)^{\frac{3}{2}}}\partial_r \mathcal{R}
\right] + \frac{f}{\sqrt{1+\drv^2}} \left[2^{2/3}\frac{r^2}{\rhov^3}
\tilde{f}^{1/3} \left( \frac{\tilde{f}^2}{f^2} \tom^2-\tk^2 \right)
 - \ell_2(\ell_2+1) \tilde{f} \right]\mathcal{R} \nn
&&\qquad  \qquad = 6\left[ \frac{r^2}{\rhov^8}\sqrt{1+\drv^2}
\left( 1-\frac{8R_v^2}{\rhov^2}\right) -\partial_r\left(
\frac{r^2 \rv \drv}{\rhov^8 \sqrt{1+\drv^2}}\right) \right]
\mathcal{R} \labell{eomRadd4d6} \eeqar
for $\dr$.  The dimensionless constant $\tom$ is related to $\omega$
via
\beq
\omega^2 =\tom^2 \frac{u_0}{L^3} = \tom^2 \left(\frac{4\pi}{3}\right)^2 {T^2}
 =\tom^2 \left(\frac{4\pi}{3}\right)^2 \frac{\mbar^2}{m} \, , \labell{tomd4d6}
\eeq
and analogously for $\tk$.

We solved \reef{eomRadd4d6} and \reef{eomPhid4d6} numerically and
determined the eigenvalues $\tom$ using a shooting method, as was
done in the D3/D7 case.  The masses are given by $M^2=\omega^2$ in
the frame in which the three momentum vanishes: ${\bf k}=0$.  The
spectra $M^2/\mbar^2$ versus $T/\mbar$ for the angular fluctuations
$\dphi$ and the radial fluctuations $\dr$ are presented in
figs.~\ref{d4d6mesPhi} and \ref{d4d6mesRad}, respectively (both for
$\ell=0$), and are qualitatively the same as those for the D3/D7
system: the $\dr$ and $\dphi$ modes become degenerate in the
zero-temperature limit, reflecting supersymmetry restoration; in
general the meson masses decrease as the temperature increases,
especially near the critical solution; and the results for $\dr$
fluctuations suggest that a new mode becomes tachyonic at each
turn-around of the curves.
\FIGURE[h!]{
\includegraphics[width=0.9 \textwidth]{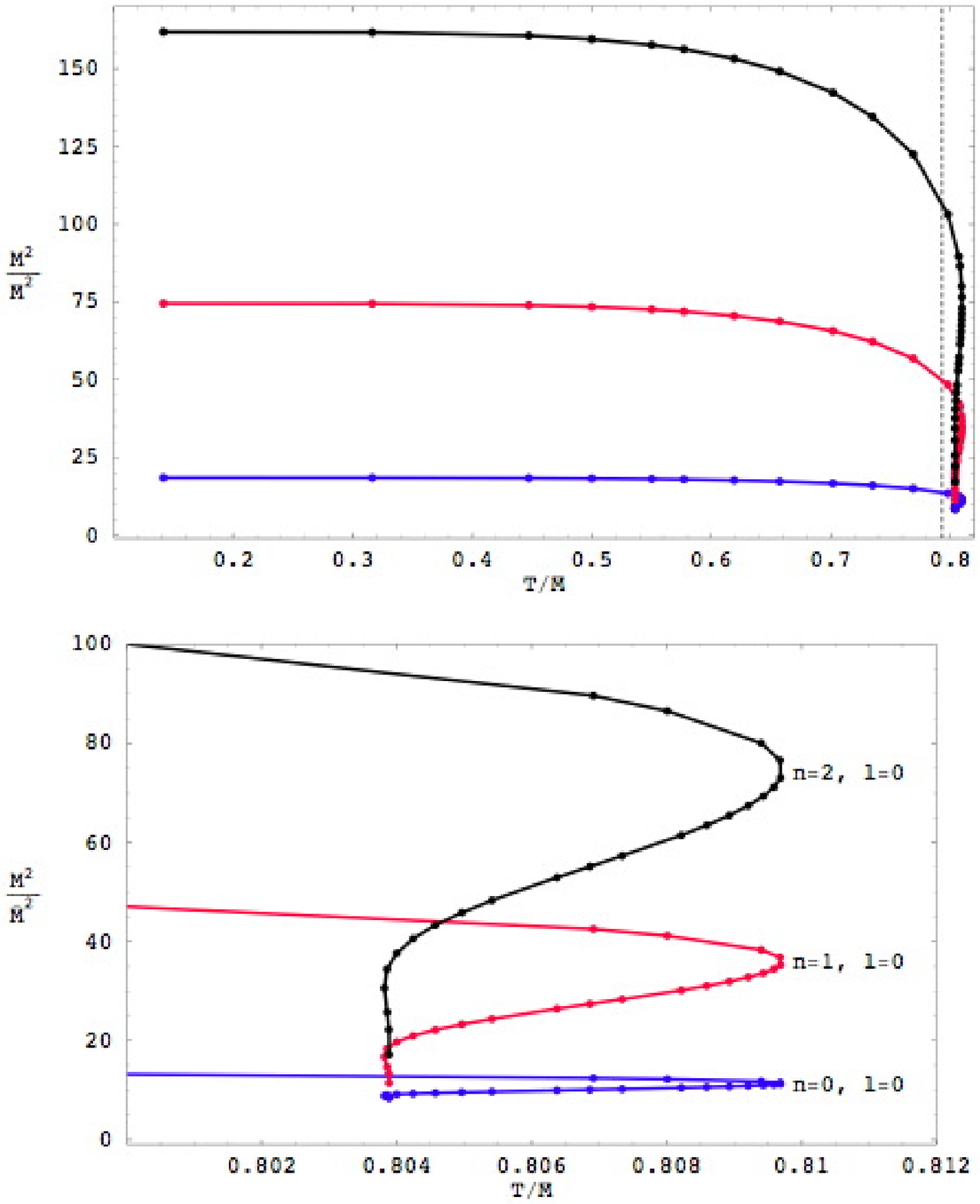} 
\caption{Mass spectrum $M^2=\omega^2|_{k=0}$ for the  $\dphi$
fluctuations for Minkowski embeddings in the D4/D6 system. The
dashed vertical line marks the phase transition.  }
\label{d4d6mesPhi} }

\FIGURE[h!]{
\includegraphics[width= 0.9 \textwidth]{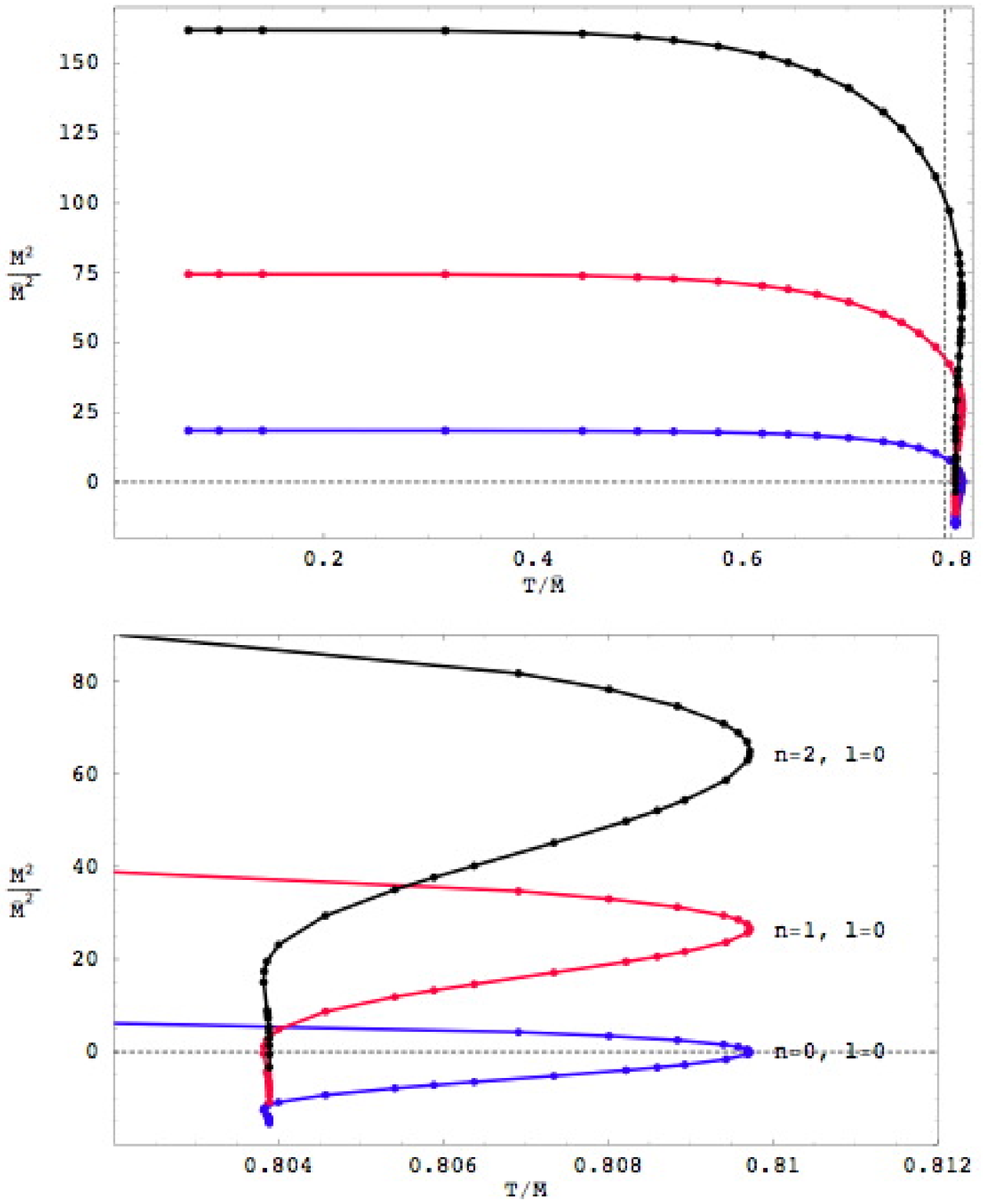} 
\caption{Mass spectrum $M^2=\omega^2|_{k=0}$ for the  $\dr$
fluctuations for Minkowski embeddings in the D4/D6 system. The
dashed vertical line marks the phase transition while the dotted
horizontal line marks $M^2=0$. Note that some modes become
tachyonic.} \label{d4d6mesRad} }

%%%%%%%%%%%%%%%%%%%%%%%%%%%%%%%%%%%%%%%%%%%%%%%%%%%%%%%%%%
%%%%%%%%%%%%%%%%%%%%%%%%%%%%%%%%%%%%%%%%%%%%%%%%%%%%%%%%%%
%%%%%%%%%%%%%%%%%%%%%%%%%%%%%%%%%%%%%%%%%%%%%%%%%%%%%%%%%%

\section{Discussion} \label{discuss}
%R few changes throughout discussion

We have shown that, in a large class of strongly coupled gauge
theories with fundamental fields, this fundamental matter undergoes
a first order phase transition at some high temperature $\tf\sim
\mbar$, where $\mbar$ is a scale characteristic of the meson
physics. As well as giving the mass gap in the meson spectrum
\cite{us-meson}, $1/\mbar$ is roughly the characteristic size of
these bound states \cite{size,holomeson2}. In our models, the gluons
and other adjoint fields were already in a deconfined phase at
$\tf$, so this new transition is not a confinement/deconfinement
transition.
%RT: added sentence and footnote
%D
Neither is it a chiral symmetry-restoration phase transition, since the chiral condensate 
$\cc \propto c$ that breaks the axial $\ua$ symmetry does not vanish above $\tf$.
\footnote{The large-$\nc$ theories under consideration enjoy an exact $\ua$ symmetry, just like QCD at $\nc =\infty$. However, unlike QCD, they do not possess a non-Abelian 
$SU(\nf)_\mt{L} \times SU(\nf)_\mt{R}$ chiral symmetry. Recall also that lattice simulations indicate that, in $\nc=3$ QCD with real-world quark masses, deconfinement and chiral symmetry restoration do not occur with a phase transition but through a smooth cross-over \cite{crossover}.}
The most striking feature of the new phase
transition is the change in the meson spectrum and so we refer to it
as a `dissociation' or `melting' transition.

In the low-temperature phase, below the transition, the mesons are
deeply bound and the spectrum is discrete and gapped. To leading
order in the large-$\nc$ expansion these states are absolutely
stable, but at higher orders they may decay into other mesons of
lower mass or glueballs. The leading channel is one-to-two meson
decay and after examining the interactions in the effective action
\cite{us-meson}, we find that parametrically the width of a typical
state is given by $\mq/(\nc\,\lam^{3/2})\simeq
M_\mt{gap}/(\nc\,\sqrt{\lam})$. Recall that this is not a confining phase
and so we can also introduce free quarks into the system. Of course,
such a quark is represented by a fundamental string stretching
between the D7-branes and the horizon. At a figurative level, in this
phase, we might describe quarks in the adjoint plasma as a
`suspension'. That is, when quarks are added to this phase, they
retain their individual identities.

Above the phase transition (\ie at $T>\tf$), the meson spectrum is
continuous and gapless. The excitations of the fundamental fields
would be characterised by a discrete spectrum of quasinormal modes
on the black hole embeddings \cite{melt,spectre}. Investigations of
the spectral functions \cite{spectre} show that some interesting
structure remains near the phase transition. Some of these
excitations may warrant an interpretation in terms of quasiparticle
excitations but in any event, there are only a few such states in
contrast with the (nominally) infinite spectrum of mesons found in
the low temperature phase. An appropriate figurative
characterisation of the quarks in this high temperature phase would
be as a `solution'. If one attempts to inject a localised quark
charge into the system, it quickly spreads out across the entire
plasma and its presence is reduced to diffuse disturbances of the
supergravity and worldvolume fields, which are soon damped out
\cite{melt,spectre}.

%RCM next two para's changed to incorporate range Tc ~ 151 -- 192 MeV
The physics above is potentially interesting in connection with QCD,
since lattice simulations indicate that heavy-quark mesons indeed
remain bound in a range of temperatures above $\td$. For example,
for the lightest charmonium states, the melting temperature may be
conservatively estimated to be around $1.65 \td \simeq 249$ to $317$
MeV \cite{ccbar,summary}, depending on the precise value of $\td$
\cite{deconfT}. Some other studies suggest that the $J/\psi(1S)$
state may persist to $\sim 2.1\td \simeq 317$ to $403$ MeV
\cite{newcc}. In the D3/D7 model, we see from fig.~\ref{freeEnD3D7}
that quark-antiquark bound states melt at $\tf \simeq 0.766 \mbar$.
The scale $\mbar$ is related to the mass $M^\ast=M_\mt{gap}$ of the
lightest meson in the theory at zero temperature through
eq.~\eqn{mbarD3D7}. Therefore we have \mbox{$\tf (M^\ast) \simeq
0.122 M^\ast$}. For the charmonium states above, taking $M^\ast
\simeq 3000$ MeV gives $\tf (c\bar{c}) \simeq 366$ MeV. Similarly,
for the D4/D6 system we have \eqn{mbarD4D6} which yields $\mbar
\simeq 0.233 M^*$. The transition temperature in this case is then
$\tf \simeq 0.793 \bar{M} \simeq 0.186 M^\ast$, which gives $\tf
(c\bar{c}) \simeq 557$ MeV. Hence it is gratifying that these
comparisons lead to a qualitative agreement with the lattice
results.

Of course, these comparisons must be taken with some caution, since
meson bound states in  Dp/Dq systems are deeply bound, \ie $M^\ast
\ll 2M_q$, whereas the binding energy of charmonium states is a
small fraction of the charm mass, \ie $M_{c\bar{c}} \simeq 2 M_c$.
It might then be more appropriate to compare with lattice results
for $s\bar{s}$ bound states which are also seen to survive the
deconfinement transition. For the $\phi$-meson, whose mass is
$M_\phi \simeq 1020$ MeV, the formulas above yield $\tf (s\bar{s})
\simeq 124$ MeV (D3/D7) and $\tf (s\bar{s}) \simeq 188$ MeV (D4/D6).
Lattice simulations suggest that the melting temperature is around
$1.4 \td \simeq 211$ to $269$ MeV \cite{ssbar,summary}. While again
we have qualitative agreement, one must observe that at least for
the D3/D7 calculation, our result lies below even the lowest
estimate for $\td \simeq 151$ MeV.

An additional caveat is that here we have identified the melting
temperature with $\tf$, above which the discrete meson states
disappear. However, the spectral function of some two-point meson
correlators in the holographic theory still exhibit some broad peaks
in a regime just above $\tf$, which suggests that a few bound states
persist just above the phase transition \cite{spectre}. This is
quite analogous to the lattice approach where similar spectral
functions are used to examine the existence or otherwise of the
bound states. Hence using $\tf$ above should be seen as a (small)
underestimate of the melting temperature.

%RCM new paragraph
Before leaving this discussion of comparisons with QCD, we reiterate
that the present holographic calculations are examining exotic gauge
theories and so any agreements above must be regarded with a
skeptical eye. However, we would also like to point out one simple
physical parallel between all of these systems. The question of
charmonium bound states surviving in the quark-gluon plasma was
first addressed by comparing  the size of the bound states to the
screening length in the plasma \cite{first}. While the original
calculations have seen many refinements (see, \eg \cite{refine}),
the basic physical reasoning remains sound and so we might consider
applying the same argument to the holographic gauge theories.
Considering first the ${\cal N}=2$ SYM theory arising from the D3/D7 system,
the size of the mesons can be inferred from the structure functions
in which the relevant length scale which emerges is
$\sqrt{\lambda}/\mq$ \cite{size}. Holographic studies of Wilson
lines  in a thermal bath \cite{wilson} show that the relevant
screening length of the SYM plasma is order $1/T$. In fact, the same
result emerges from a field theoretic scheme of hard-thermal-loop
resummation applied to SYM theories
\cite{pseudo}. In any event, combining these results, the argument
that the mesons should dissociate when the screening length is
shorter than the size of these bound states yields
$T\sim\mq/\sqrt{\lam}$. Of course, the latter matches the results of
our detailed calculations in section \ref{D3D7phase}. The same
reasoning can be applied to the D4/D6 system where the meson size is
$O(\leff(\mq)/\mq)$ \cite{holomeson2} and the screening length is
again $O(1/T)$ \cite{wilson2}. Hence this line of reasoning again
leads to a dissociation temperature in agreement with the results of
section \ref{D4D6phase}. Therefore we see that the same physical
reasoning which was used so effectively for the $J/\psi$ in the QCD
plasma can also be used to understand the dissociation of the mesons
in the present holographic gauge theories.

One point worth emphasising is that there are two distinct processes
that are occurring at $T\sim\mbar$. If we consider, \eg the entropy
density in fig.~\ref{freeEnD3D7}, we see that the phase transition
occurs in the midst of a cross-over signalled by a rise in $S/T^3$.
We may write the contribution of the fundamental matter to entropy
density as
\beq S_\mt{fun} = \frac{1}{8}\lambda\, \nf\,\nc\,T^3\
H\!\left(\frac{T^2}{\mq^2/\lambda}\right)\labell{entro}\eeq
where $H(x)$ is the function plotted in fig.~\ref{freeEnD3D7}. $H$
rises from 0 at $x=0$ to 2 as $x\rightarrow\infty$ but the most
dramatic part of this rise occurs in the vicinity of $x=1$. Hence
it seems that new degrees of freedom, \ie the fundamental
quarks, are becoming `thermally activated' at $T\sim\mbar$. We might
note that the phase transition produces a discontinuous jump in
which $H$ only increases by about 0.07, \ie the jump at the phase
transition only accounts for about 3.5\% of the total entropy
increase. Thus the phase transition seems to play an small role in
this cross-over and produces relatively small changes in the thermal
properties of the fundamental matter, such as the energy and entropy
densities.

As $\mbar$ sets the scale of the mass gap in the meson spectrum, it
is tempting to associate the cross-over above with the thermal
excitation of mesonic degrees of freedom. However, the pre-factor
$\lambda\, \nf\,\nc$ in \reef{entro} indicates that this reasoning
is incorrect. If mesons provided the relevant degrees of
freedom,\footnote{In fact we will find a contribution proportional
to $\nf^2$ for the mesons coming from the fluctuation determinant
around the classical D7-brane configuration. One can make an analogy
here with the entropy of the adjoint fields of $N=4$ SYM on $S^3$
below the deconfinement transition. In this case, the classical
gravity saddle-point yields zero entropy and one must look at the
fluctuation determinant to see the entropy contributed by the
supergravity modes, \ie by the gauge-singlet glueballs.} we should
have $S_\mt{fun}\propto\nf^2$. Instead the factor of $\nf\nc$ is
naturally interpreted as counting the number of degrees of freedom
associated with free quarks, with the factor $\lambda$ demonstrating
that the contribution of the quarks is enhanced at strong coupling.
A complementary interpretation of \reef{entro} comes from
reorganizing the pre-factor as:
\beq \lambda\,\nf\,\nc=(\gym^2\,\nf)\,\nc^2\,.\labell{reorg}\eeq
The latter expression makes clear that the result corresponds to the
first order correction of the adjoint entropy due to loops of
fundamental matter. As discussed in \cite{viscosity}, we are working
in a `not quite' quenched approximation, in that thermal
contributions of the D7-branes represent the leading order
contribution in an expansion in $\nf/\nc$, and so fundamental loops
are suppressed but not completely. In \cite{viscosity}, it was shown
that the expansion for the classical gravitational back-reaction of
D7-brane is controlled by $\lambda\nf/\nc=\gym^2\,\nf$. Hence this
expansion corresponds to precisely the expansion in loops of
fundamental matter. However, naively the fundamental loops would be
suppressed by factors of $T^2/\mq^2$ coming from the quark
propagators. So from this point of view, the strong coupling
enhancement corresponds to the fact that such factors only appear as
$\lambda\,T^2/\mq^2$ in eq.~\reef{entro}.

Hence the strongly coupled theory brings together these two
otherwise distinct processes. That is, at strong coupling, the
dissociation of the bound states and the thermal activation of the
fundamental matter happen at essentially the same temperature.
While our discussion above focused on the D3/D7 system, the D4/D6
results exhibit the same behaviour. Hence this seems to be a
universal feature of the holographic gauge theories described by
Dp/Dq systems.

The preceding behaviour might be contrasted with that which is
expected to occur at weak coupling. In this regime, one expects that
the melting of the mesons would also be a cross-over rather than a
(first-order) phase transition. Moreover, the temperature at which
the mesons dissociate would be $\tf \sim E_\mt{bind} \sim
g_\mt{eff}^4 \mq$. On the other hand, the quarks would not be
thermally activated until we reach $T_\mt{activ} \sim 2\mq$, at
which point free quark-antiquark pairs would be readily produced. Of
course, the thermal activation would again correspond to a
cross-over rather than a phase transition. The key point, which we
wish to emphasise, is that these two temperatures are widely
separated at weak coupling.

\FIGURE[h!]{
\includegraphics[width= 0.4 \textwidth]{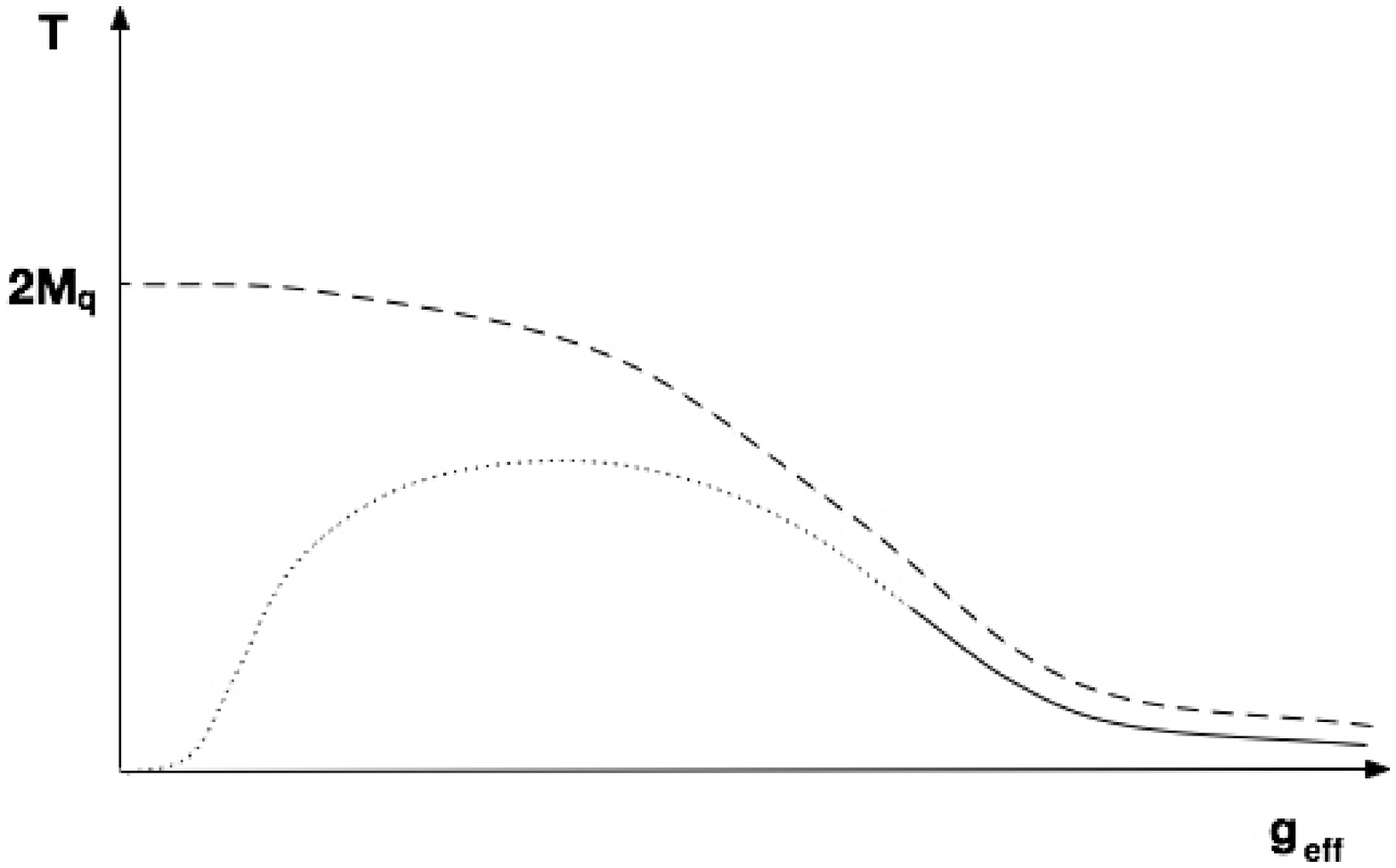}
\caption{A qualitative representation of the simplest possibility
interpolating between the weak and the strong coupling regimes. The
solid and the dotted lines correspond to $T=\tf$. At strong coupling
this corresponds to a first-order phase transition (solid line),
whereas at weak coupling it corresponds to a cross-over (dotted
line). The dashed line corresponds to $T=T_\mt{activ}$. At
strong-coupling this takes place immediately after the phase
transition, whereas at weak coupling it is widely separated from
$\tf$.} \label{weak-strong}
}
Fig.~\ref{weak-strong} is an `artistic' representation of the
simplest behaviour which would interpolate between strong and weak
coupling. One might expect that the melting point and the thermal
activation are very close for $\leff\gg1$. The line of first order
phase transitions must end somewhere and so one might expect that it
terminates at a critical point around $\leff\sim1$. Below this
point, both processes would only represent cross-overs and their
respective temperatures would diverge from one another, approaching
the weak coupling behaviour described above.

There are two aspects to enhancement of the thermal densities
discussed above. First, at strong coupling the fundamental matter
has a stronger effect on the nonabelian plasma than might have been
otherwise guessed and second the effect is a positive one. That is,
\eg the energy and entropy densities are raised. Because we are
working with $\nf/\nc\ll1$, the enhancement we observe is a small
correction to the overall properties of the plasma. In fact, it can
be added to a list of such correction terms, with others arising as
finite-$\lambda$ \cite{up} and finite-$\nc$ effects.
Both\footnote{At finite $\nc$, the classical black hole would be
surrounded by a gas of Hawking radiation which would increase both
the entropy and energy.} of these types of corrections are expected
to raise the entropy and energy densities of the plasma, as well.

As our calculations were also performed in the limit $\nc, \lambda
\rightarrow \infty$ (with $\nf$ fixed), it is natural to ask how the
detailed results of our paper depend on this approximation. First of
all, the fact that the phase transition is first order implies that
it should be stable under small perturbations and so its order and
other qualitative details should hold within a finite radius of the
$1/\nc, 1/\lambda$ expansions. Of course, finite-$\nc$ and
finite-$\lambda$ corrections may eventually modify the behaviour
uncovered here. For example, at large but finite $\nc$ the black
hole will Hawking-radiate and each bit of the brane probe will
experience a thermal bath at a temperature determined by the local
acceleration. This effect becomes more and more important as the
lower part of a Minkowski brane approaches the horizon, and may
potentially blur the self-similar, scaling behaviour found here.
However, at the phase transition, the minimum separation of brane
embeddings and the horizon is not parametrically small. For example,
$R_0 = 1.1538$ at the transition for D3/D7 system. Hence while the
Hawking radiation can be expected to interfere with the self-similar
behaviour near the critical embedding, it should not disturb the
phase transition for large but finite $\nc$.

Finite 't Hooft coupling corrections correspond to higher-derivative
corrections both to the supergravity action and the D-brane action.
These may also blur the details of the structure discussed above.
For example, higher-derivative corrections to the D-brane equation
of motion are likely to spoil the scaling symmetry of eq. \eqn{eom},
and hence the self-similar behaviour. These corrections will again
become important near the critical solution, for both Minkowski and
black hole embeddings, since the (intrinsic) curvature of the brane
becomes large there. However, the phase transition should remain
robust for large but finite $\lambda$ because at this point, the
separation of brane embeddings from the critical solution is not
parametrically small. We illustrated this for the Minkowski
embedding at the phase transition of D3/D7 above but here we can add
the same is true for the black hole embedding at this point, which
has $\chi_0 = 0.9427$.

Another significant set of corrections come from the gravitational
backreaction of the D7-branes (or more generally the probe
Dq-branes) on the background spacetime or from fundamental loops in
the gauge theory. As indicated above, these are dual descriptions of
the same expansion. Our results only represent the first
contribution in an infinite series of terms, whose magnitudes are
controlled by the ratio $\nf/\nc$. Given that low energy QCD has
$\nf/\nc=1$, it is of particular interest to study holographic
theories in Veneziano's limit of $\gym\rightarrow0$,
$\nc\rightarrow\infty$ with both $\lambda$ and $\nf/\nc$ finite
\cite{gabriel}. A variety of attempts have been made to construct
gravitational backgrounds describing gauge theories in this limit
\cite{finiteq}.

The D2/D6 system provides one interesting background where this
limit was studied at finite temperature
\cite{brandeis}.\footnote{The meson spectrum at $T=0$ including the
backreaction of the D6-branes has been studied in \cite{d2d6meson}.}
In particular, it was found that the energy density scales as $F
\sim \nf^{1/2} \nc^{3/2} T^3$, which obviously differs from \eqn{F}
with $p=2, d=2$. This discrepancy is not at all a contradiction and
has the same origin as the discrepancy found for the meson spectrum
\cite{holomeson,holomeson2}. This is the fact that the calculation
in \cite{brandeis} applies in the far infrared of the gauge theory,
whereas that presented here applies at high temperatures, \ie at
$T\gg\gym^2$.

We close with a few more observations. Ref.~\cite{greatballs} argued
for the existence of plasma balls in a broad class of confining
large-$\nc$ theories, which undergo first order deconfinement phase
transitions. That is, in these theories, one could form metastable,
localised lumps of deconfined gluon plasma. Their dual description
should consist of black holes localised along some gauge theory
directions. One may imagine an analogous construction for the
fundamental matter, based on the first order phase transition
discussed here. That is, near $\tf$ one should be able to construct
inhomogenous brane configurations in which only a localised region
on the branes has fallen through the black hole horizon, \ie the
induced brane metric would contain a localised black hole. The dual
gauge theory interpretation would be in terms of a localised bubble
inside of which the fundamental matter has melted. Such bubbles may
be of interest for understanding how the melting transition actually
occurs in a dynamical context.

Finally, we comment on the `quark condensate' at high temperatures.
If one examines fig.~\ref{cond} for example, it is tempting to infer
that, since $c$ approaches zero as $T\rightarrow\infty$, the quark
condensate vanishes in this limit. This vanishing would then be in
agreement with the intuition that at high temperatures the thermal
fluctuations should destroy any coherent condensate. However,
vanishing $c$ is not enough to ensure that $\langle {\cal O}_m
\rangle$ also vanishes. In fact, if we combine eqs.~\eqn{cqD3D7} and
\eqn{cond1}, we see that at high temperatures the condensate
actually grows as
\be \langle {\cal O}_m \rangle \sim \nc \nf \mq T^2 \,.\labell{help}
\ee
At this point, it is important to recall the form of the full
operator ${\cal O}_m$ given in eq.~\reef{oops}. The first two terms
are dimension-three operators and so in the high temperature limit
we can expect the magnitude of typical fluctuations in these to be
$O(T^3)$. Further these operators do not have a definite sign and so
presumably their expectation value vanishes when averaging over all
fluctuations in the disordered high-temperature system. This, of
course, is the basis of the intuition that $\langle\bar{\psi}
\psi\rangle\rightarrow0$. Now the last term in eq.~\reef{oops} is
only a dimension-two operator and so we expect thermal fluctuations
to be of $O(T^2)$. The key difference in this case is that the
operator only takes positive real values and so averaging over all
fluctuations we expect $\langle q^\dagger q\rangle \propto T^2$.
Hence our calculations make a precise prediction for this
expectation value in the high-temperature phase. Note though that
this is a thermal expectation value and not a coherent
(zero-momentum) condensate, which we expect that we are observing
with $\langle{\cal O}_m\rangle\ne0$ at low temperatures.

Hence it is interesting that the high temperature phase seems to
display two distinct regimes of behaviour. At very high
temperatures, the physics is dominated by incoherent thermal
fluctuations of the fundamental fields, as expected. However, there
is also a regime just above the phase transition where the system
can support a coherent condensate. This regime would correspond to
the region where $|c|/T^3$ is still growing in fig.~\ref{cond}. Of
course, there is a cross-over between these two regimes and so there
is only a rough boundary. It may be natural to define the latter as
the point where $c$ is extremized, \ie $T\simeq1.2\mbar$. Again,
this seems to be a universal property of the broad class of
holographic theories described by Dp/Dq systems. For example,
fig.~\ref{d4d6cPlot} indicates the same behaviour for D6-branes in a
D4 background.

The above seems to be one more facet of the rich phenomenology which
these holographic theories display at finite temperature. However,
this phenomenology presents several puzzles, such as why
$\tf\sim\mbar$ rather than $\mq$ is the scale at which the bound
states melt or at which the free quarks are thermally excited. For
example, the former seems counterintuitive in view of the fact that,
in the regime of strong coupling considered here, this temperature
is much lower than the binding energy of the mesons:
\be E_\mt{bind} \sim 2 \mq - \mbar \sim 2\mq \,. \ee
However, this intuition relies on the expectation that the result of
melting a meson is a free quark-antiquark pair of mass $2\mq$. The
gravity description makes it clear that this is not the case at
strong coupling. In fact, the constituent quark mass vanishes when
the branes fall into the horizon -- see appendix \ref{constitution}.
Rather, in this regime the system is better thought of as a strongly
coupled liquid of both adjoint and fundamental fields.

In any event, it is gratifying that the holographic description of
these gauge theories with fundamental matter provides once more an
extremely simple, geometric interpretation of some complicated,
strong-coupling physics, such as the existence or otherwise of
stable quark-antiquark bound states above the deconfinement
temperature. Other well known examples include the geometric
characterisations of confinement \cite{witten,conf} and chiral
symmetry breaking \cite{johanna,us,recent8,sugimoto}.

\acknowledgments We thank Alex Buchel, Sean Hartnoll, Chris Herzog,
Andreas Karch, Pavel Kovtun, Stephen Sharpe, Andrei Starinets and
Laurence Yaffe for helpful discussions and comments. Research at the
Perimeter Institute is supported in part by funds from NSERC of
Canada and MEDT of Ontario. We also acknowledge support from NSF
grant PHY-0244764 (DM), NSERC Discovery grant (RCM) and NSERC
Canadian Graduate Scholarship (RMT). Research at the KITP was
supported in part by the NSF under Grant No. PHY99-07949. RCM would
also like to thank the String Group at the \'Ecole Polytechnique for
their hospitality in the early stages of this work. Research there
is supported by the Marie Curie Excellence Grant,
MEXT-CT-2003-509661.

\appendix

\section{Embeddings for high and low temperatures for D3/D7}\label{approxSol}

\subsection{High temperatures (black hole embeddings)}

Consider the limit $T/\mbar \gg 1$.  This corresponds to black hole
embeddings with $m = \mbar/T \ll 1$. As usual, we use the
$\chi,\rho$ coordinates.  Note that the equatorial D7-brane
embedding, $\chi=0$, is an exact solution of the equation of motion
\reef{psieom}.  To study nearby solutions we expand the bulk portion
of the D7-brane action \reef{act2s} to quadratic order in $\chi$
\beq
{I_\mt{bulk} \over \N} \simeq \int_1 ^\infty d\rho
\left(1-{1\over \rho^8} \right) \rho^3 \left(1 -{3 \over 2}
 \chi^2 +{1\over 2} \dot{\chi}^2 \right) \,,
\eeq
thus obtaining the linearised equation of motion:
\beq
\partial_\rho \left[ \left(1-{1\over \rho^8} \right)
\rho^5 \dot{\chi} \right]=-3 \left(1-{1\over \rho^8} \right)  \rho^3
\chi\,. \eeq
To solve this equation, it is useful to make the change of variables
$x=\rho^2$ so that it becomes:
\beq
x(x^4-1)(4 x \chi'' + 2\chi') +2 x(5x^4+3)\chi' +3 (x^4-1) \chi =0
\eeq
where $\chi ' =d\chi/d x$.  The solution of this equation satisfying
the boundary condition $\chi ' |_{x=1}=0$ is
\beq \tilde{\chi} = {4 \over 45 \left[\Gamma\left({1\over
4}\right)\right]^2} \left[9\Gamma\left({5 \over 4}\right)
\Gamma\left({9\over 4}\right)x^{1/2} F\left({1\over4},{1\over
2};{3\over 4};x^4 \right)-5 \left[\Gamma\left({7\over 4}\right)
\right]^2 x^{3/2} F\left({1\over 2},{3\over 4};{5\over 4};x^4
\right) \right] \,, \labell{psitilde} \eeq
where $F(a,b;c;z)$ is the hypergeometric function satisfying
\beq
z(1-z)F''+[c-(a+b+1)z]F'-abF=0.
\eeq
The overall normalization of the solution is arbitrary since, we are
solving a linear equation.  In the above, we have chosen the
normalization such that
\beq
\tilde{\chi} \simeq 1/x^{1/2}+\tilde{c}/x^{3/2}\, ,\quad  x \to \infty
\eeq
where
\beq
\tilde{c} = {\Gamma\left({-1 \over 4}\right)
\Gamma\left({3 \over 4}\right)^2 \over \sqrt{2}
\pi \Gamma\left({1 \over 4}\right)} \simeq -0.456947. \labell{tildec}
\eeq
The tilde on the solution $\tilde{\chi}$ and condensate $\tilde{c}$
indicate that these is the solution for unit mass.  The general
solution for arbitrary small mass (or equivalently, high
temperatures) is simply $\chi = m \tilde{\chi}$ and the condensate
is given by
\be c=m \tilde{c} \,. \labell{cond1} \ee

\subsection{Low temperatures (Minkowski embeddings)}

Low-temperature solutions correspond to Minkowski embeddings in
which the D7 probe is very far from the horizon: $R_0 \gg 1$ or,
equivalently, $m = \mbar /T \gg 1$. In this case, we expect the
brane profile to be nearly flat, \ie $R(r)$ is approximately
constant. This motivates the ansatz $R(r) = R_0 +\delta R(r)$, where
$R_0$ is a large constant. Substituting into eq.~\reef{eomR} and
expanding to linear order in $\delta R(r)$ gives:
\beq
\partial_r\left[ r^3 \left(1-{1\over (r^2+R_0^2)^4} \right)
\partial_r(\delta R) \right]  = 8 {r^3 R_0 \over (r^2 +R_0^2)^5}  \,.
\labell{eomdeltaR}
\eeq
Integrating \reef{eomdeltaR} and requiring $\partial_r (\delta
R)|_{r=0}=0$ we obtain
\beqa
\delta R (r) &\simeq & -{R_0 \over 3} \int_0 ^r dx {1\over x^3}
\left(1-{1 \over(x^2+R_0^2)^4} \right)^{-1}
\left[{R_0^2 +4 x^2 \over (R_0^2+x^2)^4} - {1 \over R_0^6} \right]
\nonumber \\
&=& -{1 \over 24R_0^5} \left[2(3R_0^4-1)
\arctan\left({r^2 \over 1 +R_0^2(R_0^2+r^2)}\right)
 +(-1-2R_0^2-3R_0^4) \log \left(1+{r^2 \over R_0^2-1}
\right) \right. \nonumber \\
&&+ \left. (1-2R_0^2+3R_0^4) \log
\left(1+{r^2 \over R_0^2+1} \right)+
2 R_0^2 \log \left({1+(r^2 +R_0^2)^2 \over 1+R_0^4}
\right)     \right].
\eeqa
Note that $\delta R|_{r =0}=0$ while the limit $r \to \infty$
yields:
\beqa
\delta R|_{r \to \infty} &\simeq& \left( \frac{1}{12 R_0^5}
- \frac{1}{4 R_0} \right) \left( \frac{\pi}{2} - \arctan(R_0^2) \right)
+ \frac{1}{12 R_0^3} \log \left( \frac{R_0^4+1}{R_0^4-1} \right) \nonumber \\
&& \quad + \frac{1}{8} \left(\frac{1}{3R_0^5} + \frac{1}{R_0}
\right) \log \left( \frac{R_0^2+1}{R_0^2-1} \right)-
\frac{1}{6R_0^5}\frac{1}{r^2} + \cdots  \labell{asympdeltaR} \eeqa
Recall that for these embeddings $R_0 \gg 1$ so indeed $\delta R \ll
R_0$.  Note that $R(r \to \infty) \simeq m + c/r^2$ so that in the
large $R_0$, $m$ limit one has $m \simeq R_0 + \delta R|_{r\to \infty}$ and
$c \simeq -(6R_0^5)^{-1}$. For very large values of $R_0$ we can
expand \reef{asympdeltaR} further to give $m \simeq R_0 + 1/2 R_0^7$
as an approximate expression for the quark mass. Inverting this
relation yields
\be R_0 \simeq m - \frac{1}{2 m^7} \,.
\labell{inverted} \ee
We will apply this result in our discussion of the constituent quark
mass in appendix \ref{constitution}.

\section{Computation of the D7-brane entropy}\label{entropy7}

In order to evaluate the expression for the D7-brane entropy density,
\beq
S = -\frac{\partial F}{\partial T} = -\pi L^2 \frac{\partial F}{\partial u_0} \,,
\labell{entropy-app}
\eeq
we split the free energy into a bulk and a boundary contribution. We
also write pertinent expressions in terms of the dimensionful
variables
\beq \trho = u_0\, \rho  \, , \quad \tc = u_0^3\, c \, , \quad \tm=
u_0\, m \, , \eeq
to explicitly show the dependences on $u_0$, or, equivalently, the
temperature $T$.

From eq.~\reef{bact2},
\beq \fct = T\,\ict =-{\pi^2\over8} T_\mt{D7}\left[
(\trhomax^2-\tm^2)^2-4\tm \tc \right] \labell{freeb} \eeq
so that the boundary contribution to the entropy is
\beq
S_\mt{bound}= -{\pi^3 \over 2} L^2 T_\mt{D7} \tm
\frac{\partial \tc}{\partial u_0}  \,, \labell{sbound}
\eeq
as the quark condensate is the only factor in eq.~\reef{freeb} which
depends on the position of the horizon $\om$. Note that both of the
divergent regulator contributions in eq.~\reef{freeb} have been
eliminated by this differentiation. The bulk contribution to the
free energy is given by
\beq F_\mt{bulk} = T\,I_\mt{bulk} = {\pi^2 \over 2} T_\mt{D7}\om^4
\int^{\trhomax/\om}_{\trhomin/\om}d\rho\,\rho^3
\left(1-{1\over\rho^8}\right)\left(1-\chi^2\right)
\left(1-\chi^2+\rho^2\,\dot{\chi}^2\right)^{1/2}. \labell{freek}
\eeq
When we differentiate this expression with respect to $\om$
following eq.~\reef{entropy-app}, the derivative will act in three
places: i) the overall factor of $\om^4$; ii) the explicit (and
implicit in $\trhomin$) appearance of $\om$ in the end-points of the
integration; and iii) the field $\chi$ which is implicitly a
function of the background mass $\om$.
We consider each of these contributions in turn.  First one has:
\beq
S_i = -2 \pi^3 L^2 T_\mt{D7}\om^3 \int^{\trhomax/\om}_{\trhomin/\om}d\rho\,\rho^3
\left(1-{1\over\rho^8}\right)\left(1-\chi^2\right)
\left(1-\chi^2+\rho^2\,\dot{\chi}^2\right)^{1/2}. \labell{sone}
\eeq
Note that this contribution by itself is divergent in the limit
$\trhomax\rightarrow\infty$.

Next consider the contributions from the end-points. At the lower
end-point, there are two possibilities depending on whether the
brane ends on the horizon or closes off above the horizon. If the
brane ends on the horizon, $\trhomin=\om$ and hence this contribution
vanishes since $\partial_\om(\trhomin/\om)=0$. (The integrand also vanishes
when evaluated at $\rho_{min}=\trhomin/\om=1$.) If the brane closes off
above the horizon, $\partial_\om(\trhomin/\om)$ is nonvanishing but this
contribution vanishes because $\chi=1$ at the end-point. Hence only
the upper end-point at $\trhomax$ makes a contribution:
\beqa
S_{ii}&=&-  {1\over2}\pi^3 L^2T_\mt{D7}\om^4 \left[\rho^3
\left(1-{1\over\rho^8}\right)\left(1-\chi^2\right)
\left(1-\chi^2+\rho^2\,\dot{\chi}^2\right)^{1/2}
\right]_{\trhomax/\om}\times\left(-{\trhomax\over\om^2}\right)\nonumber\\
&=&{1\over2\om}\pi^3L^2T_\mt{D7}\left(\trhomax^4-\tm^2\trhomax^2\right)\ .
\labell{stwo}
\eeqa
where we have substituted the asymptotic expansion \reef{asymp} for
$\chi$ in the second expression.

Finally, we consider the contributions from the dependence of $\chi$
on $u_0$.  In this case, $\partial_\om \chi$ inside the integral can
be considered as a variation $\delta \chi$.  Hence after an
integration by parts, this derivative yields the bulk equation of
motion for $\chi$ inside the integral and a boundary term coming
from the integration by parts. Since $\chi$ solves the equation of
motion, only the boundary term contributes to the entropy with
\beq
S_{iii}=- {1\over2}\pi^3
L^2T_\mt{D7}\om^4\left[\rho^5\left(1-{1\over\rho^8}\right){1-\chi^2\over
\left(1-\chi^2+\rho^2\,\dot{\chi}^2\right)^{1/2}}\dot{\chi}\partial_\om\chi
\right]_{\trhomin/\om}^{\trhomax/\om}\ .\labell{sthree1}
\eeq
Arguments similar to those above show that the contribution at the
lower endpoint vanishes. If the brane ends on the horizon, the
second factor inside the brackets vanishes and also $\dot{\chi}$
vanishes at the horizon. If the brane closes off above the horizon,
$\chi=1$ at the lower end-point and so the numerator in the third
factor vanishes and also $\partial_\om\chi=0$. Hence again, only the
upper end-point contributes to the entropy. In order to correctly
evaluate this expression, we express the asymptotic expansion
\reef{asympD7} in terms of $\tm,\tc$:
\beq
\chi = {\tm /\om \over \rho} + {\tc/\om^3 \over \rho^3} +
\cdots \labell{asymp2}
\eeq
Then in eq.~\reef{sthree1}, we have
\beqa
\dot{\chi}= \partial_\rho \chi &=& -{\tm /\om \over \rho^2}
-3 {\tc/\om^3 \over \rho^4} + \cdots \nonumber \\
\partial_\om \chi &=&  -{\tm /\om^2 \over \rho}
-3 {\tc/\om^4 \over \rho^3}
+{\partial_\om \tc / \om^3\over \rho^3} + \cdots \labell{junk}
\eeqa
Note that it would be incorrect to evaluate $\partial_\om\chi
\simeq\partial_\om(\tm/\rho)=0$ because in the integral we have
assumed that $\partial_\om\rho=0$ and so $\partial_\om\trho\not=0$.
Inserting these expansions in \reef{sthree1} yields
\beq
S_{iii}=- {1\over2\om}\pi^3
L^2T_\mt{D7}\left[\tm^2\trhomax^2+6\tm\tc-\tm^4-\om(\partial_\om \tc)\tm
\right]\ .\labell{sthree2}
\eeq
Finally, gathering all the entropy contributions yields:
\beqa S&=&S_i+S_{ii}+S_{iii}+S_\mt{bound} \nonumber\\
&=&- {1\over2\om}\pi^3
L^2T_\mt{D7}\left[4\om^4\int_{\pmin/\om}^{\pmax/\om} d\rho\,\rho^3
\left(1-{1\over\rho^8}\right)\left(1-\chi^2\right)
\left(1-\chi^2+\trho^2\,\dot{\chi}^2\right)^{1/2}\right.\nonumber\\
&&\qquad\qquad\qquad\qquad\left.-\pmax^4+2\tm^2\pmax^2+6\tm\tc-\tm^4
\vphantom{{1\over\trho^8}}\right]\ .\labell{sss5}
\eeqa
Note that the boundary terms have provided precisely the correct
$\pmax$ terms to regulate the integral. Hence using \reef{N},
\reef{junk9} and \reef{integral}, we can express the final result
for the entropy as
\beq
{S \over \N} = -4 G(m)+(\rhomin^2-m^2)^2 -6mc \,.
\eeq

\section{Positivity of the entropy}
\label{kinks}

Here we present an analytic proof that the plot of the Dq-brane
probe Euclidean action $\ibk$ versus $m$ must exhibit mathematical
kinks and not just rapid turn overs.\footnote{For the sake of this
discussion it is irrelevant whether we plot $\ibk$ versus $m$ or
versus $1/m$, as in the main text.}  Recall that this is necessary
for the entropy $S=-\prt F/\prt T$ to be positive. We focus here on
the case of black hole embeddings of the D7-brane in the D3-brane
background for concreteness, but the analogous arguments applies to
Minkowski embeddings and to other Dp/Dq systems.

The argument proceeds by thinking of the plot $\ids(m)$ as a
parametric plot $(m(\chi_0),\ids(\chi_0))$, where $\chi_0$, which
plays the role of the parameter along the curve, is the value of
$\chi$ at the `horizon' $\rho=1$.  This is in fact the way we
construct the plot: We choose $\chi_0$ as a boundary condition at
the horizon and we integrate the differential equation `outwards',
thus obtaining a solution $\psi (\rho; \chi_0)$, from whose
asymptotic behaviour we read off $m(\chi_0)$ and $c(\chi_0)$.
Substituting the solution into the D7-brane action we then obtain
$\ids(\chi_0)$.

Now the key observation is that if the tangent vector to the curve
never vanishes, then there can be no kinks. In order to have a kink
there must be a point at which both $m' \equiv \partial m / \partial
\chi_0$ and $I' \equiv \partial \ids / \partial \chi_0$ vanish
simultaneously. We know that there are certainly an infinite number
of points at which $m'=0$, because close to criticality the function
$m(\chi_0)$ is an oscillatory function with both maxima and minima.
We will now see that at each of these points we also have $\ids'
=0$.

The renormalised D7-brane action is $\ids = I_\mt{bulk}+ \ict$, with
\be I_\mt{bulk} = \int_{\rhomin}^{\rhomax} d\rho \, {\cal L} (\chi,
\dot{\chi} )= \int_{\rhomin}^{\rhomax} d\rho \left( 1-
\frac{1}{\rho^{8}} \right)
 \rho^3\, {(1-\chi^2)} \sqrt{ 1-\chi^2 + \rho^2 \dot{\chi}^2  } \,,
\labell{i} \ee
and
\be
\ict = - \frac{1}{4} \left(\rhomax^4 -2 m^2 \rhomax^2 - 4 m c + m^4 \right) \,,
\ee
where we have set ${\cal N}=1$ for simplicity. Using the equation of
motion, we see that the derivative of $\ids$ is
\be \ids' = \frac{\partial \ids}{\partial \chi_0} = \left[
\frac{\partial {\cal L}}{\partial \dot{\chi}} \frac{\partial
\chi}{\partial \chi_0} \right]_{\rhomin}^{\rhomax} = \left[  \left(
1- \frac{1}{\rho^{8}} \right) \rho^3\, (1-\chi^2) \frac{\rho^2
\dot{\chi}}{\sqrt{ 1-\chi^2 + \rho^2 \dot{\chi}^2  }} \frac{\partial
\chi}{\partial \chi_0} \right]_{\rhomin}^{\rhomax} \,.
\labell{iprime} \ee
The contribution at $\rho=\rhomin$ clearly vanishes because
$\rhomin=1$ and $\dot{\chi} (1) =0$. Asymptotically we have
\be \chi = \frac{m}{\rho} + \frac{c}{\rho^3} + {\cal O}(\rho^{-4})
\,, \ee
and therefore
\be \frac{\partial \chi}{\partial \chi_0} = \frac{m'}{\rho} +
\frac{c'}{\rho^3} + {\cal O}(\rho^{-4}) \,. \ee
Substituting this into \eqn{iprime} we find
\be
\ids' = \left[ -m m' \rho^2 + m^3 m' -3c m' -m c' + {\cal O}(\rho^{-1})
\right]_{\rho=\rhomax} \,.
\ee
The derivative of the boundary action is just
\be
\ict' = m m' \rhomax^2  + m c' + c m' - m^3 m' \,,
\ee
so adding everything together we arrive at a simple result in the
limit in which the regulator is removed:
\be {\ids}' \equiv \frac{\partial \ids}{\partial \chi_0} = -2 c m'
\,. \ee
This formula is useful for a number of reasons. First, it shows that
$\ids'$ vanishes if and only if $m'$ vanishes, as we wanted to see.
Second, applying the chain rule we find
\be
\frac{\partial \ids}{\partial m} = -2 c  \,.
\ee
Physically we expect that $c<0$ always, because the brane is
attracted to the horizon, and this is confirmed by our numerical
results. It then follows that $\ids$ is an increasing function of
$m$, or equivalently a decreasing function of $1/m$, and hence that
the entropy is positive. Third, it provides an alternative
expression for $\ids$, namely
\be \ids(\chi_0) = -\frac{1}{2} - 2 \int_0^{\chi_0} dx \, c(x) m'(x)
\,, \labell{cow}\ee
where we have imposed the boundary condition $\ids(\chi_0=0) =
-1/2$, which follows from a straightforward calculation of the
action of the equatorial embedding. This expression can be used to
evaluating $\ids$ numerically. Moreover, close to criticality one
knows the analytic form of $m(\chi_0)$ and $c(\chi_0)$, which should
allow one to compute $\ids(\chi_0)$ analytically.

%%%%%%%%%%%%%%%%%%%%%%%%%%%%%%%%%%%%%%%%%%%%%%%%%%%%%%%%%%%%%%%%%%%%%%%%
%%%%%%%%%%%%%%%%%%%%%%%%Constituent quark mass%%%%%%%%%%%%%%%%%%%%%%%%%%
%%%%%%%%%%%%%%%%%%%%%%%%%%%%%%%%%%%%%%%%%%%%%%%%%%%%%%%%%%%%%%%%%%%%%%%%

\section{Constituent quark mass in the D3/D7
system}\label{constitution}

In this section we compute the constituent quark mass $\mc$ for
temperatures below and near the critical temperature for the D3/D7
brane system. A similar analysis has already been provided in
\cite{herzog}.

Our holographic dictionary relates the quark mass $\mq$ to the
asymptotic constant $m$ with eq.~\reef{mqD3D7}. However this is the
bare mass parameter appearing in the microscopic Lagrangian of the
gauge theory. We must expect the physical or constituent mass of a
free quark in the deconfined plasma to receive thermal corrections.
Since a free quark corresponds to a string in the D3-brane geometry
hanging from a probe D7-brane (Minkowski embedding) down to the
horizon, the constituent quark mass corresponds to the energy of
this configuration.

In the notation of the metric \reef{D3geom}, \reef{met1}, the string
worldsheet is extended in the $t,R$ directions, localized at $r=0$,
with induced metric:
\beq ds^2 = -\frac{1}{2} \left(\frac{u_0 R}{L}\right)^2
\frac{f^2}{\tilde f}dt^2 + \frac{L^2}{R^2}dR^2 . \eeq
The Nambu-Goto string action then becomes
\beq I_\mt{string} = -\frac{u_0}{2\pi \ls^2} \int dt dR\, f/\sqrt{2
\tilde{f}}\, , \eeq
where, since $r=0$, $f=1-1/R^4$ and  $\tilde{f}=1+1/R^4$.
Identifying the constituent quark mass with minus the action per
unit time of this static configuration, we have
\beq
\mc = \frac{u_0}{2\pi \ls^2 \sqrt{2}} \int_1 ^{R_0}  dR
\left( 1-\frac{1}{R^4}\right) \left( 1+\frac{1}{R^4}\right)^{-1/2} =
\frac{u_0}{2\pi \ls^2 \sqrt{2}}
\left[ R_0 \sqrt{1+\frac{1}{R_0^4}} - \sqrt{2} \right] \,,
\eeq
where we recall that $R_0=R(r=0)$ is the minimal radius reached by
the probe brane. Given the definition \reef{mqD3D7} for the bare
quark mass, we find that
\beq
{\mc \over \mq} = \frac{1}{m} \left[ R_0 \sqrt{1+\frac{1}{R_0^4}}
- \sqrt{2} \right] \,. \labell{mc/mq}
\eeq
Plots of $\mc / \mq$ versus $T/\mbar$ are given in figure
\ref{constQmass}. In the vicinity of the critical solution, there
are again multiple embeddings for a fixed value of $T/\mbar$ and so
the plots of $\mc$ show an oscillatory behaviour in this regime.
From eq.~\reef{mc/mq}, it is clear that as we approach the critical
solution, \ie $R_0 \to 1$, the constituent quark mass goes to zero.
Note however that the phase transition occurs at $T/\mbar \simeq
0.7658$, which corresponds to $R_0 \simeq 1.15$ -- which is marked
with the vertical dotted line in fig.~\ref{constQmass}. Hence the
exotic behaviour in the vicinity of the critical solution will again
not be manifest in the physical system.

As the temperature goes to zero, $\mc/\mq \to 1$. This is expected,
since for small temperatures we have $m \gg 1$ and we can use the
approximate relation \eqn{inverted} in \reef{mc/mq} to find
\beq {\mc \over \mq} \simeq 1 - \frac{\sqrt{2}}{m} + \frac{1}{2 m^4}
- \frac{5}{8m^8}+ \cdots \,. \eeq
Since $m=2 \mq /\sqrt{\lambda}T$, this can be finally converted into
\be {\mc \over \mq} \simeq 1 - \frac{\sqrt{\lambda} T}{\sqrt{2}\mq}
+ \frac{1}{2} \left( \frac{\sqrt{\lambda} T}{2 \mq} \right)^4 -
\frac{5}{8} \left( \frac{\sqrt{\lambda} T}{2 \mq} \right)^8+ \cdots
\,. \ee
The same expansion appears in \cite{herzog} but here we have
provided an analytic derivation for the coefficient of the fourth
term, which was obtained in \cite{herzog} by a numerical fit. Note
that the two expansions precisely coincide, however, one must
replace $\lambda\rightarrow 2\lambda$ above because \cite{herzog}
uses a different normalization for the 't Hooft coupling. This
difference arises from the implicit normalization of the $U(\nc)$
generators: Tr($T_a\,T_b)=d\,\delta_{ab}$. The standard field theory
convention used in \cite{herzog} is $d=1/2$ while our choice is
$d=1$, as is prevalent in the D-brane literature.
\FIGURE{
\includegraphics[width=\textwidth]{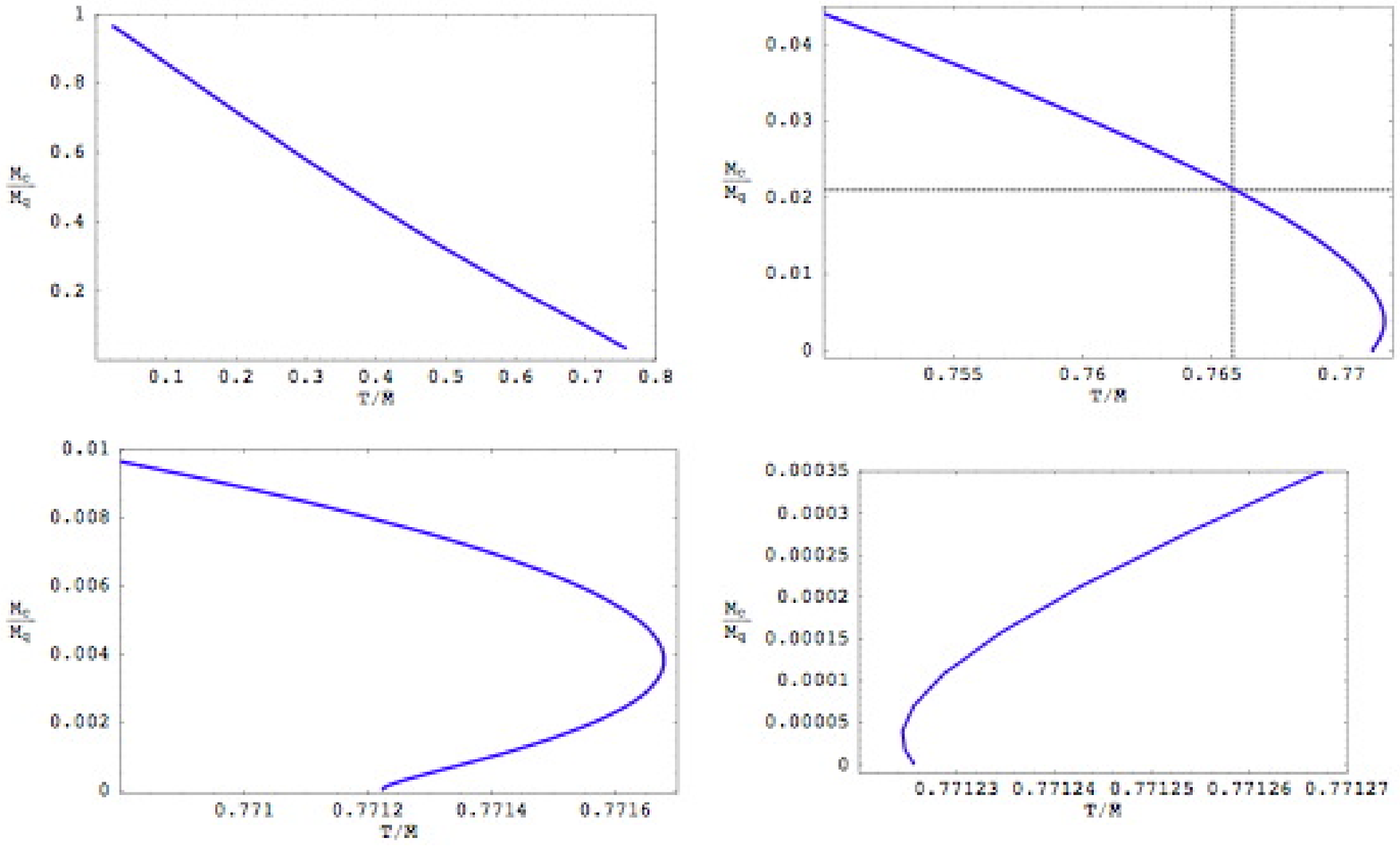} 
\caption{Constituent quark mass  $\mc / \mq$ as a function of
temperature $T/\mbar$. The vertical dotted line indicates the
temperature of the phase transition while the horizontal line
indicates that the constituent quark mass is roughly $\mc/\mq \simeq
0.0212$ at the phase transition.  Some plots zooming in on the spiral
behaviour for temperatures slightly above the transition temperature
are also shown.} \label{constQmass}}
For temperatures above the phase transition, the branes fall into
the horizon and so naively the constituent quark mass vanishes.
Rather it is probably inappropriate to speak in terms of free quarks
in this strongly coupled phase.

\section{Holographic Renormalization of the D4-brane} \label{holo4}

Gauge/gravity duality was originally extended to non-AdS backgrounds
in \cite{itz}. However, until recently the discussion of the
boundary counter-terms needed for holographic renormalization
\cite{revue} was largely limited to asymptotically AdS backgrounds.
It was shown that these techniques can also be applied in
backgrounds describing cascading gauge theory \cite{cascade}. In
principle, we believe it should be possible to extend these
techniques to general gauge/gravity dualities, in some sense by
definition to complete the holographic framework. Here we discuss
the construction of surface terms which will regulate the Euclidean
action of a black Dp-brane throat geometry. Again, while formally
this may be problematic as generally the supergravity description is
breaking down in the asymptotic region, some such approach should be
possible if we believe a gauge/gravity duality exists. We begin with
discussion on the D4-brane background since this is an interesting
place given that it lifts to an (asymptotically) AdS$_7\times S^4$
background for which the counter-terms are known. Hence in
principle, all we have to do is dimensionally reduce the latter to
express them in terms of the D4-brane description. Given our results
for the D4-brane, we make some brief comments on the general
Dp-brane backgrounds.

Let us begin by introducing the Euclidean background for a black
D4-brane:
\beqa ds^2&=&\left({r\over L}\right)^{3/2}\left(
f(r)\,d\tau^2+d\vec{x}^{\,2}\right)+\left({L\over
r}\right)^{3/2}\left({dr^2\over
f(r)}+r^2d\Omega^2_4\right) \labell{backgd}\\
&&C^{(4)}_{\tau 1234}=-i\,\left({r\over L}\right)^3\ , \qquad
e^{2\Phi}=\left({r\over L}\right)^{3/2} \,, \nonumber \eeqa
where
\beq  f(r)=1-{\om^4\over r^4} \labell{eff}\eeq
and the metric above is in string frame. Recall  that the
temperature \reef{period} and the holographic relations
\reef{relatif} for the dual five-dimensional gauge theory are given
in section \ref{D4D6phase}.

The string-frame geometry lifts to eleven dimensions as usual:
\beq ds^2_{11}=e^{-2\Phi/3}(ds^2_{10}) + e^{4\Phi/3}dz^2 \,,
\labell{mlift}\eeq
which for \reef{backgd} yields
\beqa ds^2&=&{r\over L}\left(
f(r)\,d\tau^2+d\vec{x}^{\,2}+dz^2\right)+\left({L\over
r}\right)^{2}{dr^2\over
f(r)}+L^2d\Omega^2_4\nonumber\\
&=&\left({u\over \tiL}\right)^2\left(
f(u)\,d\tau^2+d\vec{x}^{\,2}+dz^2\right)+\left({\tiL\over
u}\right)^{2}{du^2\over f(u)}+L^2d\Omega^2_4 \,,
\labell{mbackgd}\eeqa
with $r/L=(u/\tiL)^2$, $f(u)=1-(\tom/u)^6$, $\tiL=2L$ and
$\tom^2=4L\om$. For later discussion, it will be convenient to
express the throat geometry as
\beq
ds^2_{11}=e^{-2\Phi/3}\left(ds^2_{(p+2)-throat}+
e^{2\sigma}L^2d\Omega^2_{8-p}\right)
+ e^{4\Phi/3}dz^2 \,. \labell{mplift}\eeq
Here the geometry described by $ds^2_{(p+2)-throat}$ replaces the
AdS space, while the $(p+2)$-dimensional field $e^{2\sigma}$
describes the running of the internal $S^{8-p}$, and $L$ is the scale
which we will use to replace the AdS scale.

Now in the standard holographic story for AdS/CFT one refers to the
gravity action in the same dimension at the AdS space. A similar
reduction can be done for the Dp-brane throats, \ie integrate out
the internal $S^{8-p}$, however it seems more natural to think of
them as ten-dimensional geometries. Therefore we will consider the bulk
action and the Gibbons-Hawking surface term are in terms of the full
ten (or eleven) dimensions. Note that in the usual AdS$_n\times S^m$
examples, these contributions to the action are identical in $n$ and
$n+m$ dimensions. In particular, note that the sphere factor is
constant and so it does not contribute to the extrinsic curvature in
the Gibbons-Hawking term. So the relevant bosonic terms in the
Euclidean actions are:
\beqa I_{bulk}&=&-{1\over16\pi G_{11}}\int
d^{11}x\sqrt{G}\left(R(G)-{1\over2\cdot4!}(F^{(4)})^2\right)
\labell{mbulk}\\
&=&-{1\over16\pi G_{10}}\int
d^{10}x\sqrt{g}\left[e^{-2\Phi}\left(R(g)+4(\nabla\Phi)^2\right)
-{1\over2\cdot4!}(F^{(4)})^2\right] \,,
 \labell{iibulk}\eeqa
where we have only kept the terms needed to evaluate the action for
the above solution. Also we have $16\pi G_{11}=(2\pi)^8\ell_P^9$ and
$16\pi G_{10}=(2\pi)^7\ls^8\gs^2=(2\pi)^7\ell_P^9/R_{11}$. One subtlety is
that these two bulk actions are only equal up
to an integration by parts. As surface terms play an important role
in the following, we must keep track of this term. So in reducing
the M-theory action to the IIA action, one picks up an additional
surface term:
\beq -{1\over 8\pi G_{10}}\oint
d^{9}x\sqrt{h}\,{14\over3}e^{-2\Phi}\,n\cdot\nabla\Phi \,,
\labell{bow1}\eeq
where $h_{ab}$ denotes the boundary metric in string frame and $n$
is a unit radial vector. Note that the norm of the latter is fixed
by the ten-dimensional string-frame metric. Now we also need the
Gibbons-Hawking surface term, which in eleven dimensions is:
\beqa I_{GH}&=&-{1\over8\pi G_{11}}\oint d^{10}x\sqrt{H}\,K_{11}(G)
\labell{mgh}\\
&=&-{2\pi R_{11}\over8\pi G_{11}}\oint
d^{9}x\sqrt{h}\,e^{-2\Phi}\left(K_{10}(g)-{8\over3}\,n\cdot\nabla\Phi\right)
\,. \labell{iigha}\eeqa
Combining the two ten-dimensional surface terms yields
\beq I'_{GH}= -{1\over8\pi G_{10}}\oint
d^{9}x\sqrt{h}\,e^{-2\Phi}\left(K_{10}(g)
+2\,n\cdot\nabla\Phi\right) \,. \labell{iigh}\eeq
Note that for the D4 throat geometry, the internal $S^4$ varies with
the radial position, and so the full ten-dimensional geometry contributes to
$K_{10}(g)$. Hence part of the role of the additional term
proportional to the radial gradient of $\Phi$ is to cancel the
sphere contribution, as the four-sphere does not contribute in the
M-theory calculation. One can check that the `unexpected' dilaton
term in eq.~\reef{iigh} arises from transforming the standard
gravity action from Einstein to string frame.

Now the construction of the remaining boundary counter-terms
requires a Kaluza-Klein reduction from ten dimensions \cite{new}.
For the case of the D4-brane, we can in principle simply
dimensionally reduce the counter-terms for AdS$_7$, which include a
constant or volume term, as well as terms proportional to $\cal R$
(the intrinsic curvature) and ${\cal R}^2$. However, we only want to
consider the D4-brane in Poincar\'e coordinates, \ie we consider the
dual field theory in a flat background geometry. Hence the intrinsic
curvature contributions will vanish and we need only consider the
volume term. Note that the prefactor for the AdS$_7$ counter-terms
involves $(8\pi G_7)^{-1}$ and so we can think that this arose from
dimensionally reducing over the internal $S^4$. Hence we write the
counter-term as:
\beqa I_{ct}&=&{1\over8\pi
G_{11}}\int_{S^4}d^4x\sqrt{\gamma}\,\oint_{\partial(AdS_7)}d^6x\sqrt{H}\,{5\over\tiL}
\nonumber\\
&=&{1\over8\pi G_{11}}\Omega_4\,L^4\,\oint_{\partial
M}d^5x\sqrt{h}\,2\pi
R_{11}\,{5\over2L}\,\left(e^{2\sigma-2\Phi/3}\right)^{4/2}
\left(e^{4\Phi/3}\right)^{1/2} \left(e^{-2\Phi/3}\right)^{5/2}
\nonumber\\
&=&{5\over2}\,{\Omega_4\,L^3\over8\pi G_{10}}\,\oint_{\partial
M}d^5x\sqrt{h}\,e^{4\sigma}\,e^{-7\Phi/3} \,. \labell{iict} \eeqa

So now given the background \reef{backgd}, one calculates the
Euclidean action $I_E$ as the sum of the three terms above in
eqs.~\reef{iibulk}, \reef{iigh} and \reef{iict}. As usual we divide
out by the spatial volume  (see footnote \ref{foot1}), in which
case all of the thermodynamic quantities are actually densities.
In this way we arrive at
\beq I_E=-{\Omega_4 L^4\over16\pi G_{10}}{\beta\om^3\over 2L^4} =-
{2^{10}\pi^7\over3^7G_{10}}{L^9\over\beta^5}=-{2^{5}\pi^2\over3^7}\,
\lambda\,\nc^2\,T^5\,,
 \labell{eucI}\eeq
which yields the free energy density given in eq.~\reef{free4}. One
can also check that this result matches that for a planar AdS$_7$
black hole \cite{ct1}.

Now one can probably extend the counter-term above to general
Dp-brane throats. The prefactor for the $(n-1)$-dimensional
counter-terms in AdS$_n\times S^m$ examples involves $(8\pi
G_n\tiL)^{-1}$. Hence we have implicitly dimensionally reduced over
the internal $S^m$ and it seems natural that, for the Dp-branes,
the prefactor involve $\Omega_{8-p} L^{7-p}/(8\pi G_{10})\,
e^{(8-p)\sigma}=(8\pi G_{p+2}L)^{-1}\,e^{(8-p)\sigma}$. Then it
seems the general rule should be that the counter-term takes the
form
\beq I_{ct}={A\over 8\pi G_{p+2}\,L}\oint_{\partial M} d^{p+1}x
\sqrt{h}\, e^{(8-p)\sigma}\,e^{B\Phi} \,, \labell{generalpct}\eeq
where we have written the boundary metric in the string frame, as
read off from the ten-dimensional or ($p+2$)-dimensional
string-frame metric, \ie $ds^2_{(p+2)-throat}$ in eq.~\reef{mplift}.
Then $A$ and $B$ are dimensionless constants which are chosen
experimentally to cancel the relevant divergence coming from the
bulk and Gibbons-Hawking contributions to the action.

\end{document}